\documentclass{JHEP3}
\usepackage{epsfig}
\usepackage{amssymb}
\usepackage{amsmath}

\newcommand{\om}{\omega}
\newcommand{\al}{\alpha}
\newcommand{\ep}{\epsilon}

\newcommand{\lb}{\lbrack}
\newcommand{\rb}{\rbrack}

\newcommand{\msc}[1]{\mbox{\scriptsize #1}}

\newcommand{\br}{\mathbb{R}}
\newcommand{\bz}{\mbox{{\bf Z}}}

\newcommand{\bsz}{\mathbb{Z}}

\newcommand{\cA}{{\cal A}}

\newcommand{\cN}{{\cal N}}
\newcommand{\cM}{{\cal M}}

\newcommand{\cR}{{\cal R}}
\newcommand{\cC}{{\cal C}}
\newcommand{\cQ}{{\cal Q}}

\newcommand{\cH}{{\cal H}}

\newcommand{\tm}{\widetilde{m}}

\newcommand{\ket}[1]{{|#1\rangle}}
\newcommand{\bra}[1]{{\langle#1|}}
\newcommand{\dket}[1]{{\left.\left|#1\right\rangle\right\rangle}}
\newcommand{\dbra}[1]{{\left\langle\left\langle#1\right|\right.}}

\renewcommand{\Im}{\mbox{Im}}

\renewcommand{\Re}{\mbox{Re}}

\newcommand{\nn}{\nonumber\\}

\newcommand {\eqn}[1]{(\ref{#1})}

\newcommand{\bea}{\begin{eqnarray}}
\newcommand{\eea}{\end{eqnarray}}
\newcommand{\f}{\frac}
\newcommand{\dd}{\mbox{d}}
\preprint{SNUST 050601\\ UT-05-07 \\
{\tt hep-th/0507040}}
\title{D-Brane Propagation in
Two-Dimensional Black Hole Geometries}
\author{Yu Nakayama${}^{a}$,
Soo-Jong Rey ${}^{b}$ \& Yuji Sugawara${}^{a}$
 ~~~~~~~~~~~\\
${}^{a}$ Department of Physics, Faculty of Science,
University of Tokyo\\
7-3-1 Hongo, Bunkyo-ku, Tokyo 113-0033  {\rm JAPAN}\\
${}^{b}$ School of Physics \& Center for Theoretical Physics\\
Seoul National University, Seoul 151-747 {\rm KOREA}\\
~~~~~~~~~~~~~~~~\\
\email{\tt nakayama@hep-th.phys.s.u-tokyo.ac.jp, \hskip0.5cm
 sjrey@snu.ac.kr, \hskip0.5cm sugawara@hep-th.phys.s.u-tokyo.ac.jp}}
\abstract{ We study propagation of D0-brane in two-dimensional
Lorentzian black hole backgrounds by the method of boundary
conformal field theory of $SL(2, \mathbb{R})/U(1)$ supercoset at
level $k$. Typically, such backgrounds arise as near-horizon
geometries of $k$ coincident non-extremal NS5-branes, where $1/k$
measures curvature of the backgrounds in string unit and hence size
of string worldsheet effects. At classical level, string worldsheet
effects are suppressed and D0-brane propagation in the Lorentzian
black hole geometry is simply given by the Wick rotation of D1-brane
contour in the Euclidean black hole geometry. Taking account of
string worldsheet effects, boundary state of the Lorentzian D0-brane
is formally constructible via Wick rotation from that of the
Euclidean D1-brane. However, the construction is subject to
ambiguities in boundary conditions. We propose exact boundary states
describing the D0-brane, and clarify physical interpretations of
various boundary states constructed from different boundary
conditions. As it falls into the black hole, the D0-brane radiates
off to the horizon and to the infinity. From the boundary states
constructed, we compute physical observables of such radiative
process. We find that part of the radiation to infinity is in
effective thermal distribution at the Hawking temperature. We also
find that part of the radiation to horizon is in the Hagedorn
distribution, dominated by massive, highly non-relativistic closed
string states, much like the tachyon matter. Remarkably, such
distribution emerges only after string worldsheet effects are taken
exactly into account. From these results, we observe that nature of
the radiation distribution changes dramatically across the conifold
geometry $k=1$ ($k=3$ for the bosonic case), exposing the `string -
black hole transition' therein.}
\begin{document}

\section{Introduction and Summary}

An important yet unsolved problem in string theory is an {\sl ab
initio} formulation of dynamics in time-dependent closed string
background. Cosmological or black hole backgrounds are outstanding
situations of this sort. In these situations, because there is no
globally definable timelike Killing vector, there always arise
Bogoliubov excitation of the vacuum, leading to cosmological
particle production and Hawking radiation. At present, however,
full-fledged string theoretic description of the these processes is
unavailable. Open string counterpart turned out more manageable.
Decay of unstable D-brane, described by open string tachyon rolling
\cite{Sen-RT}, is amenable to exact conformal field theory approach,
thus studied extensively in recent years. Still, some issues are
left out unsettled, especially, ambiguity in prescribing timelike
conformal field theories and the open string vacuum thereof.

One may hope to find the situation better for the well-known
two-dimensional black hole, which admits exact conformal field
theory description \cite{2DBH}, and to understand physics inside the
horizon and the spacelike singularity. For such purposes, we have
learned through a variety of other situations that D-brane serves as
a better probe than closed string. Our goal is to investigate
dynamics of D0-brane propagating in the two-dimensional black hole
geometries.\footnote{Some years ago, \cite{DVV} studied extensively
closed string propagation in these backgrounds.} In this paper, we
shall take the first step toward the goal: understanding D0-brane
dynamics in the causal region outside the horizon, with particular
focus on large curvature regime, where the string worldsheet effects
become strong. The D0-brane serves as a local probe of the black
hole geometries and, in a certain sense, may be considered as an
analog of ZZ-brane in the Liouville theory.

The two-dimensional black hole is often considered as a toy model of
more realistic higher-dimensional black holes. This is not
necessarily so, since the background is intimately related to the
black NS5-branes, whose background \cite{Horowitz:1991cd} is given by
\begin{eqnarray}
&& \hspace{-5mm} \dd s^2 = -\left(1-\frac{r_0^2}{r^2}\right) \dd t^2
+ \left(1+\frac{k\al'}{r^2}\right) \left(\frac{\dd
r^2}{1-\frac{r_0^2}{r^2}} +r^2 \dd \Omega_3^2\right)+ \dd {\bf
y}^2_{\mathbb{R}^5}~,  ~~~
e^{2\Phi(r)} = g_s^2 \left(1+\frac{k\al'}{r^2}\right)
\label{blackNS5}
\end{eqnarray}
along with $k$-units of NS-NS $H_3$-flux penetrating through
$\mathbb{S}^3$. Thus, $k$ refers to the number of NS5-branes at
$r=0$, $r=r_0$ is the location of the event horizon,  ${\bf y}$ are
the spatial coordinates of the planar NS5-brane worldvolume, and
$g_s$ is the string coupling constant at infinity. Since the
NS5-brane is black, it breaks space-time supersymmetries completely
and Hawking radiates.

One hopes to gain intuition by studying classical D0-brane dynamics
near the horizon. So, consider taking various near-horizon limits of
the black NS5-brane \eqn{blackNS5}. One type of near-horizon limit
is $r_0 \rightarrow 0$ and $g_s \rightarrow 0$ independently,
leading to the `throat geometry' of extremal NS5-branes
\cite{NS5brane,CHS}:
\begin{eqnarray}
\hspace{-5mm} \dd s^2 = - \dd t^2 + k\al' \dd \rho^2 + k\al' \dd
\Omega_3^2 + \dd {\bf y}_{\mathbb{R}^5}^2 ~, \qquad {\Phi} = -\rho +
\text{constant}\ , \label{NH ext NS5}
\end{eqnarray}
where $r  = \sqrt{k\alpha'} \exp \rho$. This background is
describable by the exact conformal field theory involving linear
dilaton and $SU(2)_k$ super Wess-Zumino-Witten (WZW)
model:\footnote{Here, $k$ is the level of total current of super SU(2) WZW models and $\sqrt{\frac{2}{k}}$ is the amount of background charge for linear dilaton system.}
\bea \Big[ \mathbb{R}_t \times \mathbb{R}_{\rho , \sqrt{2\over k}}
\times SU(2)_k \Big]_\perp \times \Big[\mathbb{R}^5 \Big]_{||}.
\nonumber \eea
The first part describes the five-dimensional curved spacetime
transverse to the NS5-brane, while the second part describes the
flat spatial directions parallel to NS5-brane. The criticality
condition is satisfied for any k because
\bea \left( 1 + \frac{6}{k} + \frac{1}{2}\right) + 3 \times \left(
\frac{k-2}{k} + \frac{1}{2}\right)+6\times \left(1 + \frac{1}{2}
\right) =15 \label{crit} \eea
D-brane dynamics in this background was studied in
\cite{Kutasov,Kutasov2} via the Dirac-Born-Infeld (DBI) approach,
and observed that it strikingly resembles the rolling tachyon of
unstable D-brane in ambient flat spacetime
\cite{Sen-RT,Strominger,LNT}. This map, which we refer as
`radion-tachyon correspondence', offers a useful guide for
understanding D-brane propagating in curved spacetime geometry in
terms of known results regarding rolling tachyon dynamics.\footnote{Subsequent works along the same line include {\em e.g.}
\cite{radion-DBI}.} Introducing `tachyon' variable $X = \rho$, DBI
Lagrangian of the D0-brane propagating in the background \eqn{NH ext
NS5} is recastable to that of rolling tachyon:
\bea L_{\rm D0} = - e^{-\Phi} \sqrt{ \left( {\dd s \over \dd
t}\right)^2} = - V(X) \sqrt{1 - \dot{X}^2} \qquad \mbox{where}
\qquad V(X) = M_{0} \, e^X. \label{extremal D0} \eea
The energy is conserved, so solving $V(X) / \sqrt{1 - \dot{X}^2} =
1$, we obtain D0-brane's geodesic\footnote
 {Throughout this work, we refer by `geodesic' of D0-brane the classical 
trajectory as determined by the DBI action. Recall that equivalence 
principle does not hold in string theory due to the dilaton coupling --- 
motion of D-brane is different from that of fundamental string, the latter 
being studied for example in \cite{Bars} (See also footnote 10.). We would like to thank 
   Itzhak Bars for bringing our attention to these references.
} 
as
\bea e^X = {e^{\rho_o} \over \cosh (t - t_o)} \qquad \mbox{viz.}
\qquad e^{\rho} \cosh (t - t_o) = e^{\rho_o}. \label{extremalorbit}
\eea
This is the Lorentzian counterpart of so-called `hairpin' profile.
In the previous works \cite{NST, NPRT}, we constructed exact
boundary states of the D-brane and analyzed rolling dynamics of the
D-brane in detail.\footnote{Related analysis via boundary conformal
theory was given in \cite{Sahakyan,CLS,LapanLi}.} A technical
difficulty was that the dilaton blows up at the core, hampering
further analysis by the strong coupling singularity.

Another type of near-horizon limit is $r_0 \rightarrow 0$ and $g_s
\rightarrow 0$ while keeping the energy density above the extremal
configuration $\mu \equiv {r_0^2}/{g_s^2 \al'}$ fixed. It yields
`throat geometry' of the near-extremal NS5-branes \eqn{blackNS5}
\cite{MS-blackNS5}:
\begin{eqnarray}
&&\hspace{-5mm} \dd s^2 = - \tanh^2\rho \, \dd t^2 + k\al' \dd
\rho^2 + k\al' \dd \Omega_3^2 + \dd {\bf y}_{\mathbb{R}^5}^2 ~,
\qquad e^{2\Phi} = \frac{k}{\mu \cosh^2 \rho} ~, \label{NH black
NS5}
\end{eqnarray}
where $r= r_0\cosh \rho$. For $(t,\rho)$-subspace, the metric and
the dilaton coincide with those of the two-dimensional black hole.
This Lorentzian black hole is describable by Kazama-Suzuki
supercoset conformal field theory $SL(2; \br)_k / U(1)$ (where
$U(1)$ subgroup is chosen to be the non-compact component
(space-like direction)) of central charge $c=3(1+2/k)$. Likewise,
taking account of the NS-NS $H_3$-flux penetrating through
$\mathbb{S}^3$ which is omitted in (\ref{NH black NS5}), the angular
part $\mathbb{S}^3$ is describable by the (super) $SU(2)$-WZW model.
In this way, the string background of the nonextremal NS5-brane is
reduced to a solvable superconformal field theory system:\footnote
  {Here again, $k$ is the level of total current of super WZW models.
   Namely, $k+2$, $k-2$ are the levels of bosonic
    $SL(2)$ and $SU(2)$ currents.}
\begin{eqnarray}
\Big[ {SL(2;\br)_{k} \over U(1)} \times SU(2)_{k} \Big]_\perp \times
\Big[\, \mathbb{R}^5 \, \Big]_{||}~. \label{SCFT black NS5}
\end{eqnarray}
Here, the first part describes the five-dimensional curved spacetime
(including the time direction) transverse to the NS5-brane, while
the second part describes the flat spatial directions parallel to
the NS5-brane. The criticality condition is satisfied for any $k$ as
in \eqref{crit}.
Upon Wick rotation, Euclidean black hole has the `cigar geometry',
described by the coset conformal field theory $SL(2;\br)_{k}/U(1)
\simeq {\mathbb H}^3_+/\br$ (where $U(1)$ subgroup is the compact
component). Asymptotically, circumference of the cigar geometry is
$2\pi \sqrt{\al' k}$, and is identified with inverse of the Hawking
temperature.

Again, by introducing `tachyon' variable $Y \equiv \log \sinh \rho$,
DBI Lagrangian of the D0-brane is castable to that of rolling
tachyon:
\bea L_{\rm D0} &=& - e^{-\Phi} \sqrt{\left({\dd s \over \dd
t}\right)^2} = - V(Y) \sqrt{1 - \dot{Y}^2} \qquad \mbox{where}
\qquad
 V(Y) = M_0 \, e^Y ~. \label{nonextremal D0} \eea
We now find D0-brane's geodesic as
\bea e^Y = {\sinh \rho_o \over \cosh (t-t_o)} \qquad \mbox{viz.}
\qquad \sinh \rho \cosh (t - t_o) = \sinh \rho_o ~. \label{orbit}
\label{nonextremalorbit} \eea
Euclidean counterpart of the geodesic \eqn{orbit} describes
D1-brane profile in the Euclidean
two-dimensional black hole background. An important point is that,
in sharp contrast to the extremal background \eqn{NH ext NS5}, the
dilaton is finite everywhere. Thus, the strong coupling singularity
is now capped off by the horizon.

In both cases, it is elementary to understand classical dynamics of
the D0-brane: both by gravity and by strong string coupling
gradient, D0-brane is pulled in and finds its minimum energy and
rest mass at the location of the NS5-brane. This also fits to the
observation that the spacetime supersymmetry is completely broken as
the NS5-brane and the D0-brane preserve different combinations of
the supercharges. Eventually, D0-brane would melt into fluxes of
NS5-brane worldvolume gauge field and form a non-threshold
bound-state. As the D0-brane is pulled in, acceleration would grow
and radiate off the binding energy into closed string modes.

Interestingly, the DBI Lagrangian of D0-brane is identical for both
extremal and non-extremal NS5-brane backgrounds, see \eqn{extremal
D0} and \eqn{nonextremal D0}. Consequently, the geodesics are also
identical in $X$ or $Y$ coordinate. Does this imply that the
two-dimensional black hole is featureless and not quite black? As we
shall explain in much detail, this coincidence turns out simply an
artifact of DBI analysis and disappears in full-fledged boundary
conformal field theory description for the D0-brane boundary states.
It also hints that the radion-tachyon correspondence (based on DBI
analysis alone) would break down in non-extremal NS5-brane
background, and we will see such indications from exact boundary
conformal field theory analysis.

By construction, D-brane boundary states is obtainable from the
coupling of the D-brane to closed string modes. Semiclassically,
this can be done by overlapping the geodesic \eqn{orbit} to the
mini-superspace wave function of the closed string modes. Thus, in
section 2, after recapitulating the mini-superspace wave function in
the Euclidean black hole background, we shall construct the boundary
state of the Euclidean D1-brane \cite{RibS} by taking the inner product. Being
in Euclidean background, the construction is free from ambiguity.

We are primarily interested in D0-brane's boundary state in
Lorentzian background. Formally, the boundary state is obtainable by
Wick rotation of that for the Euclidean D1-brane, as constructed in
section 2. However, the Wick rotation is not unique and appropriate
analytic continuation has to be specified. Causal region of the
Lorentzian black hole background has four boundaries: past and
future horizons ${\cal H}^\pm$, and past and future asymptotic
infinities ${\cal I}^\pm$. See Figure 1. The specification amounts
to imposing boundary conditions at these four null infinities. In
section 3, with careful treatment of the boundary conditions, we
construct several boundary states of the D0-brane, corresponding to
physically motivated boundary conditions: (1) D0-brane emitted from
past horizon, (2) D0-brane absorbed to future horizon, and (3)
time-symmetric D0-brane. As a useful mnemonic, via the
radion-tachyon correspondence (extended to finite temperature
background), (1) and (2) are the counterpart of half S-brane, while
(3) is the counterpart of full S-brane.

\begin{figure}[htbp]
    \begin{center}
    \includegraphics[width=0.7\linewidth,keepaspectratio,clip]
      {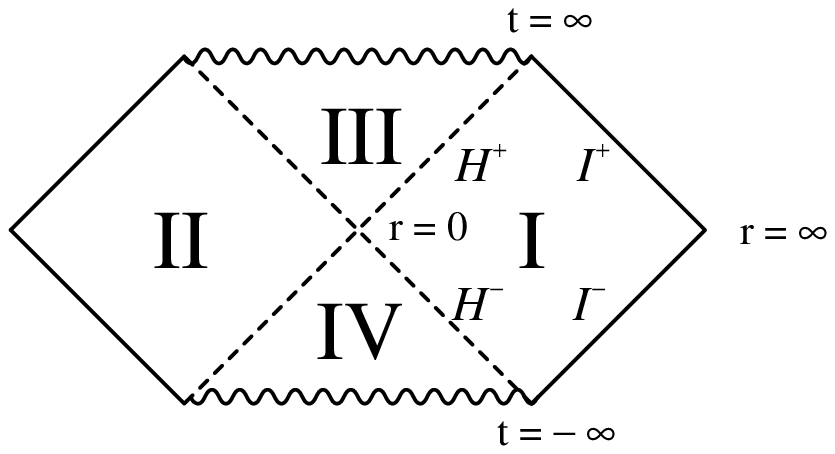}
    \end{center}
    \caption{Penrose diagram of the two-dimensional black hole.
    We focus on the causal region I outside the horizon.
    Boundaries of the region I consist of past and future asymptotic
    infinities ${\cal I}^\pm$ and past and future horizons
    ${\cal H}^\pm$.}
    \label{kruskal}
\end{figure}

The D0-brane propagating in curved spacetime emits closed string
excitations. One expects that in general radiation rate would be
proportional to spacetime curvature. In the eternal black hole
background, where the future-directed asymptotic infinities consist
of ${\cal H}^+$ and ${\cal I}^+$, the radiation emitted by the
D0-brane propagates out to either of them. In section 4, we estimate
radiation distribution for incoming part ${\cal N}_{\rm in}(M)$ (as
measured at ${\cal H}^+$) and outgoing part ${\cal N}_{\rm
out}(M)$(as measured at ${\cal I}^+$) for a fixed transverse mass
$M$. Intuitively, the radiation distribution ought to depend
sensitively on the background curvature scale,\footnote{Throughout
this work, we shall adopt the convention $\al'=2$.} set by the
level $k$. Curiously, we find that the radiation rate depends on the
Hawking temperature! This has to do with the novel feature that the
Hagedorn and the Hawking temperatures, $T_{\rm Hg}$ and $T_{\rm
Hw}$, in the two-dimensional black hole background are
both set by $k$:
\bea T_{\rm Hg} = \frac{1}{\beta_{\rm Hg}} = {1 \over 4\pi \sqrt{1 - {1 \over 2k}}} \qquad
\mbox{and} \qquad T_{\rm Hw} = \frac{1}{\beta_{\rm Hw}} =
\frac{1}{2\pi \sqrt{2k}}.  \nonumber \eea
Especially, the Hawking temperature is independent of the
non-extremality $r_0$ or $\mu$.

Notice that two temperature scales meet at $k=1$, viz. the conifold
geometry, so we anticipate some sort of cross-over or phase
transition. For $k>1$, D0-brane energy is emitted almost all to
incoming mode toward ${\cal H}^+$ and small to outgoing mode toward
${\cal I}^+$:
\bea { {\cal N}_{\rm out}(M) \over {\cal N}_{\rm in}(M)} \sim
{e^{-{1 \over 2} \beta_{\rm Hw} M } \over e^{-{1 \over 2} \beta_{\rm
Hg} M}} \ll 1. \nonumber \eea
For $k<1$, only small portion of the D0-brane energy is emitted,
equally distributed between the incoming and the outgoing parts:
\bea { {\cal N}_{\rm out}(M) \over {\cal N}_{\rm in}(M)} \sim 1.
\nonumber \eea
The distribution does not appear to be the radiation pattern one
would expect in black hole background, where the black hole would
absorb most of the radiation (as well as D0-brane). This brings
out a question: if $k<1$ background were not associated with black
hole, what would it be?\footnote{Similar question has been discussed recently in \cite{KarMS,GKRS}.}

In section 5, we argue that $k=1$ is a critical point of black hole
- string phase transition, where the branching ratio between
incoming and outgoing part of the radiation provides an `order
parameter'. Our interpretation fits nicely with recent observation
of \cite{GKRS}, where the same kind of phase transition was pointed
out for $AdS_3$ and linear dilaton backgrounds. Given that the
NS5-brane background is holographically dual to Little String Theory
(LST), it is natural to ask what the holographic dual of the
D0-brane falling into the NS5-brane. Adopting the proposal
\cite{Sahakyan} that the dual process is identifiable with decay of
a defect or soliton in the LST and taking reasonable assumption
concerning $k$-dependence of LST's decay number distribution and
density of states, we show that the defect decay rate as computed
within LST fits perfectly with the D0-brane radiation rate as
computed in the bulk (viz. black hole background). Interestingly,
the two-dimensional black hole at $k=1/2$ belongs not quite to the
black hole phase but to the far extreme of the string phase. Long
string condensation in two-dimensional string theory was recently
studied via non-singlet matrix model \cite{maldacena}. It would be
interesting to study D0-brane dynamics in this matrix model and
compare with our results.

In section 6, we extend the boundary state analysis to Ramond-Ramond
(R-R) sector by spectral flow, and, in section 7, we also analyze
the limit the NS5-brane becomes extremal, thus making contact with
our earlier results. In section 8, we return to the issue of
boundary conditions and propose yet another physically motivated
one: the Hartle-Hawking boundary condition. This boundary condition
is particularly compelling because a puzzle concerning origin/fate
of the conserved R-R charge of the D0-brane gets around, and also
fits well with the radiation rate computed in section 4.

\section{D1-brane on Euclidean Two-Dimensional Black Hole}

In this section, we study boundary state description of the D1-brane
profile on the Euclidean two-dimensional black hole geometries. We
shall begin with recapitulating aspects of the closed string
spectrum relevant for foregoing analysis.

\subsection{Mini-superspace analysis of closed strings}
Consider the Euclidean two-dimensional black hole background, known
as `cigar geometry':
\begin{equation}
\dd s^2 \equiv G_{ij} \dd x^i \dd x^j = 2k (\dd \rho^2 + \tanh^2\rho
\dd \theta^2) \qquad \mbox{and} \qquad e^{\Phi} =
\frac{e^{\Phi_0}}{\cosh\rho} ~. \label{Euclidean cigar}
\end{equation}
Recall that $k$ sets characteristic curvature radius in unit of the
string scale and hence string worldsheet effects, while $e^{\Phi_0}$
sets the maximum value of the string coupling at the tip $\rho=0$ of
the cigar geometry. We shall assume the limit $k \gg 1$ and
$e^{-\Phi_0} \gg 1$: this limit suppresses both string worldsheet
and spacetime quantum effects and facilitates to truncate closed
string spectrum to zero-modes, viz. to mini-superspace approximation.

In the mini-superspace approach, difference between bosonic strings
(with no worldsheet supersymmetry) and fermionic strings (with
${\cal N}=2$ worldsheet supersymmetry) becomes unimportant. The
closed string Hamiltonian $L_0 + \overline{L}_0$ is reduced in the
mini-superspace approximation to the target space Laplacian
$\Delta_0$, where:
\begin{eqnarray}
\Delta_0 &=& \frac{1}{e^{-2\Phi} \sqrt{G}} \partial_i
\left(e^{-2\Phi} \sqrt{G} G^{ij} \partial_j\right) \equiv
-\frac{1}{2k}[\partial_\rho^2 +2\coth2\rho\partial_\rho +
\coth^2\rho\partial_\theta^2] ~. \label{Laplacian}
\end{eqnarray}
The Hamiltonian is defined with respect to the volume element:
\begin{eqnarray}
\dd \mbox{Vol} = e^{-2\Phi}\sqrt{G} \dd \rho \dd \theta := {2k}
\sinh \rho \cosh \rho \dd \rho \, \dd \theta \equiv k \sinh 2\rho
\dd \rho \, \dd \theta~, \label{vol cigar}
\end{eqnarray}
inherited from the Haar measure on the $SL(2;\br)$ group manifold.
In the volume element, the dilaton factor $e^{-2\Phi}$ is taken into
account, as the inner product for closed string states is defined by
the worldsheet two-point correlators on the sphere. The normalized
eigenfunctions are obtained straightforwardly \cite{DVV,RibS}. They
are:
\begin{equation}
\phi_n^j(\rho,\theta) =
-\frac{\Gamma^2(-j+\frac{|n|}{2})}{\Gamma(|n|+1)\Gamma(-2j-1)}
e^{in\theta} \left[ \sinh^{|n|}\rho \cdot
F\left(j+1+\frac{|n|}{2},-j+\frac{|n|}{2};|n|+1;-\sinh^2\rho
\right)\right] ~ , \label{ef}
\end{equation}
where $F(\alpha, \beta; \gamma; z)$ is the Gaussian
hypergeometric function. These eigenfunctions correspond to the
primary state vertex operators of conformal weights
\begin{eqnarray}
&& h= \tilde{h}= -\frac{j(j+1)}{k-2} + \frac{n^2}{4k} \qquad
\mbox{or} \qquad h= \tilde{h}= -\frac{j(j+1)}{k} + \frac{n^2}{4k}
\end{eqnarray}
for bosonic\footnote{The eigenvalue is actually proportional to
$-\frac{j(j+1)}{k}+ \frac{n^2}{4k}$.} and fermionic strings,
respectively. We shall focus on the continuous series, parametrise
the radial quantum number $j$ as $j= -\frac{1}{2}+ i\frac{p}{2}$
$(p\in \br)$, and label the eigenfunctions as
$\phi^p_n(\rho,\theta)$ instead of $\phi^j_n(\rho,\theta) $. Adopt
the convention that, in the asymptotic region $\rho \sim \infty$,
the vertex operators with $p>0$ corresponds to the incoming waves
and those with $p<0$ corresponds to the outgoing waves. The
eigenfunctions \eqn{ef} are then normalized as
\begin{eqnarray}
\Big(\phi^p_n, \phi^{p'}_{n'} \Big) = \delta_{n,n'} \Big\lb 2 \pi
\delta(p-p')+ \cR_0(p',n) \, 2 \pi \delta(p+p') \Big\rb
~,\label{inner product}
\end{eqnarray}
where the inner product is defined with respect to the volume
element \eqn{vol cigar}. Here, $ \cR_0(p,n)$ refers to the
reflection amplitude of the mini-superspace analysis:
\begin{eqnarray}
\cR_0(p,n) =
\frac{\Gamma(+ip)\Gamma^2(\frac{1}{2}-\frac{ip}{2}+\frac{n}{2})}
{\Gamma(-ip)\Gamma^2(\frac{1}{2}+\frac{ip}{2}+\frac{n}{2})} \ .
\label{cref amp}
\end{eqnarray}
That is, from the definition \eqn{ef}, the reflection amplitude is
seen to obey the mini-superspace reflection relation:
\begin{eqnarray}
\phi^{-p}_n(\rho,\theta) = \cR_0(-p,|n|) \,
\phi^{+p}_n(\rho,\theta)~. \label{cref rel}
\end{eqnarray}
We shall refer $\cR_0(p,n)$ as `mini-superspace' reflection
amplitude, valid strictly within mini-superspace approximation at $k
\rightarrow \infty$, and anticipate string worldsheet effects at
finite $k$. Notice that no winding states wrapping around
$\theta$-direction are present since by definition the
mini-superspace approximation retains states with zero winding only.

Utilizing the analytic continuation formula of the hypergeometric
functions:
\begin{eqnarray}
F(\alpha,\beta;\gamma;z) &=&
\frac{\Gamma(\gamma)\Gamma(\beta-\alpha)}
{\Gamma(\beta)\Gamma(\gamma-\alpha)}
(-z)^{-\alpha}F(\alpha,\alpha+1-\gamma;
\alpha+1-\beta;1/z) \cr
&+& \frac{\Gamma(\gamma)\Gamma(\alpha-\beta)}
{\Gamma(\alpha)\Gamma(\gamma-\beta)}(-z)^{-\beta}
F(\beta,\beta+1-\gamma;\beta+1-\alpha;1/z)
\label{eq:inv} \ ,
\end{eqnarray}
the eigenfunction \eqn{ef} is decomposable into
\begin{eqnarray}
\phi^p_n(\rho,\theta) = \phi^p_{L,n}(\rho,\theta) + \cR_0(p,|n|)
\phi^p_{R,n}(\rho,\theta) ~, \label{decomp ef} \end{eqnarray}
where
\begin{eqnarray}
\phi^p_{L,n}(\rho,\theta) &\equiv& e^{in\theta} (\sinh
\rho)^{-1-ip}\,
F\Big(\frac{1}{2}+\frac{ip+n}{2},\frac{1}{2}+\frac{ip-n}{2}; 1+ip;
-\frac{1}{\sinh^2\rho} \Big) ~,\nn & \sim &
e^{-\rho}e^{-ip\rho+in\theta} \qquad \mbox{at} \qquad \rho \,
\rightarrow\, +\infty~ \label{phiL} \end{eqnarray}
and
\begin{eqnarray}
\phi^p_{R,n}(\rho,\theta) &\equiv& e^{in\theta} (\sinh
\rho)^{-1+ip}\,
F\Big(\frac{1}{2}-\frac{ip+n}{2},\frac{1}{2}-\frac{ip-n}{2}; 1-ip;
-\frac{1}{\sinh^2\rho} \Big) \nn & \sim &
e^{-\rho}e^{ip\rho+in\theta} \qquad \mbox{at} \qquad
   \rho \, \rightarrow\, +\infty
\label{phiR}
\end{eqnarray}
refer to the left- and the right-movers, respectively, at $\rho
\rightarrow +\infty$, and $\cR_0(p, |n|)$ is defined in \eqn{cref
amp}. Obviously, they are related to each other under the reflection of
radial momentum: $\phi^{+p}_{R, n} = \phi^{-p}_{L, n}$, which is
also evident from \eqn{decomp ef} and \eqn{cref amp}. These
mini-superspace wave functions \eqn{decomp ef} constitute the
starting point of constructing boundary states of D-brane in the
Euclidean two-dimensional black hole background.

We close the mini-superspace analysis with remarks concerning Wick
rotation of the results to the Lorentzian background and string
worldsheet effects present at finite $k$.
\begin{enumerate}
 \item The decomposition of $\phi^p_n$ into $\phi^p_{L,n}$ and $\phi^p_{R,n}$ is not
 globally definable over the entire cigar geometry. They
 are ill-defined around the tip $\rho =0$, and the reflection relation
\eqn{cref rel} implies that $\phi^{-p}_n$ is not independent of
$\phi^{+p}_n$. Therefore, of the continuous series, only the
eigenfunctions $\phi^p_n$ with $p>0,~ n\in \bz$ span the physical
Hilbert space of the closed strings on the Euclidean two-dimensional
black hole. On the other hand, the situation will become further
complicated once Wick rotated to the Lorentzian two-dimensional
black hole.
 \item Notice that $\phi^p_n$ is not analytic
with respect to the angular quantum number $n$ as it depends on its
absolute value, $|n|$. This leads to the ambiguity for Wick rotation
from Euclidean to Lorentzian background, under which roughly
speaking $i n$ is replaced by energy $\omega$. As for the
mini-superspace reflection amplitude $\cR_0(p, n)$, since
$\cR_0(p,-n)= \cR_0(p,n)$ holds for all $n \in \bz$, it is
unnecessary to take absolute value $|n|$ in \eqn{cref rel},
\eqn{decomp ef}. When taking Wick rotation, we will start from the
expression $\cR_0(p,|n|)$. In other words, we analytically continue
$\cR_0(p,n)$ if $n > 0$ and $\cR_0(p,-n)$ if $n<0$.
 \item It is evident that $|\cR_0(p,n)|=1$, viz, the mini-superspace
 reflection
 amplitude is purely a phase shift in the Euclidean black hole
 background. It is of
 utmost importance that, in the Lorentzian black hole background,
 $n$ is analytically continued to pure imaginary value,
 and the modulus of the reflection amplitude becomes less than unity.
 \item For the fermionic Euclidean $SL(2; \br)/U(1)$ conformal field
 theory, exact result for the reflection amplitude (i.e. taking account
 of all string worldsheet effects) is known
 \cite{Teschner-reflection,GK}. In our notations, it is
\begin{eqnarray}
\cR(j,m,\tm) = \nu(k)^{-2j-1}\,
\frac{\Gamma(1+\frac{2j+1}{k})}{\Gamma(1-\frac{2j+1}{k})}
\frac{\Gamma(2j+1)\Gamma(-j+m)\Gamma(-j-\tm)}{\Gamma(-2j-1)
\Gamma(j+1+m)\Gamma(j+1-\tm)}, \label{qref amp}\eea
where
\bea \nu(k)\equiv \frac{1}{\pi}\frac{\Gamma(1-\frac{1}{k})}
{\Gamma(1+\frac{1}{k})}~, \qquad m=\frac{kw+n}{2}~, \qquad \tm =
\frac{kw-n}{2}~.
 \nn
\end{eqnarray}
Denoting by $\Phi_{j;m,\tm}$ the vertex operator with conformal weights
$h= \frac{m^2-j(j+1)}{k}$, $\widetilde{h}= \frac{\tm^2-j(j+1)}{k}$,
the exact reflection relation reads
\begin{eqnarray}
\Phi_{-(j+1);m,\tm} = \cR(-(j+1),m,\tm) \Phi_{j;m,\tm}~, \label{qref rel}
\end{eqnarray}
The mini-superspace reflection amplitude $\cR_0(p,n)$ is
then related to the exact one $\cR(j,m,\tm)$
by taking the $k\,\rightarrow\,\infty$ limit as mentioned above
(up to overall constant):
\begin{eqnarray}
\cR_0(p,n) = \lim_{k\rightarrow + \infty}\,
\cR(j=-\frac{1}{2}+\frac{ip}{2}, m=\frac{n}{2}, \tm=-\frac{n}{2})~.
\end{eqnarray}
\end{enumerate}

\subsection{Boundary state of Euclidean D1-brane}
We shall now study D1-brane in the Euclidean two-dimensional black
hole background. Classically, profile of the D1-brane follows the
geodesic curve
\begin{eqnarray}
 \cos (\theta-\theta_0) \sinh \rho = \sinh \rho_0~,
\label{hairpin}
\end{eqnarray}
and is known as the `hairpin brane'. Here, $\theta_0$, $\rho_0$ are
free parameters characterizing the geodesic curves. The `hairpin
brane' is obtainable as a descendant of the Euclidean $AdS_2$-brane
\cite{BP} in the Euclidean $AdS_3$ space, described by $SL(2;\br)$
($\mathbb{H}^3_+$) Wess-Zumino-Witten model.
Correspondingly, exact boundary state of the D1-brane was
constructed in \cite{RibS} from the boundary conformal field theory
analysis of the $SL(2;\br)$ Wess-Zumino-Witten model \cite{PST}. See
also the closely related works {\em e.g.}
\cite{LVZ,ES-L,ASY,IPT2,FNP}, and \cite{Nakayama} for a review. For
the case of the Euclidean $SL(2; \br)/U(1)$ supercoset conformal
field theory, the relevant boundary state of the NS-NS sector is
given by
\begin{eqnarray}
{}_{\rm D1}\bra{B;\rho_0,\theta_0} &=& \int_0^{\infty} \frac{\dd
p}{2\pi}\, \sum_{n\in \bsz}\, \Psi_{D1}(\rho_0,\theta_0;p,n) \,
\dbra{p,n}~, \nn
\Psi_{\rm D1}(\rho_0,\theta_0;p,n) &=& {\cal
N}(k)\frac{\Gamma(ip)\Gamma\left(1+\frac{ip}{k}\right)}
{\Gamma\left(\frac{1}{2}+\frac{ip+n}{2}\right)
\Gamma\left(\frac{1}{2}+\frac{ip-n}{2}\right)} \, e^{in\theta_0}
\Big[e^{-ip\rho_0} + (-1)^n e^{ip\rho_0}\Big]~.
\label{hairpin D1}
\end{eqnarray}
Here, $\dbra{p,n}$ refers to the Ishibashi state constructed over
the primary state whose mini-superspace wave function is given by
$\phi^p_n(\rho,\theta)$. Also, ${\cal N}(k)$ is a normalization
factor. Since it would not affect foregoing analysis, we will set it
to $2\pi$ for simplicity. One can readily check that the boundary
wave function \eqn{hairpin D1} is consistent with the exact
reflection amplitude \eqn{qref amp}.

The result \eqn{hairpin D1} can be understood intuitively as
follows. A D-brane boundary wave function is the weighted sum of the
wave function of closed string states restricted to the location of
the D-brane. In the mini-superspace approximation, as is implicit in
\cite{RibS}, the weighted sum equals to the overlap between the
mini-superspace wave function and the delta function constraint
enforcing $(\rho, \theta)$ coordinates over the hairpin trajectory
\eqn{hairpin} (with respect to the volume element \eqn{vol cigar}).
The result is
\begin{eqnarray}
&& \int_0^\infty \sinh \! \rho \, \dd \sinh \! \rho
\int_{-\frac{\pi}{2}+\theta_0}^{\frac{\pi}{2}+\theta_0} \dd\theta\,
\delta \Big(\cos (\theta-\theta_0) \sinh \rho-\sinh \rho_0 \Big)
\phi^p_{n}(\rho,\theta) = \int_{-\frac{\pi}{2}}^{\frac{\pi}{2}}
\!\dd \theta' \, \frac{\sinh \rho_0}{\cos^2 \theta'}
\phi^p_{n}(\widehat{\rho}(\rho_0,\theta'),\theta') e^{i n \theta_0}
, \nonumber
\end{eqnarray}
where $\theta'= (\theta-\theta_0)$ and
$\widehat{\rho}(\rho_0,\theta')$ refers to the solution of $\cos
\theta' \sinh \rho= \sinh \rho_0$. Using the decomposition
\eqn{decomp ef}, we are then to evaluate integrals:
\begin{eqnarray}
 && \int_{-\frac{\pi}{2}}^{\frac{\pi}{2}} \dd \theta \,
\frac{\sinh \rho_0}{\cos^2 \theta} \,
\phi^p_{L,n}(\hat{\rho}(\rho_0,\theta),\theta) =
\frac{2\pi\Gamma(ip)} {\Gamma\left(\frac{1}{2}+\frac{ip+n}{2}\right)
\Gamma\left(\frac{1}{2}+\frac{ip-n}{2}\right)} \, e^{-ip\rho_0} ~,
\nn
 && \int_{-\frac{\pi}{2}}^{\frac{\pi}{2}} \dd \theta \,
\frac{\sinh \rho_0}{\cos^2 \theta} \,
\phi^p_{R,n}(\hat{\rho}(\rho_0,\theta),\theta) =
\frac{2\pi\Gamma(-ip)}
{\Gamma\left(\frac{1}{2}-\frac{ip+n}{2}\right)
\Gamma\left(\frac{1}{2}-\frac{ip-n}{2}\right)} \, e^{+ ip\rho_0} ~.
\label{evaluation overlap phi}
\end{eqnarray}
Details of the computation are relegated in Appendix B. Using the
mini-superspace reflection amplitude \eqn{cref amp}, we then obtain
\begin{eqnarray}
&&\Psi^{(0)}_{\rm D1}(\rho_0,\theta_0;p,n) =  \frac{2\pi\Gamma(ip)}
{\Gamma\left(\frac{1}{2}+\frac{ip+n}{2}\right)
\Gamma\left(\frac{1}{2}+\frac{ip-n}{2}\right)} \, e^{in\theta_0}
\left(e^{-ip\rho_0} + (-1)^n e^{+ip\rho_0}\right)~. \label{hairpin
D1 classical}
\end{eqnarray}
We see the result \eqn{hairpin D1 classical} reproduces the exact
result \eqn{hairpin D1} {\sl modulo the factor}
$\Gamma\left(1+i\frac{p}{k}\right)$. Importantly, this missing
factor depends on $k$ (measured in string unit) and hence
corresponds precisely to the corrections due to string worldsheet
effects. The mini-superspace approximation sets $k \rightarrow
\infty$, so this factor is consistently dropped out. Equivalently,
this missing factor can be reinstated to the D-brane boundary wave
function by demanding consistency of the wave function in the
mini-superspace approximation with the exact reflection amplitude
\eqn{qref amp}.

\section{D0-Brane in Lorentzian Two-Dimensional Black Hole}

\subsection{Analytic continuation of boundary states}

In this section, we shall construct the exact boundary state
describing the D0-brane moving in the Lorentzian two-dimensional
black hole background. Recall that the Lorentzian two-dimensional
black hole (`Lorentzian cigar') background is obtainable by the Wick
rotation $\theta = it$ of the Euclidean one \eqn{Euclidean cigar}
\begin{equation}
\dd s^2 =-2k(\dd \rho^2 - \tanh^2\! \rho  \, \dd t^2) \qquad
\mbox{and} \qquad e^{\Phi} = \frac{e^{\Phi_0}}{\cosh\rho} ~.
\label{Lorentzian cigar}
\end{equation}
Wick-rotating the geodesic of the Euclidean D1-brane, we found the
geodesic of the Lorentzian D0-brane in \eqn{nonextremalorbit} as\footnote
{Another familiar parametrization of the two-dimensional
black hole is the analogue of the Kruskal coordinates
$$
u= \sinh \rho e^t ~, ~~~ v=-\sinh \rho e^{-t} ~, ~~~ \dd s^2 = -{2k}
\frac{\dd u \dd v}{1-uv}~,
$$
  and the geodesic \eqn{trajectory D0} is just a straight line
  in these coordinates. This is also pointed out in \cite{Yogendran}.
}
\begin{eqnarray}
 \cosh(t-t_0) \sinh \rho = \sinh \rho_0~,
\label{trajectory D0}
\end{eqnarray}
where $t_0$, $\rho_0$ are free parameters. Notice that the D0-brane
reaches the horizon $\rho = 0$ at $t \rightarrow \pm \infty$
irrespective of the values of $\rho_0$ and $t_0$. Thus, formally, the
Lorentzian D0-brane boundary state is obtainable by Wick rotation of
the Euclidean D1-brane boundary state \eqn{hairpin}.\footnote{Some
classical analysis of D-brane dynamics was attempted in
\cite{Yogendran} within the Dirac-Born-Infeld approach.}

Reconstructing boundary states of the Lorentzian D-brane from those
of the Euclidean D-brane is generically not unique. Rather, the
following potential subtleties need to be faced:
\begin{itemize}
 \item The Euclidean momentum $n$ along the asymptotic circle
of cigar is quantized, while the corresponding quantum number in the
Lorentzian theory ({\em i.e.} the energy) takes a continuous value.
 \item The Wick rotations of primary states are not necessarily
unique. Often, appropriate boundary conditions should be specified.
\end{itemize}
As for the first point, which has to do with Matsubara formulation,
we can formally avoid the difficulty of quantized momentum by the
following heuristic consideration. Suppose the boundary wave
function $\hat{f}(n,\al)$ ($n\in \bz$ is the quantized Euclidean
energy, and $\al$ denotes the remaining quantum numbers not touched
here) is given by the Fourier transform of a periodic function
$f(x+2\pi, \al)=f(x, \al)$. We then obtain
\begin{eqnarray}
\hspace{-1cm} \bra{B}=
\sum_{\al}\sum_{n\in\bsz}\,\widetilde{f}(n,\al)\dbra{n,\al} &=&
\sum_{\al}\sum_{n\in\bsz}\, \frac{1}{2\pi}\int_{-\pi}^{\pi} \dd x\,
f(x,\al) e^{in x} \, \dbra{n,\al} \nn &=&  \sum_{\al}
\int_{-\infty}^{\infty} \frac{\dd q}{2\pi}\, \int_{-\infty}^{\infty}
\dd x\, f(x,\al) e^{iqx} \, \dbra{q,\al}~, \label{formal extension}
\end{eqnarray}
where we used the identity $ \sum_{n\in \bsz}\, \delta(q-n) =
\sum_{m\in \bsz}\, e^{2\pi i m q} $ in obtaining the last
expression. Assuming that $f(x, \al)$ is analytic along the entire
real $x$ axis, the Wick rotation can be performed. Often, $f(x,
\al)$ is non-analytic over the real $x$ axis, and the integral in
the last expression is ill-defined. This turns out to be the case
for the boundary wave function of the Euclidean D1-brane
\eqn{hairpin D1}: in the coordinate space, the wave function has
branch cuts and singularities along the real $x$-axis. In such
cases, the best we can do is to adopt the slightly deformed
integration contour $\cC$ in $x$-space\footnote
   {To be more precise, we should allow to use
  some decomposition
$$
 f(x,\al) = f_1(x,\al)+ f_2(x,\al)+\cdots~,
$$
and to take the different contours for each piece $f_i(x,\al)$. } to
render the Fourier integral well-defined:
\begin{eqnarray}
&& \bra{B'} \Big\vert_{\rm Euclidean} :=
\sum_{\al}\int_{-\infty}^{\infty} \frac{\dd q}{2\pi}\, \int_{\cC}
\dd x\, f(x,\al) \, e^{iqx} \, \dbra{q,\al} ~. \label{formal
extension 2}
\end{eqnarray}
Likewise, disk one-point function of vertex operator $\Phi^{\rm
Euclidean}_{q,\al}$ (associated with the Ishibashi state
$\dbra{q,\al}$) is evaluated as the deformed contour integral:
\begin{eqnarray}
&& \Big< \Phi^{\rm Euclidean}_{q,\al} \Big>_{\msc{disk}}
   = {}_{\rm E}\!\langle B' \vert q, \al \rangle\rangle = \int_{\cC} \dd x\, f(x,\al) e^{iqx}~.
\label{formal disk amp 1}
\end{eqnarray}
Assuming sufficient analyticity, one then defines Wick rotation of
the states \eqn{formal extension 2} by the contour deformation of
$\cC$ accompanied by the continuation $q\,\rightarrow\, i\om, x
\rightarrow \, i t$;
\begin{eqnarray}
\bra{B'} \Big\vert_{\rm Lorentzian} :=
\sum_{\al}\int_{-\infty}^{\infty} \frac{i \dd \om}{2\pi}\,
\int_{-\infty}^{\infty} i\dd t\, f(it, \al) e^{-i\om t} \,
\dbra{i\om,\al} ~. \label{Wick rotation 0}
\end{eqnarray}
This is essentially the procedure  \cite{NST}. Of course, we
potentially have an ambiguity in the choice of the contour $\cC$,
and the correct choice should be determined by the physics under
study.

In the present case $\bra{B}$ corresponds to \eqn{hairpin D1} and
$\bra{B'}$ is given by
\begin{eqnarray}
{}_{\rm D1}\bra{B';\rho_0,\theta_0} = \int_0^{\infty} \frac{\dd
p}{2\pi}\, \int_{-\infty}^{\infty} \frac{\dd q}{2\pi}\, \Psi'_{\rm
D1} (\rho_0,\theta_0;p,q) \, \dbra{p,q} ~, \end{eqnarray}
where
\begin{eqnarray} \Psi'_{\rm D1}(\rho_0,\theta_0;p,q) &=& \frac{\sinh(\pi p)}
{\left|\cosh\Big(\pi\frac{p+iq}{2}\Big)\right|^2} \,
\frac{\pi\Gamma(ip)\Gamma\Big(1+\frac{ip}{k}\Big)}
{\Gamma\Big(\frac{1}{2}+\frac{ip+q}{2}\Big)
\Gamma\Big(\frac{1}{2}+\frac{ip-q}{2}\Big)} \, e^{iq\theta_0}
\left[e^{-ip\rho_0} +\frac {\cosh\left(\pi\frac{p-i|q|}{2}\right)}
{\cosh\left(\pi\frac{p+i|q|}{2}\right)}
 e^{ip\rho_0}\right] \nn
\hspace{2cm} &\equiv&  B\left(\frac{1}{2}-\frac{ip-q}{2},
\frac{1}{2}-\frac{ip+q}{2}\right) \Gamma\left(1+\frac{ip}{k}\right)
\, e^{iq\theta_0}\left[e^{-ip\rho_0}
+\frac{\cosh\left(\pi\frac{p-i|q|}{2}\right)}
{\cosh\left(\pi\frac{p+i|q|}{2}\right)} e^{ip\rho_0}\right].
\label{D1'}
\end{eqnarray}
\begin{figure}[htbp]
    \begin{center}
    \includegraphics[width=0.5\linewidth,keepaspectratio,clip]
      {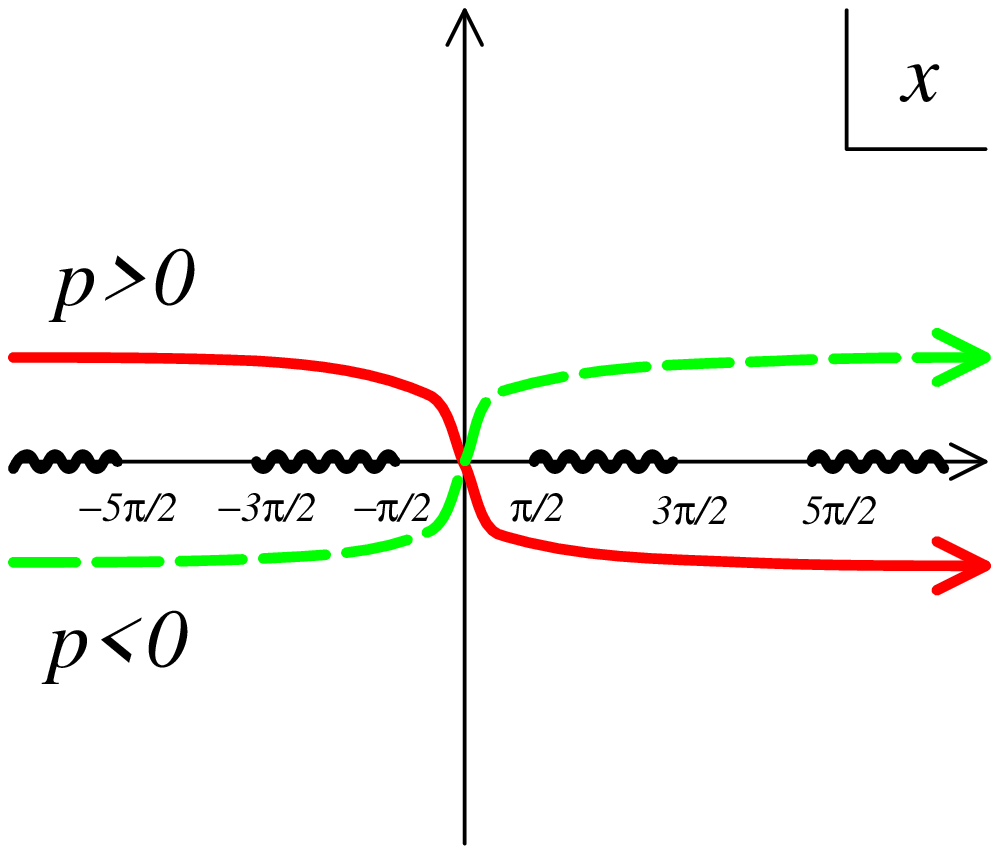}
    \end{center}
    \caption{The red (green broken) line
is the contour $\cC^+$ for $p > 0$
         ($\cC^-$ for $p <0$). The imaginary $x$-axis corresponds
      to the Lorentzian time. Notice that an infinite number
      of branch cuts repeats in the Euclidean time: $\frac{\pi}{2}+2n\pi < x <
       \frac{3\pi}{2} +2n\pi$, $(n\in \bz)$ along the real $x$-axis.}
    \label{c-array}
\end{figure}
Here $B(p,q) \equiv \Gamma(p)\Gamma(q)/\Gamma(p+q)$ denotes Euler's
beta function. The integration contour $\cC$ we choose is shown in
Figure \ref{c-array} \cite{NST}. As in \eqn{evaluation overlap phi},
we separately evaluated the integrals of $\phi^{p}_{L,q}$ and
$\phi^p_{R,q}$ based on the decomposition \eqn{decomp ef}. For the
convergence of integrals, we choose the contour $\cC^+$ for
$\phi^{p}_{L,q}$ ($p>0$ sector) and $\cC^-$ for  $\phi^{p}_{R,q}$
($p<0$ sector). Such choice of integration contours rendered an
extra damping factor $\sinh(\pi p) /
{|\cosh\left(\pi\frac{p+iq}{2}\right)|^2}$, which improves the
ultraviolet behavior of the wave function and makes it possible to
take the Wick rotation sensibly. The non-trivial phase factor $
{\cosh\left(\pi\frac{p-i|q|}{2}\right)}/
{\cosh\left(\pi\frac{p+i|q|}{2}\right)} $ in the second term
originates from the reflection amplitude,
and it reduces to $(-1)^n$ when $q=n\in \bz$.

The second subtlety implies that $\dbra{i\om, \al}$ is not uniquely
defined in \eqn{Wick rotation 0}. This is the issue that arises in a
background with horizon, equivalently, non-existence of globally
definable timelike Killing vector. As such, this subtlety did not
arise for the extremal NS5-brane geometry (described asymptotically
by free linear dilaton theory \cite{NS5brane,CHS}) considered in
\cite{NST}. In the next section, within the mini-superspace analysis
for the Lorentzian two-dimensional black hole, we shall clarify this
subtlety.

An alternative, sensible prescription of the analytic continuation
is to define the disk one-point correlator {\em directly\/} via the
Lorentzian Fourier transform:
\begin{eqnarray}
&& \Big< \Phi^{\msc{Lorentzian}}_{\om,\al} \Big>_{\msc{disk}}
   = \int_{-\infty}^{\infty} \dd t\, f(it,\al) e^{-i\om t}~.
\label{formal disk amp 2}
\end{eqnarray}
This is {\em not\/} always equivalent to the the former method
elaborated above. In fact, the latter method does not necessarily
assert that the boundary state constructed so is expandable in terms
of the Lorentzian Ishibashi states that are analytically continued
from the Euclidean ones.

\subsection{Lorentzian mini-superspace wave functions}

The Wick rotation of the mini-superspace eigenfunctions in the
Euclidean cigar geometry \eqn{ef} is not so trivial. Fortuitously,
the Lorentzian eigenfunctions are already classified thoroughly in
\cite{DVV}. The complete basis for waves outside the black hole
horizon are spanned by the following four types of
eigenfunctions\footnote{Here we adopt slightly different
normalization  from \cite{DVV}.} of the Lorentzian
Klein-Gordon operator. For those with the eigenvalue
$\frac{p^2}{4k}-\frac{\om^2}{4k}+ \frac{1}{4k}$ of the Klein-Gordon
operator, the four eigenfunctions are
\begin{eqnarray}
U^p_{\om}(\rho,t) &=& -
\frac{\Gamma^2(\nu_+)}{\Gamma(1-i\om)\Gamma(-ip)} e^{-i\om t} (\sinh
\rho)^{-i\om} F(\nu_+,\nu^*_-;1-i\om;-\sinh^2\rho) \nn &\sim&
e^{-i\om t - i\om \ln \rho} \qquad \mbox{as} \qquad
\rho\,\rightarrow\,0~,
\label{U} \\
V^p_{\om}(\rho,t) &=&  -
\frac{\Gamma^2(\nu^*_+)}{\Gamma(1+i\om)\Gamma(ip)} e^{-i\om t}
(\sinh \rho)^{i\om} F(\nu^*_+,\nu_-;1+i\om;-\sinh^2\rho) \nn &\sim&
e^{-i\om t + i\om \ln \rho} \qquad \mbox{as} \qquad
\rho\,\rightarrow\,0~,
\label{V} \\
L^p_{\om} (\rho,t ) &=& e^{-i\om t} (\sinh \rho)^{-1-ip} F(\nu^*_+,
\nu^*_-;1+ip; -\frac{1}{\sinh^2 \rho}) \nn &\sim& e^{-\rho}
e^{-ip\rho -i\om t} \qquad \mbox{as} \qquad
\rho\,\rightarrow\,\infty~,
\label{L} \\
R^p_{\om} (\rho,t )
&=& e^{-i\om t} (\sinh \rho)^{-1+ip} F(\nu_+, \nu_-;1-ip;
-\frac{1}{\sinh^2 \rho}) \nn &\sim& e^{-\rho} e^{+ip\rho -i\om t}
\qquad \mbox{as} \qquad \rho\,\rightarrow\,\infty \label{R}
\end{eqnarray}
with the notations
\begin{eqnarray}
\nu_{\pm} = \frac{1}{2} - i\left(\frac{p}{2}\pm
\frac{\om}{2}\right)~. \nonumber
\end{eqnarray}
These eigenfunctions are defined by the following analytic
continuations of the mini-superspace Euclidean eigenfunctions:
\begin{eqnarray}
&& U^p_{\om}(\rho,t)=
\left\{
\begin{array}{ll}
\phi^p_{n=+i\om}(\rho,\theta = +it)~ & ~~ (\om>0,~n<0)  \\
\phi^p_{n=-i\om}(\rho,\theta = -it)~ & ~~ (\om<0,~n>0)
\end{array}
\right. \nn
&& V^p_{\om}(\rho,t)=
\left\{
\begin{array}{ll}
\phi^{-p}_{n=-i\om}(\rho,\theta = - it)~ & ~~ (\om>0,~n<0)  \\
\phi^{-p}_{n=+i\om}(\rho,\theta = +it)~ & ~~ (\om<0,~n>0)
\end{array}
\right. \nn && L^p_{\om}(\rho,t)= \phi^p_{L,n=i\om} (\rho,
\theta=+it) \nn && R^p_{\om}(\rho,t)= \phi^p_{R,n=i\om} (\rho,
\theta=+it) ~, \label{ac UVLR}
\end{eqnarray}
where the $n<0$ and $n>0$ ranges are mapped to $\om>0$ and $\om<0$,
respectively.

\begin{figure}[htbp]
    \begin{center}
    \includegraphics[width=0.6\linewidth,keepaspectratio,clip]
      {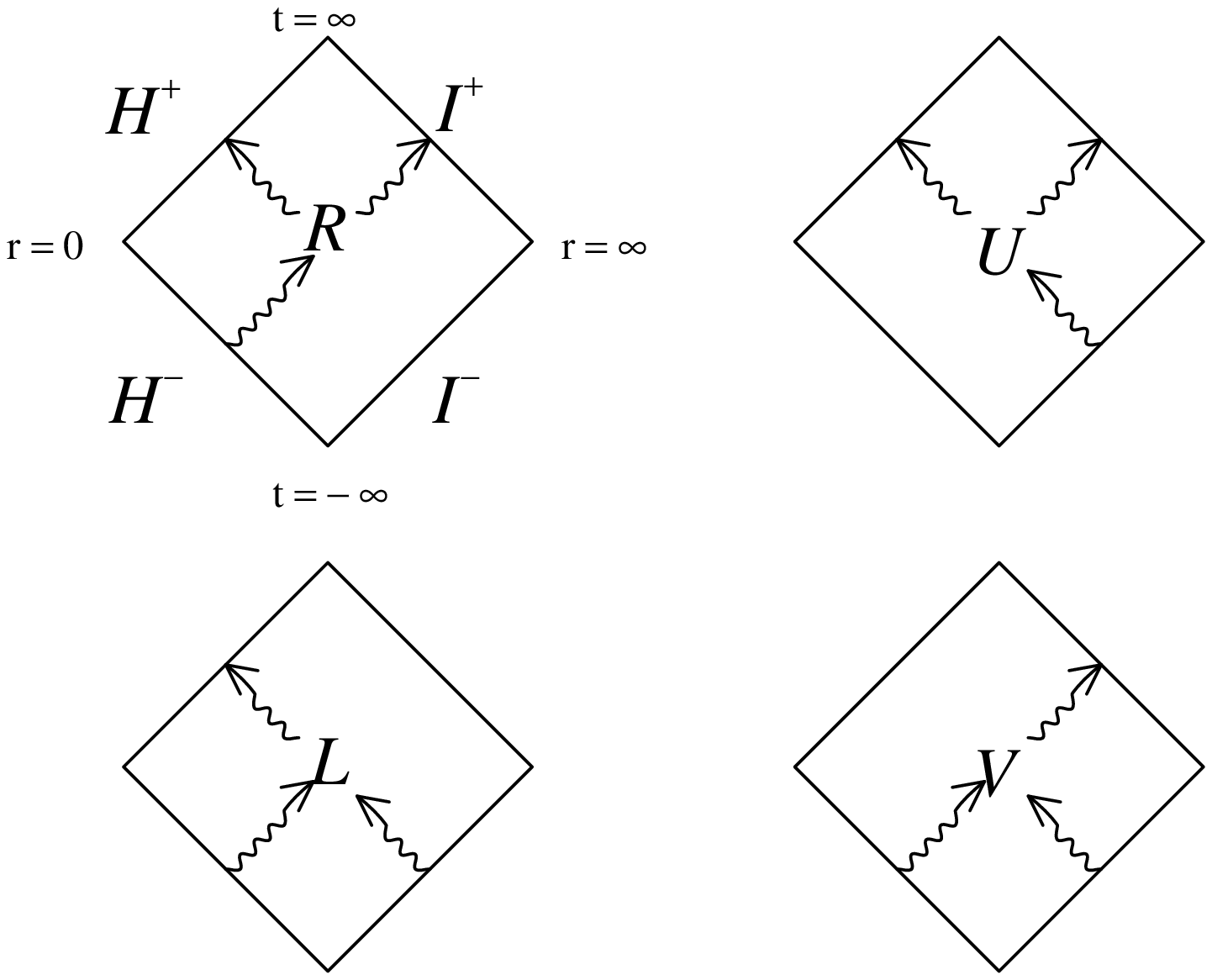}
    \end{center}
    \caption{The boundary conditions of the Lorentzian
       eigenfunctions ($\om>0$ sector). For $\om<0$, the
        figures for $L$ and $R$ should be interchanged.}
    \label{wave}
\end{figure}

As discussed in \cite{DVV}, only two out of the four eigenfunctions
are linearly independent. In particular,
\begin{eqnarray}
V_\omega^p(\rho, t) = U^{p*}_\omega(\rho, -t) \qquad \mbox{and}
\qquad R_\omega^p(\rho, t) = L^{p*}_\omega (\rho, -t)\ . \nonumber
\end{eqnarray}
The reason why we introduce the above four eigenfunctions is because
they encode four possible boundary conditions (We here assume $p>0$)
in the Lorentzian black hole background. Recall that, for the region
outside the horizon of the eternal black hole, the boundaries
consist of four segments: `future (past) horizon' $t=+\infty, \,
\rho=0$ ($t=-\infty, \, \rho=0$) by ${\cal H}^+$ (${\cal H}^-$), and
the `future (past) infinity' $t=+\infty, \, \rho=+\infty$
($t=-\infty, \, \rho=+\infty$) by ${\cal I}^+$ (${\cal I}^-$). The
four eigenfunctions $U, V, L, R$ are the ones obeying boundary
conditions:
\begin{eqnarray}
&& U^p_{\om} = 0 ~~ \mbox{at}~ {\cal H}^-~, ~~~ V^p_{\om} = 0 ~~
\mbox{at}~~ {\cal H}^+~, ~~~ L^p_{\om} = 0~ (R^p_{\om} = 0)~~
\mbox{at}~~ {\cal I}^+~, ~~~ R^p_{\om} = 0 ~(L^p_{\om} = 0) ~~
\mbox{at}~~ {\cal I}^- \nonumber
\end{eqnarray}
for $\om>0$ ($\om <0$). See Figure \ref{wave}.


By Wick rotating the mini-superspace reflection relations \eqn{cref
rel}, we obtain linear relations among the Lorentzian
eigenfunctions:
\begin{eqnarray}
&& U^p_{\om} = L^{p}_{\om} + \cR_0(p,\om) R^p_{\om} \qquad
\mbox{and} \qquad V^p_{\om} = R^{p}_{\om} + \cR^*_0(p,\om)
L^p_{\om}~. \label{decomp ef 2}
\end{eqnarray}
Equivalently,
\begin{eqnarray}
&& L^p_{\om} = \frac{1}{1-|\cR_0(p,\om)|^2} \left\{ U^p_{\om} -
\cR_0(p,\om) V^p_{\om} \right\} \quad \mbox{and} \quad R^p_{\om} =
\frac{1}{1-|\cR_0(p,\om)|^2} \left\{ V^p_{\om} - \cR^*_0(p,\om)
U^p_{\om} \right\}. \label{decomp ef 2-2}
\end{eqnarray}
Here, the mini-superspace reflection amplitude $\cR_0(p,\om)$ in
Lorentzian theory is given by
\begin{eqnarray}
&&\cR_0(p,\om) = \frac{\Gamma(+ip)\Gamma^2(\nu_+)}
{\Gamma(-ip)\Gamma^2(\nu^*_-)}
 \equiv - \frac{B(\nu_+,\nu_-)}{B(\nu_+^*,\nu_-^*)} \cdot
\frac{\cosh \pi \left(\frac{p-\om}{2}\right)}
{\cosh \pi \left(\frac{p+\om}{2}\right)}~.
\label{cref amp 2}
\end{eqnarray}
Notice that, in sharp contrast to the Euclidean black hole, the
reflection amplitude is less than unity due to the second factor:
\bea |\cR_0(p,\om)|^2 = {\cosh^2 \pi \left({p-\om \over 2}\right)
\over \cosh^2 \pi \left({p+\om \over 2} \right)} \leq 1\ .
\label{inequality} \eea
The inequality is saturated at $p=\omega = 0$. The inequality
\eqn{inequality} shall play a prominent role for understanding string
dynamics in the Lorentzian black hole background. The
mini-superspace reflection relations for $U^p_{\om}$, $V^p_{\om}$
are also expressible in a form similar to the Euclidean ones.
Recalling that $\cR_0(-p, \om) \cR_0(+p, \om) = 1$,
\begin{eqnarray}
&& U^{-p}_{\om}(\rho,t)= \cR_0(-p,\om) U^{p}_{\om}(\rho,t) \qquad
\mbox{and} \qquad V^{-p}_{\om}(\rho,t)= \cR^*_0(-p,\om)
V^{p}_{\om}(\rho,t)~, \label{cref rel UV}
\end{eqnarray}
while $L^p_{\om}$ and $R^p_{\om}$ are simply related by reflection:
\begin{eqnarray}
&& L^{-p}_{\om} (\rho,t) = R^{+p}_{\om} (\rho,t)~. \label{cref rel
LR}
\end{eqnarray}
Moreover, $U^p_{\om}$ and $V^p_{\om}$ are linearly independent
except for the special kinematic regime, $\om=0$. Notice also, in
the relation \eqn{inequality}, the reflection amplitude involves the
mini-superspace contribution only, not the full-fledged stringy one.

Before proceeding further, we shall here collect explicitly
relations among inner products of Lorentzian primary fields, where
the inner product is defined with respect to the Lorentzian measure
$\dd v_L = k\sinh 2\rho \dd \rho \dd t$. Taking quantum
numbers $p$, $\om$ fixed and dropping off delta function factors
$2\pi\delta(p-p')$, $2\pi\delta(\om-\om')$ for notational
simplicity, we have
\begin{eqnarray}
 && (U^p_{\om}, U^p_{\om}) = (V^p_{\om}, V^p_{\om})
   = N_0(p,\om)~, \qquad N_0(p,\om)
\equiv \frac{1+|\cR_0(p,\om)|^2}{2}
\nn
 && (U^p_{\om},V^p_{\om}) = \cR_0^*(p,\om)~,
\nn
 && (L_{\om}^p,L_{\om}^p)= (R^p_{\om}, R^p_{\om})= \frac{1}{2}~,
 \qquad \quad (L^p_{\om}, R^p_{\om})= 0 ~, \nn
 && (U^p_{\om},L^p_{\om}) = (V^p_{\om}, R^p_{\om}) =\frac{1}{2}~,
 \qquad \quad (R^p_{\om},U^p_{\om}) = (V^p_{\om}, L^p_{\om})
   =\frac{\cR_0(p,\om)}{2}~.
\label{inner product UVLR}
\end{eqnarray}
The inner products involving $L^p_\om$ and $R^p_\om$ are readily
evaluated since dominant contributions are supported in the
asymptotic region $\rho \gg 0$, yielding the volume factor
$2\pi\delta(0)$. The remaining inner products are then extractable
from the linear relations \eqn{decomp ef 2}, \eqn{decomp ef 2-2}.\footnote
{We checked these inner products numerically using
MATHEMATICA.} We also fixed the overall normalization factors from
consistency with the Euclidean inner product \eqn{inner product}
under the $\om\,\rightarrow\, 0$ limit. Notice also that
\begin{eqnarray}
 N_0(-p,\om) = \left|\cR_0(-p,\om)\right|^2 \, N_0(+p,\om)~,
\end{eqnarray}
as is consistent with the mini-superspace reflection relation
\eqn{cref rel UV}.

It is easy to construct the exact string vertex operators or primary
states corresponding to the mini-superspace eigenfunctions $U$, $V$,
$L$, $R$.
To be specific, we shall consider primarily the fermionic $SL_k (2,
\br)/U(1)$ supercoset conformal field theory.\footnote
    {For the bosonic $SL(2;\br)_{\kappa}/U(1)$ coset conformal field theory,
  we instead have
$h= \tilde{h}= \frac{p^2}{4(\kappa-2)}-\frac{\om^2}{4\kappa}
 +\frac{1}{4(\kappa-2)}$, and
$\cR(p,\om)\equiv
\cR_0(p,\om) \frac{\Gamma\left(1+\frac{ip}{\kappa-2}\right)}
{\Gamma\left(1-\frac{ip}{\kappa-2}\right)}$.
}
The primary states $\ket{U^p_{\om}}$, $\ket{V^p_{\om}}$ are the ones
of conformal weights $h= \tilde{h}= \frac{p^2}{4k}-\frac{\om^2}{4k}
+\frac{1}{4k}$ and obey the exact reflection relations
\begin{eqnarray}
&& \ket{U^{-p}_{\om}}= \cR(-p,\om) \ket{U^{p}_{\om}}~, ~~~
\ket{V^{-p}_{\om}}= \cR^*(-p,\om) \ket{V^{p}_{\om}}~, \label{qref rel
UV}
\end{eqnarray}
and the exact reflection amplitude is given by
\bea \cR(p,\om)\equiv \cR_0(p,\om)
\frac{\Gamma\Big(1+\frac{ip}{k}\Big)}
{\Gamma\Big(1-\frac{ip}{k}\Big)}\ . \label{exactra} \eea
Notice that the string worldsheet effect entering through the
$1/k$-correction is a pure phase. Thus, the exact reflection
probability $\vert {\cal R} (p, \om) \vert^2$ remains unmodified
from the mini-superspace approximation result $\vert {\cal R}_0(p,
\om) \vert^2$ given in \eqn{inequality}. We shall normalize the
primary states $\ket{U^p_{\om}}$, $\ket{V^p_{\om}}$ ($p>0$) as
\begin{eqnarray}
&& \bra{U^p_{\om}} U^{p'}_{\om'}\rangle = \bra{V^p_{\om}}
V^{p'}_{\om'}\rangle = N(p,\om) \, 2\pi\delta(p-p')
2\pi\delta(\om-\om') ~, \nonumber \\
&& \bra{V^p_{\om}} U^{p'}_{\om'} \rangle = \cR^* (p,\om) \,
2\pi\delta(p-p') 2\pi\delta(\om-\om') ~, \label{norm UV}
\end{eqnarray}
where the new normalization factor $N(p,\om)$ is simply defined by
replacing $\cR_0$ with $\cR$ in $N_0(p,\om)$. The primary states
$\ket{L^p_{\om}}$, $\ket{R^p_{\om}}$ are also definable by using the
linear relations \eqn{decomp ef 2} or \eqn{decomp ef 2-2} but now
with $\cR_0$ replaced by $\cR$. Notice that $\ket{U^p_{\om}}$,
$\ket{V^p_{\om}}$ are the ones analytically continuable to the
Euclidean primary states $\ket{\phi^{\pm p}_n}$, so often referred
as the `Hartle-Hawking vacua'. On the other hand, the states
$\ket{L^p_{\om}}$, $\ket{R^p_{\om}}$ does not have Euclidean
counterparts. Recall that, over the Euclidean black hole background,
$\phi^p_{L,n}$, $\phi^p_{R,n}$ behave badly in the vicinity of $\rho
= 0$ and hence ill-defined.


We also find it useful to introduce the dual basis
$\widehat{\bra{U^p_{\om}}}$, $\widehat{\bra{V^p_{\om}}}$ ($p,p'>0$)
with inner products
\begin{eqnarray}
 && \widehat{\bra{U^p_{\om}}} U^{p'}_{\om'} \rangle
   = \widehat{\bra{V^p_{\om}}} V^{p'}_{\om'} \rangle
   = 2\pi\delta(p-p')2\pi\delta(\om-\om')~, \qquad
 \widehat{\bra{U^p_{\om}}} V^{p'}_{\om'} \rangle
   = \widehat{\bra{V^p_{\om}}} U^{p'}_{\om'} \rangle
   = 0~.
\label{hat U V}
\end{eqnarray}
Explicitly, they are given by
\begin{eqnarray}
&& \widehat{\bra{U^p_{\om}}} = \frac{2}{1-\left|\cR(p, \om)
\right|^2} \left\{ \bra{L^p_{\om}} - \cR^* (p, \om) \bra{R^p_{\om}}
\right\}~, \qquad \widehat{\bra{V^p_{\om}}} =
\frac{2}{1-\left|\cR(p, \om) \right|^2} \left\{ \bra{R^p_{\om}} -
\cR(p, \om) \bra{L^p_{\om}} \right\}~. \label{def hat U V}
\end{eqnarray}
As such, these dual basis obey the following exact reflection
relations:
\begin{eqnarray}
 &&\widehat{\bra{U^{-p}_{\om}}} = \cR(p,\om)
 \widehat{\bra{U^{p}_{\om}}} \qquad \mbox{and} \qquad
\widehat{\bra{V^{-p}_{\om}}} = \cR(p,\om)^*
 \widehat{\bra{V^{p}_{\om}}}~.
\label{ref hat U V}
\end{eqnarray}

A remark is in order. The dual basis $\widehat{\bra{U^{p}_{\om}}}$,
$\widehat{\bra{V^{p}_{\om}}}$ are {\em not\/} Wick rotatable to the
Euclidean dual basis $\bra{\phi^{+p}_n}$, $\bra{\phi^{-p}_n}$, since
$|\cR(p,\om)|=1$ for $\om \in i\br$. The correct procedure would be
that we first define Wick rotations for the `ket' states, and then
define their dual states within the Lorentzian Hilbert space.
Nevertheless, one-point correlators in the Lorentzian theory, from
which a set of physical observables can be computed, ought to be
always analytically continuable to the one-point correlators in the
Euclidean theory. Roughly speaking, ambiguities inherent to the Wick
rotation of dual states drop out upon taking inner product.

Having obtained the Lorentzian primary states, we shall now
construct several interesting class of boundary states for a
D0-brane propagating in the black hole background. We have seen that
the D0-brane propagates along the trajectory \eqn{trajectory D0}.
The two-dimensional black hole is eternal, so, in addition to the
past and the future asymptotic infinities, the causal propagation
region has the past horizon ${\cal H}^-$ surrounding the white hole
singularity and the future horizon ${\cal H}^+$ surrounding the
black hole singularity. As such, by taking variety of possible
boundary conditions, we can construct interesting class of boundary
states.
\subsection{Boundary state of D0-brane absorbed to future horizon}
Consider first the boundary state obeying the boundary condition
$\psi(\rho,t)\,\rightarrow\, 0$ at the past horizon ${\cal H}^-$,
viz. the primary states $\ket{U^p_{\om}}$.
This boundary condition is relevant for scattering of a D0-brane off
the black hole, since the condition represents absorption only and
no emission of the D0-brane by the black hole. D0-brane boundary
state obeying such absorbing boundary condition is then expanded
solely by the Ishibashi states ${}^{\widehat{U}}\dbra{p,\om}$,
$\dket{p,\om}^U$ that are associated with the primary states
$\widehat{\bra{U^p_{\om}}}$, $\ket{U^p_{\om}}$:
\begin{eqnarray}
 &&
{}_{\msc{absorb}}\!\bra{B;\rho_0,t_0} = \int_0^{\infty}\frac{\dd
p}{2\pi} \int_{-\infty}^{\infty}\frac{\dd \om}{2\pi}\,
  \Psi_{\msc{absorb}}(\rho_0,t_0;p,\om) \,
{}^{\widehat{U}}\!\dbra{p,\om}~, \nn
  &&
\ket{B;\rho_0,t_0}_{\msc{absorb}} = \int_0^{\infty}\frac{\dd
p}{2\pi} \int_{-\infty}^{\infty}\frac{\dd \om}{2\pi}\,
  \Psi^*_{\msc{absorb}}(\rho_0,t_0;p,\om)
\, \dket{p,\om}^U~. \label{falling D0 0}
\end{eqnarray}
The boundary wave function $\Psi_{\msc{absorb}}(\rho_0,t_0;p,\om)$
is then interpreted as the disk one-point correlators:
\begin{eqnarray}
\Psi_{\msc{absorb}}(\rho_0,t_0;p,\om) &=& \langle U^p_{\om}
\rangle_{\msc{disk}} \equiv {}_{\msc{absorb}}\!\bra{B;\rho_0,t_0}
U^p_{\om} \rangle~,
\label{falling disk}
\end{eqnarray}

The boundary wave function \eqn{falling disk} is then obtained by
taking the Wick rotation $q\,\rightarrow\, i\om$ ($q\,\rightarrow\,
-i\om$) for $q<0$ ($q>0$)
in \eqn{D1'} (recall \eqn{ac UVLR}):\footnote
        {In reality, there is a further overall factor $i$,
          but, for notational simplicity, we will absorb it to the definition
          of the Ishibashi states.}
\begin{eqnarray}
&& \hspace{-1cm} \Psi_{\msc{absorb}}(\rho_0,t_0;p,\om) = B(\nu_+,
\nu_-) \Gamma\Big(1+\frac{ip}{k}\Big) \, e^{-i\om t_0}\left[ e^{-ip
\rho_0} -  \frac{\cosh\left(\pi \frac{p-\om}{2}\right)}
{\cosh\left(\pi \frac{p+\om}{2}\right)} e^{ip\rho_0 } \right]~,
\label{falling D0}
\end{eqnarray}
The relative minus sign in the second term of
$\Psi_{\msc{absorb}}(\rho_0,t_0;p,\om)$ originates from the fact
that the contour rotation defining the Wick rotation has opposite
directions for $\cC^+$ (suitable for $p>0$) and $\cC^-$ (suitable
for $p<0$). See figure \ref{c-array}. This boundary wave function
\eqn{falling D0} satisfies the exact reflection relation
\begin{eqnarray}
\Psi_{\msc{absorb}}(\rho_0,t_0;-p,\om)= \cR(-p,\om) \,
\Psi_{\msc{absorb}}(\rho_0,t_0;p,\om)~. \label{ref falling D0}
\end{eqnarray}

With such boundary condition, the boundary wave function
\eqn{falling D0} would have no overlap with D0-brane's trajectory
\eqn{trajectory D0} in the far past region $t\ll t_0$. In fact, the
trajectory \eqn{trajectory D0} starts from the past horizon ${\cal
H}^-$ at $t=-\infty$, reaches the time-symmetric point $\rho =
\rho_0$ at $t = t_0$, and then falls back the future horizon ${\cal
H}^+$ at $t=+\infty$, while the wave function $U^p_{\om}$ does not
have any component outgoing from ${\cal H}^-$. We thus interpret
that the boundary state \eqn{falling D0} describes the future half
of the classical trajectory \eqn{trajectory D0}. We shall hence call
it the `absorbed D-brane'.

By utilizing the radion-tachyon correspondence, the rolling radion
(as described by the boundary state \eqn{falling D0}) is also
interpretable as the rolling tachyon. In the latter interpretation,
the D0-brane absorbed to the future horizon is the counterpart of
the future-half S-brane \cite{GS1,Strominger,LNT}, in which the
tachyon rolls down the potential hill at asymptotic future $t
\rightarrow + \infty$ and emits radiation.

\subsection{Boundary state of D0-brane emitted from past horizon}
Consider next the boundary condition: $\psi(\rho,t)\,\rightarrow\,
0$ at ${\cal H}^+$, viz. use the basis $\dket{p,\om}^V$,
${}^{\widehat{V}}\!\dbra{p,\om}$ instead of $\dket{p,\om}^U$,
${}^{\widehat{U}}\!\dbra{p,\om}$. Utilizing the reflection relation,
we can first rewrite \eqn{D1'} as the form which only includes the
$p<0$ Ishibashi states by means of the reflection relation. Then, we
can analytically continue the states $\ket{\phi^{-p}_q}$ ($p>0$)
into $\ket{V^p_\om}$.
The resultant boundary state is obtained by
simply replacing $p\,\rightarrow\,-p$,
$\om\, \rightarrow\, -\om$
in \eqn{falling D0};
\begin{eqnarray}
&& {}_{\msc{emitted}}\!\bra{B;\rho_0,t_0} = \int_0^{\infty}\frac{\dd
p}{2\pi} \int_{-\infty}^{\infty}\frac{\dd \om}{2\pi}\,
  \Psi_{\msc{emitted}}(\rho_0,t_0;p,\om) \,
{}^{\widehat{V}}\!\dbra{p,\om}~. \nn
&& \ket{B;\rho_0,t_0}_{\msc{emitted}} = \int_0^{\infty}\frac{\dd
p}{2\pi} \int_{-\infty}^{\infty}\frac{\dd \om}{2\pi}\,
  \Psi^*_{\msc{emitted}}(\rho_0,t_0;p,\om) \,
\dket{p,\om}^V ~. \label{emitted D0} \eea
where
\bea \Psi_{\msc{emitted}}(\rho_0,t_0;p,\om) = B(\nu^*_+, \nu^*_-)
\Gamma\left(1-\frac{ip}{k}\right) \, e^{-i\om t_0}\left[
e^{ip\rho_0} - \frac{\cosh\left(\pi \frac{p-\om}{2}\right)}
{\cosh\left(\pi \frac{p+\om}{2}\right)} e^{-ip\rho_0} \right]~.
\nonumber
\end{eqnarray}
Obviously, the emitted D0-brane wave function is the time-reversal
of the absorbed D0-brane wave function \eqn{falling D0}:
\bea \Psi_{\msc{emitted}}(\rho_0,t_0;p,\om) =
\Psi^*_{\msc{absorb}}(\rho_0,-t_0;p,\om) ~. \nonumber \eea
Namely, it describes the D0-brane emitted from the past horizon at
asymptotic past $t=-\infty$. By the choice of the boundary
condition, this boundary state \eqn{emitted D0} describes only the
past half of the classical D0-brane trajectory \eqn{trajectory D0}.

The exact reflection relation has the form
\begin{eqnarray}
&& \Psi_{\msc{emitted}}(\rho_0,t_0;-p,\om)= \cR^*(-p,\om) \,
\Psi_{\msc{emitted}}(\rho_0,t_0;p,\om)~. \label{ref emiited D0}
\end{eqnarray}

Again, in light of the radion-tachyon correspondence, the D0-brane
emitted from the past horizon is the counterpart of the past-half
S-brane in tachyon rolling. The radiation creeps up the tachyon
potential hill from past infinity and forms an unstable D-brane.

\subsection{Boundary state of time-symmetric D0-brane}

The third possible boundary state is obtainable by {\em directly\/}
taking the analytic continuation in the disk one-point amplitudes,
as we already mentioned. Recalling \eqn{ac UVLR}, we shall
analytically continue the disk amplitudes as (assume $p>0$)
\begin{eqnarray}
  \langle \phi^{+p}_q \rangle_{\msc{disk}}~\longrightarrow~
  \langle U^p_{\om} \rangle_{\msc{disk}} \qquad \mbox{and} \qquad
  \langle \phi^{-p}_q \rangle_{\msc{disk}}~\longrightarrow~
  \langle V^p_{\om} \rangle_{\msc{disk}}~.
\label{ac disk amp}
\end{eqnarray}
The Euclidean one-point amplitudes $\langle \phi^{\pm p}_q
\rangle_{\msc{disk}}$ are given in \eqn{D1'}, and can be expressed
in contour integrals as in \eqn{formal disk amp 1}. Recall that
$\langle \phi^p_{L,q}   \rangle_{\msc{disk}}$, $\langle
\phi^{p}_{R,q} \rangle_{\msc{disk}}$ are prescribed by the contour
integrals over $\cC^+$, $\cC^-$ in figure \ref{c-array}. We shall
thus analytically continue them to the real time axis (imaginary
$x$-axis). In this way, we extract the Lorentzian disk one-point
amplitudes as
\begin{eqnarray}
&& \langle U^p_{\om} \rangle_{\msc{disk}} = \langle U^p_{\om}
\rangle_{\msc{disk}}^{(\msc{absorb})} \qquad \mbox{and} \qquad \langle V^p_{\om}
\rangle_{\msc{disk}} = \langle V^p_{\om} \rangle_{\msc{disc}}^{(\msc{emitted})}~,
~~~ \label{rel disc amp}
\end{eqnarray}
where the right-hand sides are simply the amplitudes associated with
the `absorbed' and `emitted' D0-branes considered in the previous
subsections and explicitly given in \eqn{falling D0} and
\eqn{emitted D0}. Since $U^p_{\om}$ and $V^p_{\om}$ constitute the
complete set of basis for Lorentzian primary fields, the amplitudes
\eqn{rel disc amp} would yield yet another Lorentzian D0-brane
boundary states. As is obvious from the above construction, this
state keeps the time-reversal symmetry manifest and reproduces the
entire classical trajectory \eqn{trajectory D0}, that is, it
describes a D0-brane emitted from the past horizon and reabsorbed to
the future horizon. From the viewpoint of the boundary conformal
theory, this would be considered the most natural one since it
captures the entire classical trajectory of the D0-brane. In the
radion-tachyon correspondence, this state is the counterpart of the
full S-brane \cite{GS1,Sen-RT}.

Explicitly, the time-symmetric boundary states are given by
\begin{eqnarray}
&& {}_{\msc{symm}}\!\bra{B;\rho_0,t_0} =
{}_{\msc{absorb}}\!\bra{B;\rho_0,t_0} +
{}_{\msc{emitted}}\!\bra{B;\rho_0,t_0} \nn && \hspace{2cm} =
\int_0^{\infty}\frac{\dd p}{2\pi} \int_{-\infty}^{\infty}\frac{\dd
\om}{2\pi}\, \left[
  2\Psi_{\msc{symm}}(\rho_0,t_0;p,\om)
\, {}^L\!\dbra{p,\om}
  +
  2\Psi^*_{\msc{symm}}(\rho_0,-t_0;p,\om) \, {}^R\!\dbra{p,\om}
\right]  \nn
&& \ket{B;\rho_0,t_0}_{\msc{symm}} =
\ket{B;\rho_0,t_0}_{\msc{absorb}} +
\ket{B;\rho_0,t_0}_{\msc{emitted}}  \nn && \hspace{2cm} =
\int_0^{\infty}\frac{\dd p}{2\pi} \int_{-\infty}^{\infty}\frac{\dd
\om}{2\pi}\, \left[
  2\Psi^*_{\msc{symm}}(\rho_0,t_0;p,\om)
\, \dket{p,\om}^L
  +
  2\Psi_{\msc{symm}}(\rho_0,-t_0;p,\om) \, \dket{p,\om}^R
\right] ~, \label{symmetric D0} \eea
where
\bea \Psi_{\msc{symm}}(\rho_0,t_0;p,\om) = B(\nu_+, \nu_-)
\Gamma\left(1+\frac{ip}{k}\right) \, e^{-ip\rho_0-i\om t_0}
\nonumber
\end{eqnarray}
and ${}^L\!\dbra{p,\om}$, $\dket{p,\om}^L$, ${}^R\!\dbra{p,\om}$,
$\dket{p,\om}^R$ are the Ishibashi states constructed over the
primary states $\bra{L^p_{\om}}$, $\ket{L^p_{\om}}$,
$\bra{R^p_{\om}}$, $\ket{R^p_{\om}}$,\footnote{The extra factor of
`2' was introduced for convenience. Recall \eqn{inner product
UVLR}.} respectively. One can readily check that the second lines
in \eqn{symmetric D0} are indeed correct by evaluating the disk
one-point amplitudes from them. For instance, using \eqn{inner
product UVLR}, we obtain
\begin{eqnarray}
 \langle U^p_{\om} \rangle_{\msc{disk}}^{(\msc{symm})}
&=& {}_{\msc{symm}}\!\bra{B;\rho_0,t_0} U^p_{\om} \rangle \nn &=&
\Psi_{\msc{symm}}(\rho_0,t_0;p,\om) +\cR(p,\om)
\Psi^*_{\msc{symm}}(\rho_0,-t_0;p,\om) \nn &=&  B(\nu_+,\nu_-)
\Gamma\left(1+\frac{ip}{k}\right) e^{-i\om t_0} \, \left[
e^{-ip\rho_0} - \frac{\cosh\left(\pi \frac{p-\om}{2}\right)}
{\cosh\left(\pi \frac{p+\om}{2}\right)} e^{ip\rho_0} \right] \nn &=&
\langle U^p_{\om} \rangle^{(\msc{absorb})}_{\msc{disk}} \equiv
{}_{\msc{absorb}}\!\bra{B;\rho_0,t_0}U^p_{\om}\rangle ~. \nonumber
\end{eqnarray}
Other one-point amplitudes can be checked analogously.

Two remarks are in order. First, notice that, though the disk
one-point amplitudes are, the symmetric boundary states
\eqn{symmetric D0} by themselves are {\em not\/} analytically
continuable to the Euclidean boundary state \eqn{D1'}. This should
not be surprising as the Lorentzian Hilbert space is generated by
{\sl twice} as many generators as the Euclidean theory.
In other words,
the Lorentzian bases $\ket{U^p_{\om}}$, $\ket{V^p_{\om}}$ correspond
to $\ket{\phi^p_n}$, $\ket{\phi^{-p}_n}$ in the Euclidean theory,
which were however linearly dependent due to the reflection
relation.
Nevertheless, the
boundary state \eqn{symmetric D0} is a consistent one and yields
disk one-point amplitudes that can be correctly continued to the
Euclidean ones. Second, the full Lorentzian Hilbert space is
decomposed as
\begin{eqnarray}
 \cH = \cH^U \oplus \cH^V  \qquad \mbox{and} \qquad
 \widehat{\cH} = \widehat{\cH^U} \oplus \widehat{\cH^V}~,
\label{decomp Hilb}
\end{eqnarray}
where $\cH^U$ ($\cH^V$) is spanned by $\ket{U^p_{\om}}$ , ($\,
\ket{V^p_{\om}}\, $) and their descendants. The dual space
$\widehat{\cH^U}$ ($\widehat{\cH^V}$) is similarly spanned by
$\widehat{\bra{U^p_{\om}}}$, ($\widehat{\bra{V^p_{\om}}}$). Here,
the Hilbert subspaces $\cH^{U}$, $\widehat{\cH^{U}}$ ($\cH^{V}$,
$\widehat{\cH^{V}}$) correspond to the boundary condition
$\psi(\rho,t)\,\rightarrow\, 0$ at ${\cal H}^{-}$ (${\cal H}^+$).
The `absorbed' and `emitted' D0-brane boundary states \eqn{falling
D0}, \eqn{emitted D0} are consistent {\sl only} in the subspaces
$\cH^U$, $\cH^V$ ($\widehat{\cH^U}$, $\widehat{\cH^V}$), while the
`symmetric' D0-brane boundary state \eqn{symmetric D0} is
well-defined in the entire Hilbert space $\cH$ ($\widehat{\cH}$). We
thus have simple relations
\begin{eqnarray}
\ket{B;\rho_0,t_0}_{\msc{absorb}} = P_U \,
\ket{B;\rho_0,t_0}_{\msc{symm}} \qquad &\mbox{and}& \qquad
{}_{\msc{absorb}}\! \bra{B;\rho_0,t_0} =
{}_{\msc{symm}}\!\bra{B;\rho_0,t_0}\,\widehat{P_U}~,\nn
\ket{B;\rho_0,t_0}_{\msc{emitted}} = P_V \,
\ket{B;\rho_0,t_0}_{\msc{symm}} \qquad &\mbox{and}& \qquad
 {}_{\msc{emitted}}\!\bra{B;\rho_0,t_0} =
{}_{\msc{symm}}\!\bra{B;\rho_0,t_0}\,\widehat{P_V}~,
\label{proj symmetric D0}
\end{eqnarray}
where $P_{U, V}$ ($\widehat{P_{U,V}}$) denotes projection of the
Hilbert space ${\cal H}$ to $\cH^{U,V}$ ($\widehat{\cH^{U, V}}$).

\section{Radiation out of D0-Brane Rolling in the Black Hole Background}

In the background of the black hole, the D0-brane moves along the
geodesic and we have constructed a variety of boundary states
describing the geodesic motion, specified by appropriate boundary
conditions.

Both by gravity and by strong string coupling gradient, the
D$p$-brane is pulled in and finds its minimum energy and mass at the
location of the NS5-brane. The D$p$-brane is supersymmetric in flat
spacetime, but preserves no supersymmetry in black NS5-brane
background. Even in extremal NS5-brane background, until the
D$p$-brane dissociates into the NS5-brane and form a non-threshold
bound-state, the spacetime supersymmetry is completely broken. In
these respects,the D$p$-brane propagating in the NS5-brane
background is much like excited D$p$-brane (many excited open
strings attached on it) in flat spacetime. Decay of the latter via
closed string emission was studied extensively for $p=1$
\cite{excitedDdecay}: the decay spectrum was found to match exactly
with the Hawking radiation of the non-extremal black hole made out
of these excited D-branes, and the effective temperature of excited
open string modes agrees exactly with the Hawking temperature. In
this section, we shall find certain analogous results for the closed
string radiation off the rolling D0-brane, though special features
also arise.

As the D0-brane is pulled in, acceleration would grow and radiate
off the binding energy into closed string modes. Details of the
radiation spectra would differ for different choice of the boundary
conditions, viz. for different boundary states of the D0-brane. In
this section, as a probe of the black hole geometry and D-brane
dynamics therein, we shall analyze spectral distribution of the
closed string radiation off the rolling D0-particle.

By applying the optical theorem, the radiation rate during the
radion-rolling process is obtainable as the imaginary part of the
annulus amplitude in the closed string channel.\footnote{For the
tachyon rolling process in flat spacetime background, the amplitude
was evaluated first in \cite{LLM,KLMS}.} Denote the differential
number density $\dd {\cal N}(p, M)$ of the radiation at a fixed
value of the radial momentum $p$ and the mass-level $M$. By the
definition of the D-brane boundary state, the radiation number
density $\dd \cN$ is then given in terms of the boundary wave
functions:
\begin{eqnarray}
\dd \cN (p,M)   &:=& {\dd p \over 2 \pi} {\dd M \over (2 \pi)^d}
\int{\dd \om} \, \Big< \Psi(\om, p, M) \Big\vert \delta(L_0 +
\overline{L}_0) \Big\vert \Psi (\om, p, M) \Big>
\nonumber \\
&=& \frac{\dd p}{2 \pi} \frac{\dd M}{(2 \pi)^d} \frac{1}{2
\om(p,M)}\, \Big| \Psi(p,\om(p,M))\Big|^2 ~. \label{radiation rate
0}
\end{eqnarray}
Here, $\om, p$ are the energy and the radial momentum in
two-dimensional Lorentzian background, $M$ is the total mass
(conformal weight) of the remaining subspaces of dimension $d$
(including mass gap), $\Psi(\om, p, M)$ is the boundary wave
function (including that of the remaining subspace), and
$\om(p,M)(>0)$ is the on-shell energy of the radiated closed string
state determined by the on-shell condition $L_0 + \overline{L}_0 =
0$ including the ghost contribution. From the kinematical
consideration, it is obvious that the differential number density
\eqn{radiation rate 0} is nonzero only when the D-brane is rolling.
Of particular physical interest is the spectral distribution in the
phase-space, as measured by the independent moments, {\em e.g.}
\bea \Big< \om^m M^n \Big> &=& \int \frac{\dd p}{2 \pi} \frac{\dd
M}{(2 \pi)^d} \om^m(p, M) M^n \frac{1}{2 \om(p, M)}  \Big|\Psi(p,
\om(p,M))\Big|^2 \nonumber \eea
for $m, n =0, 1, 2, \cdots$. We shall evaluate these spectral
observables by first evaluating the integral over the radial
momentum $p$ by saddle-point approximation. In doing so, we pay
particular attention to the asymptotic behavior as the mass-level
$M$ becomes asymptotically large. We shall then evaluate the
integral over the mass-level (conformal weight) $M$, and extract the
spectral observables.

Consider the
boundary state \eqn{falling D0} describing a D0-brane absorbed by
the future horizon. The radiation emitted by the D0-brane is
decomposable into `incoming' (toward the horizon) and `outgoing'
(toward the null infinity) components in the far future. The
positive energy sector is expanded by the wave function $U^p_{\om}$,
and has the following asymptotic behavior at $t\rightarrow +\infty$:
\begin{eqnarray}
 && U^p_{\om}(\rho,t) \sim e^{- i\om \ln \rho -i \om t}
  + d(p, \om) e^{-\rho} e^{+ ip \rho -i\om t} \qquad \mbox{where}
  \qquad  |d(p, \om)| \sim e^{-\pi p}~.
\label{as U}
\end{eqnarray}
Here, we assumed $\om \sim M \gg 0$. The first and the second terms
correspond to the incoming wave supported around $\rho=0$ and the
outgoing wave supported in the region $\rho\sim +\infty$,
respectively. The damping factor $d(p)$ originates from the exact
reflection amplitude $\cR(p,\om)$. (See \eqn{decomp ef 2},
\eqn{decomp ef 2-2}.) To obtain the radiation number density, we
need to evaluate $\left|\Psi(p,\om)\right|^2 \times
|U^{p}_{\om}(\rho,t)|^2$. At far future infinity, the interference
term in $|U^p_\om|^2$ drops off upon taking the $p$-integral.
Therefore, after integrating over the radial momentum $p$, the
partial radiation distribution is seen to consist of the `incoming'
and `outgoing' parts:
\begin{eqnarray}
 \cN(M)_{\msc{in}} &\equiv& \int_0^M \dd M {\dd \cN_{\msc{in}}
 \over \dd M} = \int_0^{\infty} \frac{\dd p}{2 \pi} \frac{1}{2\om(p,M)}
 \Big|\Psi(p,\om(p,M))\Big|^2 \nn
\cN(M)_{\msc{out}} &\equiv& \int_0^M \dd M {\dd \cN_{\msc{out}}
\over \dd M} = \int_0^{\infty} \frac{\dd p}{2
\pi}\frac{1}{2\om(p,M)} \Big|d(p) \Big|^2
\Big|\Psi(p,\om(p,M))\Big|^2~. \label{radiation rate 1}
\end{eqnarray}
We shall now evaluate the branching ratio between the two radiation
rates \eqn{radiation rate 1} with emphasis on possible string
worldsheet effects. To this end, consider the conformal field theory
defined by $SL(2;\br)/U(1) \times \cM$, where $SL(2;\br)/U(1)$
denotes the (super)coset model and $\cM$ denotes a unitary
(super)conformal field theory of central charge $c_\cM$. Such
(super)conformal field theory covers a variety of interesting string
theory backgrounds. For the fermionic string, superconformal
invariance asserts that the central charge ought to be critical:
\begin{eqnarray}
 3\Big(1 + \frac{2}{k}\Big) + c_\cM =15~, \nonumber
\end{eqnarray}
where $k$ denotes the level of the super $SL(2;\br)$ current
algebra. If the background describes a stack of black NS5-branes,
$\cM= SU(2)_{k} \times \br^5$ where $k$ equals to the NS5-brane
charge. Likewise, for the bosonic string case, conformal invariance
asserts that the central charge should take the critical value:
\begin{eqnarray}
 2 + \frac{6}{\kappa-2} + c_{\cM} =26~,
\end{eqnarray}
where now $\kappa$ refers to the level of the bosonic $SL(2;\br)$
current algebra. For the background describing the black hole in
two-dimensional string theory, $\cM$ is empty and $\kappa$ should be
set to $9/4$.

It would be illuminating to analyze the branching ratio for the `rolling closed
string', viz. a closed string state of fixed transverse mass $M$ and
radial momentum $p$ propagating in black hole geometry. The
branching ratio is simply given by the reflection amplitude (see
\eqn{exactra}):
\bea \left. {{\cal N}_{\rm out}(p, \om) \over {\cal N}_{\rm in } (p,
\omega)} \right|_{\rm closed \, string} = |{\cal R}(p, \om)|^2 =
{\cosh^2 \pi \left(\om-p \over 2 \right) \over \cosh^2 \pi \left(
{\om+p \over 2}\right)}~. \label{tachyon} \eea
As emphasized below \eqn{exactra}, string worldsheet effects are
present for the reflection amplitude ${\cal R}$ itself but, being an
overall phase, it drops out of \eqn{tachyon}. The $k$-dependence
enters in the branching ratio \eqn{tachyon} only through the
on-shell dispersion relation $\omega = \sqrt{p^2 + 2 k M^2}$. For
two-dimensional case, first studied in \cite{DVV} and \cite{GKPS},
$k=1/2$, $M=0$ and $\omega = p$, so the scattering probability is
exponentially suppressed as the energy increases.

For a fixed transverse mass $M$ {\sl and} the forward radial
momentum $p$, the reflection probability of the infalling D0-brane
is given precisely by the same result as \eqn{tachyon}:
\bea \left. {{\cal N}_{\rm out}(p, \om) \over {\cal N}_{\rm in } (p,
\omega)} \right|_{\rm D0-brane} = |{\cal R}(p, \om)|^2 = {\cosh^2
\pi \left(\om-p \over 2 \right) \over \cosh^2 \pi \left( {\om+p
\over 2}\right)}~. \eea
This is simply because back-scattering of the boundary wave function
originates from that of the closed string wave function: roughly
speaking, the boundary wave function is defined by overlap of the
closed string wave function with the classical trajectory of the
D0-brane.

Radiation out of the falling D0-brane is coherent, so we integrate
over the radial momentum $p$ as in \eqn{radiation rate 1} in
extracting the branching ratio. We shall first analyze the partial
radiation distribution at large mass-level, $M \rightarrow \infty$.
More precisely, we shall examine asymptotic behavior of ${\cal N}
(M)$ multiplied by the phase-space `degeneracy factor' $\rho(M)\sim
e^{\frac{1}{2}M \beta_{\rm Hg}}$, where $\beta_{\rm Hg}$ denotes
inverse of the Hagedorn temperature. The closed string states that
couple to the boundary states are left-right symmetric, so we need
to take the square root of the usual degeneracy factor in the closed
string sector. Here, inverse of the Hagedorn temperature is given by
\begin{eqnarray}
\beta_{\rm Hg} = 4\pi \sqrt{1-\frac{1}{2k}}~, ~~~ \label{Hagedorn
super}
\end{eqnarray}
for the superstring theory, and
\begin{eqnarray}
\beta_{\rm Hg} = 4\pi \sqrt{2-\frac{1}{2(\kappa-2)}}~, ~~~
\label{Hagedorn bosonic}
\end{eqnarray}
for the bosonic string theory, where the $1/k $
$(1/\kappa)$-correction is interpreted as the string worldsheet
effects of the two-dimensional background. These results are
derivable from the Cardy formula with the `effective central charge'
$c_{\msc{eff}}=c-24 h_{\msc{min}}
$ \cite{KutS}, where $h_{\msc{min}}$ refers to the lowest conformal
weight of normalizable primary states.

\subsection{Radiation distribution in superstring theory}
\label{radiation super}
Begin with the spectral distribution in
superstring theories. We shall focus exclusively on the NS-NS sector
of the radiation and defer the analysis of the R-R sector to section
6. The on-shell condition of closed string state in NS-NS sector is
given by
\begin{eqnarray}
&& -\frac{\om^2}{4k} + \frac{p^2}{4k}+ \frac{1}{4k}+ \Delta_\cM =
\frac{1}{2}~, \label{on-shell super}
\end{eqnarray}
where $\Delta_\cM$ denotes the conformal weight of the $\cM$-part.
The on-shell energy is given by
\begin{eqnarray}
\om \equiv \om(p,M) = \sqrt{p^2+2k M^2} \qquad \mbox{where} \qquad
M^2 \equiv 2 \left(\Delta_\cM + \frac{1}{4k} - \frac{1}{2} \right)~.
\nonumber \end{eqnarray}
Consider now a D0-brane propagating outside the black hole and
absorbed into the future horizon. The relevant boundary wave
function was constructed in \eqn{falling D0} and, from them, the
differential radiation number distributions \eqn{radiation rate 1}
can be computed. At large $\omega$ and $p$, using Stirling's
approximation, we find that
\begin{eqnarray}
\cN (M)_{\msc{in}} &=& \int_0^{\infty}\frac{\dd p}{2 \pi} \frac{1}
{2\om(p,M)}\, \Big| \Psi_{\rm absorb} (\rho_0,t_0;p,\om(p,M))\Big|^2
\nonumber
\\
&\sim& {1 \over M} \int_0^{\infty}{\dd p}\,
e^{+\pi\left(1-\frac{1}{k}\right)p - \pi\sqrt{p^2+2k M^2}}~ \label{N
M super in}\\
\nonumber \\
\cN (M)_{\msc{out}} &=& \int_0^{\infty}\frac{\dd p}{2 \pi} \,
\Big|d(p,\omega(p,M))\Big|^2 \, \frac{1} {2 \om(p,M)} \Big|
\Psi_{\rm absorb} (\rho_0,t_0;p,\om(p,M))\Big|^2 \nonumber \\
&\sim& {1 \over M} \int_0^{\infty} {\dd p} \, e^{-\pi
\left(1+\frac{1}{k}\right)p - \pi\sqrt{p^2+ 2k M^2}}~. \label{N M
super out}
\end{eqnarray}
In the second lines, we have taken $M$ large, viz. $\omega \gg p \gg
1$, and keep the leading terms only. Thus, for each fixed but large
$M$, the partial number distributions take the forms:
\begin{eqnarray}
\cN (M)_{\msc{in}} \sim \int_0^{\infty} \dd p \,
\sigma_{\msc{in}}(p) e^{-\frac{1}{2}\beta_{\rm Hw} M} \qquad
\mbox{and} \qquad N(M)_{\msc{out}} \sim \int_0^{\infty} \dd p \,
\sigma_{\msc{out}}(p) e^{-\frac{1}{2}\beta_{\rm Hw} M}~,
\label{grey body}
\end{eqnarray}
where
\begin{eqnarray}
 \beta_{\rm Hw} = 2\pi \sqrt{2k}~,
\label{Hawking temp super}
\end{eqnarray}
is the inverse Hawking temperature of the fermionic two-dimensional
black hole. As discussed above, the radiation off the D-brane in
NS5-brane background is analogous to the decay of excited D-brane in
flat ambient spacetime.
Indeed, asymptotic expression \eqn{grey
body} suggests that open string
excitations of energy $M$ on
the rolling D0-brane are
populated as the distribution function $\exp
(-{1 \over 2} \beta_{\rm Hw} M)$
and decay into closed string
radiation.
In this interpretation, the distribution function encodes
change of available states for open string excitations on the
D0-brane after emitting radiations of energy $M$.
Curiously,
`effective temperature'
of the excited closed strings is set by the
Hawking temperature of the nonextremal NS5-brane, not that of a
black hole that would have been made
of the D0-brane. It is tempting
to interpret this as indicating
that the D0-brane represents a class
of possible excitation modes of the black NS5-brane. The closed
string states of energy $M$ emitted
by the D0-brane are certainly coherent,
but according to this interpretation, they still can be
recasted in effective thermal distribution set by the Hawking
temperature of the two-dimensional black hole.
In the next subsection, we shall account for the origin of such
effective thermal behavior of the rolling D0-brane from the
viewpoints of Euclidean cylinder amplitude between D1-brane,
extending the argument of \cite{DVV} for the Hawking radiation of
closed strings in the black hole background.

The functions
$\sigma_{\rm in}$ and $\sigma_{\rm out}$ are
interpretable as the black hole `greybody' factors
for incoming and outgoing parts of the
radiation. The factor 1/2 in the exponent of the Boltzmann
distribution function reflects the fact that only left-right
symmetric closed string states can appear
in the boundary states and the radiated closed string modes.

The `greybody factors' $\sigma_{*}(p)$ depend on the radial momentum
$p$ exponentially, so the radiation distribution would be modified
{\sl once} the radial momentum $p$ is integrated out. Below, we
shall show this explicitly. We are primarily interested in keeping
track of string worldsheet effects set by the value of the level
$k$. We shall consider different ranges of the level $k$ separately,
and focus on the asymptotic behaviors at large $M$ via the saddle
point methods.

\begin{description}
 \item[(i) \underline{$k > 1$}: ] \hfill\break

This is the case for the black NS5-brane background. Consider first
the incoming part. Since $1- \frac{1}{k}> 0$, the dominant
contribution in the $p$-integral arises from the saddle point:
\begin{eqnarray}
p \sim p_* = \frac{k-1}{\sqrt{1-\frac{1}{2k}}} M~.\nonumber
\end{eqnarray}
Substituting this to \eqn{N M super in}, we obtain
\begin{eqnarray}
\cN (M)_{\msc{in}} \sim e^{-2\pi M \sqrt{1-\frac{1}{2k}}} =
e^{-\frac{1}{2}M \beta_{\rm Hg}}~, \label{eq:in}
\end{eqnarray}
up to pre-exponential powers of $M$. Taking account of the density
of states $\rho(M) \sim e^{\frac{1}{2}M \beta_{\rm Hg}}$, we find
that $\rho(M) \cN(M)_{\msc{in}}$ scales with powers of $M$, and is
independent of $k$. More explicitly, for the black NS5-brane $\cM=
SU(2)_{k} \times \mathbb{R}^5$, the incoming radiation
distribution of the D$p$-brane parallel to the NS5-brane yields
\begin{eqnarray}
\cN(M)_{\msc{in}} &\sim& {1 \over M} \int {\dd^{5-p} {\bf k}_{\perp}
\over (2 \pi)^{5-p}} \int_0^\infty {\dd p} \, e^{\pi
(1-\frac{1}{k})p-\pi\sqrt{p^2+2k(M^2+{\bf k}_{\perp}^2)}}
\nonumber \\
&\sim& M^{2-\frac{{p}}{2}}\, e^{-2\pi M\sqrt{1-\frac{1}{2k}}} \ . \nonumber
\end{eqnarray}
Taking  account of the density of states $\rho(M) \sim
M^{-3}e^{2\pi M\sqrt{1-\frac{1}{2k}}}$, the average radiation number
distribution is given by
\begin{eqnarray}
\frac{\overline{\cN}_{\msc{in}}}{V_p} \sim \int^{M_{\rm D}} {\dd M
\over M} \, M^{-\frac{p}{2}} \qquad \mbox{where} \qquad M_{\rm D} \sim {\cal
O}({1 \over g_{\rm st}}) \ . \label{conL}
\end{eqnarray}
This result coincides with the computations of \cite{LLM,KLMS}, and
corroborates with the radion-tachyon correspondence. Interestingly,
the incoming part of the radiation number distribution in the the
nonextremal NS5-brane background is exactly the same as the
distribution in the extremal NS5-brane background. Later, we shall
examine carefully taking the extremal limit and its consequence in
section \ref{ext}. As in the extremal case, \eqn{conL} implies that
nearly all the D0-brane potential energy is released into closed
string radiations before it falls into the black hole.

On the other hand, for the outgoing radiation, the far infrared $p
\sim 0$ dominates the momentum integral. We thus obtain
\begin{eqnarray}
\cN (M)_{\msc{out}} \sim e^{-2\pi M \sqrt{\frac{k}{2}}} =
e^{-\frac{1}{2}M \beta_{\rm Hw}}~, \nonumber
\end{eqnarray}
displaying effective thermal distribution set by the Hawking
temperature. Taking account of the density of states,
\begin{eqnarray}
\rho(M) \cN (M)_{\msc{out}} \sim e^{\frac{1}{2}M \left(\beta_{\rm
Hg}-\beta_{\rm Hw}\right)} = e^{2\pi M
\left(\sqrt{1-\frac{1}{2k}}-\sqrt{\frac{k}{2}} \right)}~. \nonumber
\end{eqnarray}
This is ultraviolet finite for any $k$ since
\begin{eqnarray}
\left(1-\frac{1}{2k}\right) -\frac{k}{2} = -\frac{1}{2k}
\left(k-1\right)^2 < 0~. \label{eq:ini}
\end{eqnarray}

We thus conclude that the radiation number distribution is mostly in
the incoming part:
\bea \left. {{\cal N}_{\rm out}(M) \rho (M) \over {\cal N}_{\rm
in}(M) \rho (M)} \right\vert_{\rm falling \,\, D0} \sim {e^{-{1
\over 2} \beta_{\rm Hw} M } \over e^{-{1 \over 2} \beta_{\rm Hg} M}}
= e^{2 \pi M \left( \sqrt{1 -{1 \over 2 k}} - \sqrt{k \over 2}
\right)} \ll 1\ . \nonumber \eea
Intuitively, this may be understood as follows: for the absorbed
boundary state, the boundary condition is such that the D0-brane
flux is directed from past null infinity to the future horizon. This
also corroborates the observation that $T_{t\rho}$-component of
D0-brane's energy-momentum tensor is nonzero and increases
monotonically as the D0-brane approaches the future horizon. The
outgoing part of the distribution is exponentially small compared to
the incoming part and exhibits effective thermal distribution at the
Hawking temperature. Notice that, despite being so, this outgoing
part has nothing to do with the Hawking radiation of the black hole.
The latter is the feature of the background by itself. A priori, the
outgoing radiation could be in a distribution characterized by a
temperature different from the Hawking temperature. As mentioned
above, it is tempting to interpret coincidence of the two
temperatures as a consequence of maintaining equilibrium between the
black NS5-brane and the D0-brane.

\item[(ii) \underline{$\frac{1}{2} < k  \leq 1$}: ]
\hfill\break

This is the regime which includes the conifold geometry at $k=1$.
Since $1-\frac{1}{k} \leq 0$, the dominant contribution to the
momentum integral is from $p \sim 0$, not only for the outgoing
radiation but also for the incoming one. We thus obtain
\begin{eqnarray}
\cN(M)_{\msc{in}} \sim \cN (M)_{\msc{out}} \sim e^{-2\pi M
\sqrt{\frac{k}{2}}} \equiv e^{-\frac{1}{2}M \beta_{\rm Hw}}~,
\end{eqnarray}
viz. both are in effective thermal distribution set by the Hawking
temperature. All spectral moments are manifestly ultraviolet finite
since, at large $M$, exponential growth of the density of the final
closed string states is insufficient to overcome the suppression by
the distribution. Thus,
\bea {{\cal N}_{\rm out}(M) \rho(M) \over {\cal N}_{\rm in} (M) \rho
(M)} \Big|_{\rm falling \, D0} \sim 1\ . \nonumber \eea
We interpret this as indicating that the D0-brane does not radiate
off most of its energy before falling into the horizon.

\item[(iii) \underline{$k=\frac{1}{2}$} : ]
\hfill\break

This special case corresponds to empty $\cM$. The two-dimensional
background permits no transverse degrees of freedom of the string.
The physical spectrum includes massless tachyon only, with $M=0$ and
$\rho(M)=1$. We now have a crucial difference from the previous
cases for the on-shell configurations. The radial momentum $p$ is
fixed by the on-shell condition as $\om = \pm p$, so it should not
be integrated over for the final states. Consequently, we cannot
decompose the radiation distribution into incoming and outgoing
radiations,
and only the total distribution is physically relevant.

We thus obtain the following large $\om$ behavior of the radiation
distribution:
\begin{eqnarray}
\cN(\om)\sim  e^{-2\pi \om} \equiv e^{-\om \beta_{\rm Hw}}~.
\label{radiation 2D BH super}
\end{eqnarray}
Again, we have found effective thermal distribution at the Hawking
temperature! Notice the absence of extra 1/2-factor in contrast to
the previous regimes. This is not a contradiction. In the present
case, the transverse oscillators are absent and the string behaves
as a point particle. Again, the D0-brane does not radiate off most
of its energy before falling across the black hole horizon.
\end{description}


\subsection{Radiation distribution in bosonic string theory}

The analysis for the bosonic string case proceeds quite the same
route. The boundary state for the infalling D0-brane includes the
string worldsheet correction factor
$\Gamma\left(1+i\frac{p}{\kappa-2}\right)$, where again $\kappa$
refers to the level of bosonic $SL(2;\br)/U(1)$ coset model. The
on-shell condition now reads
\begin{eqnarray}
&& -\frac{\om^2}{4\kappa} + \frac{p^2}{4(\kappa-2)}+
\frac{1}{4(\kappa-2)} + \Delta_\cM = 1~, \label{on-shell bosonic}
\end{eqnarray}
where $\Delta_\cM$ denotes the conformal weight in the $\cM$-sector.
This is solved by
\begin{eqnarray}
&& \om \equiv \om(p,M) = \sqrt{\frac{\kappa}{\kappa-2}p^2+2\kappa
M^2} \qquad \mbox{where} \qquad M^2 \equiv 2 \left(\Delta_\cM +
\frac{1}{4(\kappa-2)} - 1\right)~.
\end{eqnarray}
The partial radiation number distribution at large $M$ limit is
given by:
\begin{eqnarray}
&& \cN(M)_{\msc{in}}
 \sim  {1 \over M} \int_0^{\infty} {\dd p} \,
e^{+\pi \left(1- \frac{1}{\kappa-2}\right)p - \pi
\sqrt{\frac{\kappa}{\kappa-2} p^2 + 2\kappa M^2}}~.
\label{N M bosonic in} \\
&& \cN(M)_{\msc{out}}
 \sim  {1 \over M} \int_0^{\infty} {\dd p} \,
e^{-\pi \left(1+ \frac{1}{\kappa-2}\right)p
- \pi \sqrt{\frac{\kappa}{\kappa-2}p^2+ 2\kappa M^2}}~.
\label{N M bosonic out}
\end{eqnarray}
Thus, as in the superstring case, there can arise several distinct
behaviors depending on how stringy the background is.


\begin{description}
 \item[(i) \underline{$\kappa > 3$}: ]
\hfill\break
Consider first the incoming radiation part. Since
$1-\frac{1}{\kappa-2}
> 0$, the dominant contribution to the momentum integral in \eqn{N M
bosonic in} is from the saddle point
\begin{eqnarray}
p \sim p_* = \frac{\kappa-3} {\sqrt{2-\frac{1}{2(\kappa-2)}}} M~.
\nonumber
\end{eqnarray}
We thus obtain, up to pre-exponential powers of $M$,
\begin{eqnarray}
\cN(M)_{\msc{in}} \sim e^{-2\pi M \sqrt{2-\frac{1}{2(\kappa-2)} } }
\, = \, e^{-\frac{1}{2} M \beta_{\rm Hg}}~, \nonumber
\end{eqnarray}
where $\beta_{\rm Hg}$ denotes the Hagedorn temperature of the
bosonic string theory \eqn{Hagedorn bosonic}.
In this way, we again find the power-law behavior of $\rho(M) \cN
(M)_{\msc{in}}$ at large $M$, independent of the level $\kappa$.

For the outgoing radiation part, again the $p\sim 0$ dominates the
momentum integral in \eqn{N M bosonic out}. The result is
\begin{eqnarray}
\cN (M)_{\msc{out}} \sim e^{-2 \pi M \sqrt{\frac{\kappa}{2}}} =
e^{-\frac{1}{2} M \beta_{\rm Hw}}~. \nonumber
\end{eqnarray}
Here,
\begin{eqnarray}
 \beta_{\rm Hw} \equiv 2\pi \sqrt{2\kappa} \nonumber
\end{eqnarray}
is the Hawking temperature of the bosonic two-dimensional black
hole. We then obtain
\begin{eqnarray}
\rho(M) \cN (M)_{\msc{out}} \sim e^{\frac{1}{2}\left(\beta_{\rm
Hg}-\beta_{\rm Hw}\right)M} = e^{2\pi M
\left[\sqrt{2-\frac{1}{2(\kappa-2)}} -\sqrt{\frac{\kappa}{2}}
\right] }~. \nonumber
\end{eqnarray}
As in the superstring case, the exponent is always negative
definite:
\begin{eqnarray}
\left(2-\frac{1}{2(\kappa-2)}\right)
-\frac{\kappa}{2} =
-\frac{\left(\kappa-3\right)^2}{2(\kappa-2)}
 \leq  0~. \nonumber
\end{eqnarray}
so the outgoing radiation distribution (as well as spectral moments)
is manifestly ultraviolet finite.

Physical interpretation of the above results is the same as the
superstring case: The D0-brane falling into the black hole has
nonzero component $T_{t\rho}$ of the energy-momentum tensor, and
entails that dominant part of the closed string radiation is
incoming toward the future horizon. The outgoing part of the
radiation is exponentially suppressed, and is in effective thermal
distribution set by the Hawking temperature. Again, this
distribution is distinct from the Hawking radiation of the
two-dimensional black hole. As for the fermionic string, the
branching ratio is exponentially suppressed.

\item[(ii) \underline{$\frac{9}{4} < \kappa \leq 3$}: ]
\hfill\break
In this regime, $1- \frac{1}{\kappa-2} < 0$ and the momentum
integrals for both incoming and outgoing radiation distributions are
dominated by $p \sim 0$:
\begin{eqnarray}
\cN(M)_{\msc{in}} \sim \cN(M)_{\msc{out}} \sim  e^{-2 \pi M
\sqrt{\frac{\kappa}{2}}} \equiv e^{-\frac{1}{2}M \beta_{\rm Hw}}~.
\nonumber
\end{eqnarray}
Both are in effective thermal distribution at the Hawking
temperature, and all spectral moments are manifestly ultraviolet
finite since, at large $M$, the growth of the density of state does
not overcome the suppression by the distribution. The branching
ratio remains order unity.

\item[(iii) \underline{$\kappa=\frac{9}{4}$} : ]
\hfill\break
This is the most familiar situation: black hole in two-dimensional
bosonic string theory, originally studied in \cite{2DBH}. The
physical spectrum of closed string consists only of the massless
tachyon, so we again need to set $M=0$ and $\rho(M)=1$. The
calculation is slightly more complicated than the supersymmetric
case: The canonically normalized energy is
\begin{eqnarray}
 E = \frac{\sqrt{2}}{3}\om = \sqrt{2}p~, \nonumber
\end{eqnarray}
so we obtain
\begin{eqnarray}
&& \cN (E) \sim  e^{\pi \left(1-\frac{1}{1/4}\right) p -\pi
\sqrt{\frac{9/4}{1/4}}p} = e^{-3 \sqrt{2}\pi E} \equiv e^{- E
\beta_{\rm Hw}}~. \nonumber
\end{eqnarray}
It again shows effective thermal distribution of the radiated closed
string modes at the Hawking temperature: $\beta_{\rm Hw} = 2\pi
\sqrt{2\kappa} = 3\pi \sqrt{2}$.

\end{description}


\subsection{Radiation distribution for emitted or time-symmetric boundary states}

The closed string radiations for the other types boundary states,
viz. the `emitted' \eqn{emitted D0} or the `symmetric'
\eqn{symmetric D0} D0-branes, can be studied analogously.

For the emitted D0-brane boundary state \eqn{emitted D0}, by the
time-reversal, we should observe the radiation distribution at the
far past: $t\sim -\infty$. The relevant decomposition corresponding
to \eqn{as U} is given by (assuming $\om>0$, $p>0$)
\begin{eqnarray}
 && V^p_{\om}(\rho,t) \sim e^{i\om \ln \rho-i\om t}
  + d^*(p,\om) e^{-\rho} e^{-ip\rho -i\om t} ~,
\label{as V}
\end{eqnarray}
where the first term is supported near the past horizon
and the second term corresponds to the incoming wave from the null
infinity.
Obviously we find precisely the same behavior of the radiation
distribution as the absorbed D0-brane once the role of `in' and
`out' states are reversed. So, for $k>1$, ${\cal N}(M)_{\rm in} \sim
\exp ( - {1 \over 2} \beta_{\rm Hw} M)$ while ${\cal N}(M)_{\rm out}
\sim \exp (-{1 \over 2} \beta_{\rm Hg} M)$ and, for $ 1 \ge k >
1/2$, ${\cal N}(M)_{\rm in}$, ${\cal N}(M)_{\rm out} \sim \exp (-{1
\over 2} \beta_{\rm Hw} M)$.

Consider next the boundary state describing D0-brane in symmetric
boundary condition \eqn{symmetric D0}. Recalling the relations
\eqn{rel disc amp}, one finds that the radiation rates are simply
obtained by adding contributions from `absorbed' and `emitted'
D0-brane boundary states. So, the radiation distributions behave as
${\cal N}(M)_{\rm in}$, ${\cal N}(M)_{\rm out} \sim \exp (-{1 \over
2} \beta_{\rm Hg} M)$ for $k>1$ and the dependence on Hawking
temperature disappeared.\footnote{Dependence on the Hawking
temperature exponentially suppressed, so completely negligible
compared to other power-suppressed subleading terms.} We then find
that the `detailed balance' $\cN(M)_{\msc{in}} = \cN(M)_{\msc{out}}$
is obeyed. This is as expected since the boundary state
\eqn{symmetric D0} is defined so that it keeps the time-reversal
symmetry and the one-particle state unitarity manifest.

\subsection{Revisit to the radiation distribution:
thermal string propagator}

We shall revisit the radiation distribution and discuss salient
features of the distribution from another different angle. Argument
we shall present here would be somewhat heuristic, but we feel it
quite helpful for grasping physical intuition and for understanding
how the effective thermal behavior of the radiation comes about.
This argument is similar to the one given in \cite{DVV}, where
thermal distribution of the Hawking radiation in the two-dimensional
black hole background was observed via the closed string thermal
propagator. Our foregoing discussion is an extension of theirs to
the open string sector.

Consider the thermal cylinder amplitude for the D1-brane on the
Euclidean cigar \eqn{hairpin D1}.\footnote{To be more precise, we consider
   the fermionic black-hole of level $k$
    and focus on the space-time bosons.
    If the space-time fermions are considered, the thermal
    Kaluza-Klein momenta should be half integer-valued
    $n\in 1/2 + \bz$ instead of being integer-valued $n\in \bz$ as for the bosons.
    This change leads to the Fermi-Dirac distribution
   $(e^{\beta_{\msc{Hw}}\om_{p,M}}+1)^{-1} $ instead of
 the Bose-Einstein $(e^{\beta_{\msc{Hw}}\om_{p,M}}-1)^{-1}$ in the
 following argument.}
Schematically, the amplitude is evaluated as (we omit the parameters
$\rho_0$, $\theta_0$ for simplicity)
\begin{eqnarray}
 \hspace{-1cm}
\cA_{\msc{cylinder}}^{(E)} &=& \int_0^{\infty} \dd T\,
 {}_{D1} \Big< B \Big\vert e^{-\pi T H^{(c)}} \Big\vert B \Big>_{D1}
 \approx  \sum_M \sum_{n\in \bsz} \int {\dd  p \over 2 \pi} \,
 \frac{1}{p^2+ \left(\frac{2\pi n}{\beta_{\msc{Hw}}} \right)^2+M^2}
 \, \sqrt{\rho_c(M)} \Big|\Psi_{D1}(p,n)\Big|^2~ \nn
& =& \beta_{\msc{Hw}} \sum_M \int {\dd p \over 2 \pi} \int {\dd q
\over 2 \pi} \, \sqrt{\rho_c(M)} \frac{\left|\Psi_{D1}
\left(p,\frac{q}{2\pi}\beta_{\msc{Hw}}\right)
\right|^2}{p^2+q^2+M^2}\, \left(1+ \sum_{m\in \bsz_{>0}}
e^{i\beta_{\msc{Hw}} m q}
 + \sum_{m\in \bsz_{>0}}e^{-i \beta_{\msc{Hw}} m q}
\right)~.
\label{thermal cylinder}
\end{eqnarray}
Here, $p$ is the radial momentum, $n$ is the Kaluza-Klein momentum
around the asymptotic circle (thermal circle) of the cigar geometry,
$M$ is again the transverse mass in the $\cM$-sector, $\rho_c(M)$ is
the density of the closed string states, and $\beta_{\msc{Hw}}\equiv
2\pi \sqrt{2k}$ is the inverse Hawking temperature. We now Wick
rotate the cylinder amplitude by contour deformation of the
$q$-integration in the manner similar to \cite{DVV}. 
By formal manipulation $q \to i \om$ \footnote
    {Here, to make comparison with \cite{DVV} easier, we normalized
    $\omega$,  $p$ as $L_0 = - \frac{1}{2} \om^2 + \frac{1}{2}p^2
   + \cdots$, rather than $L_0 = -\frac{1}{4k} \om^2 + \frac{1}{4k}
   p^2 + \cdots$.} and
$\cA^{(E)}_{\msc{cylinder}} \to i \cA^{(L)}_{\msc{cylinder}}$, we
obtain
\begin{eqnarray}
 \cA^{(L)}_{\msc{cylinder}}& \approx&
\beta_{\msc{Hw}} \sum_M \int {\dd p \over 2 \pi} \,
\sqrt{\rho_c(M)}\, \left\lb \int {\dd \om \over 2 \pi} \, \frac{
\left|\Psi_{\rm D1}\left(p,\frac{i \om}{2 \pi} \beta_{\msc{Hw}}
\right) \right|^2 }{p^2+M^2-\om^2+i\ep} -\frac{2\pi i }{\om_{p,M}}
\frac{\left|\Psi_{D1}\left(p, \frac{i \om_{p,M}}{2 \pi}
\beta_{\msc{Hw}} \right)\right|^2} {e^{\beta_{\msc{Hw}}
\om_{p,M}}-1} \right\rb ~,
 \label{evaluation AL}
\end{eqnarray}
where we denoted $\om_{p,M} \equiv \sqrt{p^2+M^2}$ for the on-shell
energy and used the identity
$$
\left|\Psi_{D1}\left(p,-\frac{i \om_{p,M}}{2 \pi} \beta_{\msc{Hw}}
\right)\right|^2 = \left|\Psi_{D1}\left(p,\frac{i \om_{p,M}}{2\pi}
\beta_{\msc{Hw}} \right)\right|^2~.
$$
Since $ \left|\Psi_{D1}\left(p,\frac{i\om}{2\pi} \beta_{\msc{Hw}}
\right) \right|^2$ is proportional to $\exp
(\frac{1}{2}\beta_{\msc{Hw}} |\om|)$, the first term (including the
Feynmann propagator) gives rise to an ultraviolet divergent
contribution. This is not surprising and reveals the reason why the
naive Wick-rotation of \eqn{hairpin D1} is not viable. The second
term ($\equiv \cA_{\msc{thermal}}^{(L)} $) exhibits an effective thermal distribution. More pertinently,
this term contributes to the imaginary part of the thermal cylinder
amplitude we are interested in. Indeed, it yields the anticipated
behavior:
\begin{eqnarray}
\Im \, \cA_{\msc{thermal}}^{(L)} &\sim&
  \frac{1}{\om_{p,M}}
\frac{1}{e^{\beta_{Hw}\om_{p,M}}-1} \, \sqrt{\rho_c(M)}
 \left|\Psi_{D1}\left(p,\frac{i \om}{2 \pi} \beta_{\msc{Hw}}
\right)\right|^2 \nonumber \\
&\sim& \frac{\sqrt{\rho_c(M)}}{\om_{p,M}}\,\sigma(p)\,
 e^{-\frac{1}{2}\beta_{\msc{Hw}} \om_{p,M}}~,
\label{thermal behavior}
\end{eqnarray}
and reproduces the previous results \eqn{grey body} including the
correct greybody factor $\sigma(p)$ and the density of the radiated
closed string states $\sqrt{\rho_c(M)} \sim \rho(M)$. Recall that,
in our construction of the Lorentzian boundary states, the damping
factor was crucial, which reads in the present conventions as $\sinh
(\pi \sqrt{2k} p) \cosh^{-1}( \pi \sqrt{\frac{k}{2}}(p+\om))
\cosh^{-1} ( \pi \sqrt{\frac{k}{2}}(p-\om) ) $. At large $\om$ or
large $p$, this damping factor shows the same asymptotic behavior as
the Boltzmann distribution function $(e^{\beta_{\msc{Hw}}\om}\pm
1)^{-1}$. In this sense, our prescription of Wick-rotating the
Euclidean boundary states would be roughly identified with the
prescription of keeping only the finite second term in \eqn{evaluation AL}.
This then explains origin of the effective thermal distribution as
derived from the Lorentzian boundary states.

As yet another viewpoint, consider the thermal cylinder amplitude
\eqn{thermal cylinder} in the open string channel. For simplicity,
concentrate on the asymptotic region $\rho \gg 0$. The hairpin
D1-brane \eqn{hairpin D1} appears just as two halves of the
$D1$-$\overline{D}1$ system, which put open strings around the
thermal circle to obey Dirichlet boundary condition (so, identified
as the `s$D$-s$\bar{D}$ system' \cite{Strominger}), as pointed out
in \cite{NST}. In this set up, for simple kinematical reasons, we
find {\em on-shell} closed string states in the cylinder amplitude,
while only {\em off-shell} states in the open string channel.  As
discussed {\em e.g.} in \cite{thermal open}, using the modular
transformation, it can be shown that the thermal distribution of
{\em physical} closed string states emitted/absorbed by the
s$D$-s$\bar{D}$ system is captured by the {\em unphysical} open
string winding modes along the thermal circle.\footnote
   {This is a simple extension of the standard argument
   concerning the thermal toroidal partition functions
    \cite{thermal closed}. For instance, the Hagedorn behavior
    is interpretable as the tachyonic instability due to the
     {\em unphysical} winding modes around the thermal circle.
     }
Especially, unit of the winding energy should determine temperature
of the thermal distribution of closed string states coupled with the
s$D$-s$\bar{D}$ system. In the present case, it is identified with
the interval of the hairpin $(= \frac{1}{2}\beta_{\msc{Hw}})$, which
is just associated to the open string stretched between
$D1$-$\bar{D}1$. (Notice that, taking suitably the GSO projection
into account, we can check that the zero winding modes, {\em i.e.}
the $D1$-$D1$ or $\bar{D}1$-$\bar{D}1$ strings, are canceled out.
See \cite{NST}.) This would be the simplest explanation for the
reason we get the effective thermal distribution
$\exp(-\frac{1}{2}\beta_{\msc{Hw}}\om_{p,M})$ from the cylinder
amplitude \eqn{thermal cylinder}.

As already noted in footnote 9, all the {\em regular} geodesics of
the $D0$-brane motion are just straight lines in the Kruskal
coordinates. Once Wick-rotated back to the hairpin profiles of
Euclidean D1-brane, this means that they all have the {\em same}
interval $\frac{1}{2}\beta_{\msc{Hw}}$ around the thermal circle.
This observation leads us again to the same effective thermal
behavior \eqn{grey body} characterized by the Hawking temperature
(before integrating $p$ out),\footnote
   {One might ask why the D0-brane motion with different
    `temperature' is not considered. Such case corresponds
to singular hairpin profiles and hence to singular Lorentzian
trajectories of the D0-brane. They cannot be solutions of the
D0-brane's DBI action because of discontinuity of the velocity at
the singular points. Quite interestingly, this feature is strikingly
similar to the original Hawking's prescription for black hole
temperature: demanding the Euclidean geometry smooth, we can fix
asymptotic periodicity of the Euclidean time and read off the
temperature characterizing the radiation from the black-hole.} as
is already pointed out.

\section{`String - Black Hole' Transition}
It has been a recurrent theme \cite{BH} that an elementary particle
or a string is a black hole: a configuration consisting of
(multiple) strings with high enough total mass is equivalent to a
black hole of the same mass and other conserved charges. This brings
a question whether a given configuration is most effectively
described in terms of strings or black holes. By the string - black
hole transition, we will refer to such change of the effective
description for a configuration involving massive string
excitations. Roughly speaking, the string is dual to the black hole
and vice versa.

An immediate, interesting question is whether the two-dimensional
black hole geometries is also subject to the string - black hole
transition and if so what precisely the dual of the geometries would
be. In this section, we shall investigate this transition by
studying rolling dynamics of a D0-brane placed on the background. If
the background undergoes the transition between the black hole and
the string configurations, propagation of a probe D0-brane would be
affected accordingly. The transition is triggered by $k$ or
$\kappa$, which measures characteristic curvature scale of the
background measured in sting unit and hence string worldsheet
effects. We shall explore a signal of the transition by examining
spectral distribution of the closed string radiation out of the
rolling D0-brane. Other physical observables associated with
D0-brane would certainly be equally viable probes. Though
straightforward to analyze, in this work, we shall not consider
them.

\subsection{Probing `string - black hole' transition via D-brane}

In the previous section, we observed that $\cN(M)_{\msc{in}} \gg
\cN(M)_{\msc{out}}$ for both the supersymmetric and bosonic string
theories in case the string worldsheet effects are weak enough, viz.
$k > 1$ and $\kappa > 3$, respectively. Obviously, such behavior is
interpretable as indicating that the background on which the
radiative process takes place is indeed a black-hole: D0-brane falls
into the horizon and subsequent radiation is mostly absorbed by the
black hole. On the other hand, the behavior that $\cN
(M)_{\msc{in}}\sim \cN (M)_{\msc{out}} \ll \rho(M)^{-1}$ for $k<1$
or $\kappa < 3$ does not seem to bear features present in the black
hole background: while D0-brane falls inward, subsequent radiation
is not mostly absorbed by the black hole but disperse away. Since
this is the regime where the string worldsheet effects are
significant, the background may be described most effectively in
terms of strings. We are thus led to conclude that the background,
whose stringy effects are controlled by the parameter $k$ or
$\kappa$, would make a phase-transition between the black hole and
the string across $k=1$ or $\kappa = 3$. In a different physical
context, this so-called `black hole-string transition' was studied
recently \cite{KarMS,GKRS}. What distinguishes our consideration and
result from \cite{KarMS,GKRS} is that we are probing possible
phase-transition of the (closed string) background by introducing a
D0-brane in it and studying open string dynamics.


Possible existence of such a phase transition was first hinted in
\cite{Kutasov:1990ua} in the closed string sector, where they
observed that the $\cN=2$ Liouville superpotential becomes
normalizable once $k>1$ and it violates the Seiberg bound. Recall
that the marginal interaction term is
\begin{eqnarray}
S^{\pm} = \psi^{\mp} e^{-\frac{1}{\cQ}(\phi \pm iY)}~,
~~~(\cQ=\sqrt{2/k})
\end{eqnarray}
for the $\cN=2$ Liouville theory, and
\begin{eqnarray}
S^{\pm} =  e^{-\frac{1}{\cQ}(\phi \pm \sqrt{1+\cQ^2}iY)}
\equiv e^{-\sqrt{\frac{\kappa-2}{2}} \phi \mp
\sqrt{\frac{\kappa}{2}} iY}~,
~~~ (\cQ=\sqrt{2/(\kappa-2)})~,
\end{eqnarray}
for the bosonic sine-Liouville theory, respectively. Both
interactions are normalizable (exponentially falling off in the
asymptotic far region) if the curvature is sufficiently small that
$k>1$ or $\kappa
>3$ is satisfied. As is well-known, $\cN=2$ Liouville or
sine-Liouville theory is T-dual to the $SL(2;\br)/U(1)$ coset theory
\cite{FZZ2,GK,HK1}, so the condition on the level $k$ or $\kappa$
ought naturally to descend to the two-dimensional black hole
description. Indeed, such aspect was discussed in \cite{KarMS}
purely in the language of the $SL(2;\br)/U(1)$ coset theory (see
also \cite{HK1}). Their reasoning is closely related to the
non-formation of the black hole in two-dimensional string theory
(see also \cite{Friess:2004tq} for the discussion concerning this
issue from the matrix model viewpoint).\footnote
   {Another interesting observation related
    to the $k=1$ transition is the following. If we consider
        a two-dimensional $U(1)$ gauge theory in the ultraviolet that
        flows to $SL(2;\br)/U(1)$ coset theory in the infrared (as was
        introduced in \cite{HK1} to prove the mirror duality to the
        $\cN=2$ Liouville theory),
  the central charge of the $U(1)$ gauge theory
  is given by $9$. Since the IR $SL(2;\br)/U(1)$ coset theory
  has a central charge $c=3(1+\frac{2}{k})$, there is an
  apparent contradiction to Zamolodchikov's $c$-theorem
  if the level $k<1$ is considered. However, we should note that
  $SL(2;\br)/U(1)$ coset theory is dilatonic so that the effective
  central charge is always given by $3$.}
In the strong curvature regime, $k<1$, the background is described
more effectively in terms of the $\cN=2$ Liouville theory as it is
weakly coupled. Evidently, the black hole interpretation of the
$SL(2;\br)/U(1)$ theory is less clear in this region, because the
classical $\cN=2$ Liouville theory does not admit an interpretation
in terms of black hole geometry in any obvious way.

We emphasize that such string - black hole transition is not likely
to arise perturbatively and could arise only from nonperturbative
string worldsheet effects. For instance, tree-level closed string
amplitudes are manifestly analytic with respect to the level $k$.
These amplitudes exhibit a finite absorption rate (thus displaying
the non-unitarity of the reflection amplitudes) regardless of the
value of $k$. In fact, finite-$k$ correction to the amplitudes yield
an irrelevant phase-factor \cite{DVV,GKPS}.


However, as was first observed in \cite{NPRT}, situation changes
drastically if we consider the closed string radiation from the
rolling D-brane in such a background. In \cite{NPRT}, it was shown
that the distribution of radiation off D0-brane in extremal
NS5-brane background becomes ultraviolet finite for $k < 1$. In the
previous section, extending the analysis of \cite{NPRT}, we have
shown that the $k=1$ transition shows up manifestly in the open
string sector in the sense that branching ratio between the incoming
and the outgoing radiation distribution (as well as spectral
moments) behaves very differently across $k=1$.
Remarkably, retaining finite $1/k$-correction, which originated from
consistency with the exact reflection relations, was crucial in
obtaining physically sensible results {\sl even for} $k\gg 1$.
Cancellation between the radiation distribution and the exponential
growth of the density of states at large $M$ is quite nontrivial,
and relied crucially on precise functional dependence on $k$.

An `order-parameter' of the transition is thus provided by the
radiation distribution of rolling D-brane. The phase transition
across $k=1$ is that while the radiation distribution from the
falling D-brane exhibits powerlike ultraviolet divergence for $k>1$,
it becomes finite for $k<1$. Thus, the rolling D-brane in the $k<1$
regime does {\it not} yield a large back-reaction unlike the $k>1$
case. This is also consistent with the assertion that black hole
cannot be formed in the two-dimensional string theory: It seems
difficult to construct two-dimensional black hole by injecting
D-branes to the linear dilaton (or usual Liouville) theory.\footnote
{Such a possibility was proposed in \cite{KarMS}.}

It is also worth mentioning that the radion-tachyon correspondence
is likely to fail in the two-dimensional string theory ($k=1/2$). In
fact, had we have such a correspondence, the rolling radion of the
D0-brane could be identified with the rolling tachyon of the
ZZ-brane in the Liouville theory. On the other hand, it is known
that the radiation distribution of the-ZZ brane exhibits a powerlike
ultraviolet divergence \cite{KMS} at leading order in string
perturbation theory, while that of the falling D0-brane does not.

\subsection{Holographic Viewpoint}

The string - black hole transition across $k=1$ also has a natural
interpretation in terms of the holographic principle, as recently
discussed in \cite{GKRS}. Adding $Q_1$ fundamental strings to $k$
NS5-branes, one obtains the familiar bulk geometry of the
$AdS_3/CFT_2$-duality. In this context, the density of states of the
dual conformal field theory is given by the naive Cardy formula
$S=2\pi\sqrt{\frac{cL_0}{6}}+2\pi\sqrt{\frac{\bar{c}\bar{L}_0}{6}}$
with $c = 6 k Q_1$ for $k>1$, but not for $k<1$. Rather, the central
charge that should be used in the Cardy formula is replaced by an
effective one $c_{\rm eff}= 6Q_1(2-\frac{1}{k})$ \cite{KutS}. The
similar effects also showed up in the double scaling limit of the
`little string theory'(LST) \cite{GK}.\footnote
   {Even though the original `little string theory' is the theory of
   NS5-brane, so $k$ should be positive integer-valued,
one can also consider models with fractional value of the level $k$,
which is less than 1 generically. This is achieved by considering
the {\em wrapped\/} NS5-brane backgrounds, or compactifications on a
Calabi-Yau threefold having rational singularity \cite{GKP}.  From
the regularized torus partition function, one can prove that there
is no normalizable massless states (corresponding to the
`Lehmann-Symanzik-Zimmerman-poles' \cite{AGK}) in such string vacua
if $k<1$, as was discussed in {\em e.g.} \cite{ES-BH,ES-con}. }
We shall now show that such change of the central charge is also
imperative for reproducing the closed string radiation distribution
correctly from the dual holographic picture.

It is an interesting attempt to reproduce the phase transition in
the radiation distribution of rolling D-brane across
$k=1$ from the holographic viewpoint. In \cite{Sahakyan}, it was
proposed that the rolling D-brane should correspond to the decay of
a certain defect in the dual LST. We shall now extend that analysis
to the $k<1$ case and explore the phase-transition. The relevant
holographic description is based on the following two assumptions.
\begin{enumerate}
\item \underline{fixed radiation number distribution}: The radiation distribution for a fixed mass $M$ is determined
by large $k$ behavior of the pressure in the far future (past). This
is equivalent to the statement that the decay of the radion is
described by a `holographic tachyon condensation'. We assume that
there is no phase transition at $k=1$ for a fixed mass $M$.\footnote
   {Theoretically, there is no reason to exclude
    a finite $1/k$ correction here.
     We only need this assumption phenomenologically
    in order to reproduce the ten-dimensional calculation
    even for $k>1$. A priori,
     the tachyon condensation (in the critical bosonic string)
    itself may receive large string worldsheet corrections. In the
    Dirac-Born-Infeld action analysis, such potential corrections
    were completely dropped out.}
In our convention, the distribution is given by
\begin{equation}
\cN(M)_{\rm LST} \sim e^{-2\pi M \sqrt{\frac{k}{2}}} \ .
\label{asum}
\end{equation}

\item \underline{change of density of states}: The final density of closed little string states
in the `holographic tachyon condensation' is given by the square
root of the full nonperturbative density of states in LST. As is
discussed in \cite{GKRS}, the full nonperturbative density of states
of the LST is believed to exhibit a phase transition at $k=1$: for
$k>1$, the density of states is related to the Hawking temperature
as
\begin{equation}
n(M)_{\rm LST} \sim e^{{4\pi} M \sqrt{\frac{k}{2}}} \ .
\label{asuma}
\end{equation}
In other words, the Hagedorn temperature in LST should
be equated with the Hawking temperature \cite{ABKS}
(see also, {\em e.g.} \cite{HarO,BerRoz,KutSah}).

On the other hand, for $k<1$, because of the non-normalizability of
the black hole excitation, the nonperturbative density of states of
the LST is equivalent to the density of states of the (dual)
perturbative string theory \cite{GKRS}:
\begin{equation}
n(M)_{\rm LST} \sim e^{4\pi M \sqrt{1-\frac{1}{2k}}} \ .
\label{asumb}
\end{equation}
\end{enumerate}
With these assumptions, we can estimate the average radiation number
of the `holographic tachyon condensation' to be
\begin{eqnarray}
\overline{\cN}_{\rm LST} = \int_0^\infty \dd M \cN(M) \,
\sqrt{n(M)_{\rm LST}} \ . \nonumber
\end{eqnarray}
Note that, in contrast to the bulk string theory calculation, we
have no integration over the radial momentum. Substituting
\eqref{asum} and \eqref{asuma} or \eqref{asumb} according to the
value of $k$, we obtain
\begin{eqnarray}
\overline{\cN}_{\rm LST} \sim \int^\infty \dd M \, e^{-{2\pi}M
{\sqrt{\frac{k}{2}}} +{2\pi}M \sqrt{\frac{k}{2}}} \nonumber
\end{eqnarray}
for $k>1$, showing powerlike ultraviolet divergent behavior because
of the complete cancellation in the exponent, and
\begin{eqnarray}
\overline{\cN}_{\rm LST} \sim \int^\infty \dd M \, e^{-{2\pi M }
{\sqrt{\frac{k}{2}}}+2\pi M \sqrt{1-\frac{1}{2k}}} \ , \nonumber
\end{eqnarray}
for $k<1$, showing exponential suppression in the ultraviolet. It is
easy to see that this holographic dual computation reproduces the
bulk computation presented in section \ref{radiation super} up to a
subleading power dependence \eqref{eq:in}, \eqref{eq:ini}.\footnote
   {The exact determination of the pre-exponential power part
   is beyond the scope of the rough estimate presented here.
    It requires the full computational ability in the LST.}

It should be noted, however, that the cancellation between the
radiation distribution and the density of states has a different
origin in the dual holographic description as compared to the bulk
side. In the holographic description, the origin of the phase
transition is the nonperturbative density of the states in LST while
the radiation distribution at a fixed mass-level $M$ keeps its
functional form unchanged. On the other hand, in the bulk theory,
origin of the cancellation was that the radiation distribution
changes at $k=1$ due to the disappearance of the non-trivial saddle
point in the integration of the radial momentum $p$, while the
density of states is always given by the same formula. Thus the
agreement between the two descriptions is quite non-trivial and we
believe that our results provide yet another evidence of the
holographic duality for the NS5-brane and blackhole physics.

Though we presented the dual description based on some assumptions,
we can turn the logic around and regard our results as a support for
such assumptions. In particular the quantum gravity phase transition
at $k=1$ in the dual theory proposed in \cite{GKRS} is crucial for
understanding the radiation distribution out of a defect decay in
the dual LST. We thus propose our discussion in this section as a
strong support for string - black hole transition.

\section{Boundary States and Radiation in the Ramond-Ramond Sector}

In the case of fermionic black hole background, the rolling D0-brane
would also radiate off closed string states in the Ramond-Ramond
(R-R) sector. In this section, we shall construct R-R boundary state
of the D0-brane and compute radiation rates. Since the worldsheet
theory corresponds to $\cN =2$ superconformal field theory,
correlation functions of the R-R sector and boundary states are
readily obtainable by performing the standard $\cN=2$ spectral flow.

We shall begin with discussion regarding properties of reflection
amplitudes for the R-R sector (see \cite{GKPS} in the context of 2D
black hole). Recall that the reflection relation was given in the
NS-NS sector as
\begin{eqnarray}
&& U^{-p}_{\om}(\rho,t)^{\rm NS}= \cR^{\rm NS}(-p,\om)
U^{p}_{\om}(\rho,t)^{\rm NS} \quad \mbox{and} \quad
V^{-p}_{\om}(\rho,t)^{\rm NS}= \cR^{\rm NS *}(-p,\om)
V^{p}_{\om}(\rho,t)^{\rm NS}~, \nonumber \end{eqnarray}
where the exact reflection amplitude $ \cR^{\rm NS}(-p,\om)$ was
defined by
\begin{eqnarray}
\cR^{\rm NS}(p,\om) = \frac{\Gamma(1+\frac{ip}{k})\Gamma(+ip)
\Gamma^2(\frac{1}{2}-i\frac{p+\omega}{2})}
{\Gamma(1-\frac{ip}{k})\Gamma(-ip)
\Gamma^2(\frac{1}{2}+i\frac{p-\omega}{2})} \ . \nonumber
\end{eqnarray}
To obtain the reflection relation of the R-R sector, we shall
perform the spectral flow by half unit of the $\cN=2$ $U(1)$
current.

In sharp contrast to the $\cN=2$ Liouville theory,
the reflection amplitude now depends on the spin structure of the
R-R sector.\footnote
  {This is because, in the $\cN=2$ Liouville theory,
   the reflection amplitudes for the momentum modes have
   a symmetry under $\omega \to -\omega$.}
Explicitly, the spectral flow is defined as $\omega \to \omega \pm
i$, where the $+$ sign corresponds to spin ($+,-$) states and $-$
sign corresponds to spin ($-,+$) states (in the
$(\frac{1}{2},\frac{1}{2})$ picture): in the $\rho \to \infty$
limit, they are described by $S^{\pm} e^{-\rho} e^{-ip \rho-i\omega
t}$ and the conformal weight is given by $h =
\frac{p^2-\omega^2+1}{4k} + \frac{1}{8}$.

Therefore, for the R-R states with spin ($+,-$), the exact
reflection amplitudes become
\begin{equation}
\cR^{\rm R+}(p,\om) = \frac{\Gamma(1+\frac{ip}{k})\Gamma(+ip)
\Gamma^2(1-i\frac{p+\omega}{2})} {\Gamma(1-\frac{ip}{k})\Gamma(-ip)
\Gamma^2(1+i\frac{p-\omega}{2})} \ . \label{refRp}
\end{equation}
Equivalently, if we take spin ($-,+$) R-R states, the exact
reflection amplitudes become
\begin{equation}
\cR^{\rm R-}(p,\om) =
\frac{\Gamma(1+\frac{ip}{k})\Gamma(+ip)\Gamma^2(-i\frac{p+\omega}{2})}
{\Gamma(1-\frac{ip}{k})\Gamma(-ip)\Gamma^2(+i\frac{p-\omega}{2})} \
. \label{refRm}
\end{equation}
It is important to notice that the latter amplitudes have a second
order zero in the light-cone direction $p = \omega >0$ (recall that
$p>0$ in our convention). Similarly, we could derive the reflection
relation for $(\pm,\pm)$ spin structure, but the resultant
amplitudes are compatible only with the analytic continuation to the
`winding time' (in the interior of the singularity), so we would not
delve into details anymore.

Consider next the boundary wave function of the R-R sector. For
definiteness, we shall take the absorbed D0-brane \eqn{falling D0}
(We focus on the $t_0=0$ case for simplicity.)
\begin{eqnarray}
{}_{\msc{absorb}}\!\bra{B,{\rm NS};\rho_0} =
\int_0^{\infty}\frac{\dd p}{2\pi} \int_{-\infty}^{\infty}\frac{\dd
\om}{2\pi}\,
  \Psi_{\msc{absorb:NS}}(\rho_0;p,\om) \,
{}^{\widehat{U}}\!\dbra{p,\om} \ ,\nonumber \eea
where
\bea \Psi_{\msc{absorb:NS}}(\rho_0;p,\om) =
\frac{\Gamma(\frac{1}{2}-i\frac{p+\omega}{2})
\Gamma(\frac{1}{2}-i\frac{p-\omega}{2})}{\Gamma(1-ip)}
\Gamma\left(1+\frac{ip}{k}\right) \, \left[ e^{-ip\rho_0} -
\frac{\cosh\left(\pi \frac{p-\om}{2}\right)} {\cosh\left(\pi
\frac{p+\om}{2}\right)} e^{+ip\rho_0} \right]~. \nonumber
\end{eqnarray}
The boundary wave functions of the R-R sector are then derived by
applying the ${\cal N}=2$ spectral flow $\omega \to \omega \pm i$:
\begin{eqnarray}
 \Psi_{\msc{absorb:R}+}(\rho_0;p,\om)
\frac{\Gamma(-i\frac{p+\omega}{2})
\Gamma(1-i\frac{p-\omega}{2})}{\Gamma(1-ip)}
\Gamma\left(1+\frac{ip}{k}\right) \, \left[ e^{-ip\rho_0} +
\frac{\sinh\left(\pi \frac{p-\om}{2}\right)} {\sinh\left(\pi
\frac{p+\om}{2}\right)} e^{+ip\rho_0} \right]~, \nonumber
\end{eqnarray}
and
\begin{eqnarray}
\Psi_{\msc{absorb:R}-}(\rho_0;p,\om) =
\frac{\Gamma(1-i\frac{p+\omega}{2})
\Gamma(-i\frac{p-\omega}{2})}{\Gamma(1-ip)}
\Gamma\left(1+\frac{ip}{k}\right) \, \left[ e^{-ip\rho_0} +
\frac{\sinh\left(\pi \frac{p-\om}{2}\right)} {\sinh\left(\pi
\frac{p+\om}{2}\right)} e^{+ip\rho_0} \right]~, \nonumber
\end{eqnarray}
for the two opposite spin structures. These boundary wave functions
are of course consistent with the exact reflection amplitudes
\eqn{refRp},\eqn{refRm}.

From these boundary wave functions, we can deduce some physical
properties of the boundary states in the R-R sector:
\begin{itemize}
\item For $k > {1 \over 2}$, in the saddle point approximation of
the radial momentum integral,
radiation distribution of the R-R sector behaves the same as that of
the NS-NS sector. In particular, the absolute value of the
reflection amplitudes behave in the similar manner. Thus, the
radiation distribution of the R-R sector is the same as that of the
NS-NS sector.
\item For $k = {1 \over 2}$, viz. the two-dimensional black hole,
considerable differences arise. Both boundary wave function and
reflection amplitudes show singularity (or zero) when we take
particular spin structure. It is not clear what the origin of these
singularities of lightlike on-shell states $p = \omega$ would be. We
note that some related discussions were given in \cite{GKPS}.
\item In the mini-superspace limit $k \to \infty$,
the mass gap in the R-R sector vanishes. Therefore, it is well-posed
to question radiation of the massless R-R states off the R-R charge.
>From the boundary states given above, we observe that, assuming $p,
\omega > 0$, there is no lightlike pole in $R+$ state while there is
a pole at $p = + \omega$ in the $R-$ state. It is also interesting
to note that, in the subleading contribution proportional to
$e^{+ip\rho_0}$, the pole from the gamma function is cancelled by
the zero in the $\sinh(\pi\frac{p-\omega}{2})$ factor.

A possible interpretation is that, roughly speaking, R-R charge is
localized on the incoming light-cone $p = \omega$.\footnote
        {This is true only in the asymptotic region
       $\rho \to \infty$ since the distribution
   near $\rho = 0$ is further related
   to the basis of Ishibashi states
   used in the expansion. In the case of `absorbed'
   basis, there is no contribution from the past horizon.
In addition, because the reflection amplitude vanishes in the $R-$
sector, an observer at $\rho \to \infty$ do not detect any outgoing
wave.}

\end{itemize}
\section{Back to Extremal NS5-Brane Background}\label{ext}

By tuning off $\mu \rightarrow 0$, we are back to the extremal
NS5-brane background. Roughly speaking, the extremal background is
described by the free linear dilaton theory, but crucial differences
from the non-extremal counterpart studied in this work are the
followings:
\begin{itemize}
 \item We have no reflection relation, and the $p>0$ and $p<0$
 states should be treated as independent states.\footnote
   {In this sense, the arguments given in \cite{NST}
    are not completely precise, although the main part of
    physical results, say, the closed string
    radiation rates, are not altered.
    }
 \item The conformal field theory description is not effective in the
 entire space-time: the string coupling diverges at the location of the
 NS5-brane. We cannot completely trace the classical trajectory of
 the D0-brane \eqn{trajectory D0} without facing strong coupling problem.
\end{itemize}
We thus have to keep it in mind that the validity of the conformal
field theory description of extremal NS5-brane is limited to the
sufficiently weak string coupling region.

For the extremal NS5-brane, since the relevant conformal field
theory involves linear dilaton and hence is a free theory, we can
introduce the basis of the Ishibashi states as $\dket{p,\om}$,
$(p,\om \in \br)$ associated with the wave function
$\psi^p_{\om}(\rho,t)\propto e^{-\rho} e^{-ip \rho-i\om t}$. Another
non-trivial difference from the non-extremal case is the volume form
of the space-time. Since we have the linear dilaton $\Phi =
\mbox{const}-\rho$ and a flat metric $G_{ij}=\eta_{ij}$, the
relevant volume form becomes
\begin{eqnarray}
 \dd \mbox{Vol}= e^{-2\Phi}\sqrt{G}\dd \rho \dd t = e^{2\rho} \dd \rho \dd t~.
\label{vol linear dilaton}
\end{eqnarray}

Now, the classical trajectory of D0-brane in the extremal NS5-brane
is given by \cite{Kutasov}:
\begin{eqnarray}
 2\cosh(t-t_0) e^{\rho} = e^{\rho_0}~.
\label{trajectory 2}
\end{eqnarray}
The boundary state describing the D0-brane moving along
\eqn{trajectory 2} ought to have the following form:
\begin{eqnarray}
 && \bra{B;\rho_0,t_0} = \int_{-\infty}^{\infty}\frac{\dd p}{2\pi}\,
\int_{-\infty}^{\infty} \frac{\dd \om}{2\pi}\, \Psi
(\rho_0,t_0;p,\om) \dbra{p,\om}~. \label{symmetric D0 extremal}
\end{eqnarray}
The boundary wave function is evaluated as
\begin{eqnarray}
 \Psi(\rho_0,t_0;p,\om) &\sim& \int \dd v\, \delta\Big(
2\cosh(t-t_0)e^{\rho}-e^{\rho_0} \Big)\, e^{-\rho-ip\rho-i\om t} \nn
&=& \int_{-\infty}^{\infty} \dd t \, e^{-ip\rho_0} e^{-i\om t}
\Big[2 \cosh(t-t_0)\Big]^{ip-1} \nn & =&
\frac{1}{2}B\left(\frac{1}{2}-i\frac{p+\om}{2},
\frac{1}{2}-i\frac{p-\om}{2} \right) \, e^{-ip \rho_0-i \om t_0}~.
\nn \end{eqnarray}
In the last expression, we used the formula \eqn{formula 2}. This is
essentially the calculation given in \cite{NST}. Finally, by
restoring the important `worldsheet correction factor'
$\Gamma\left(1+i\frac{p}{k}\right)$,\footnote
  {Since in this case we do not have the reflection relation,
   the inclusion of the factor
   $\Gamma\left(1+i\frac{p}{k}\right)$ may sound less affirmative than
   the nonextremal NS5-brane background.
   We argue that the procedure is actually justified by
   considering the limit from the non-extremal case.}
we obtain the boundary wave function
\begin{eqnarray}
 \Psi(\rho_0,t_0;p,\om) = \frac{1}{2} B(\nu_+,\nu_-)
\Gamma\left(1+i\frac{p}{k}\right)\, e^{-ip\rho_0-i\om t_0}~.
\label{symmetric D0 extremal 2} \qquad \mbox{where} \qquad \nu_{\pm}
\equiv \frac{1}{2}- i\frac{p\pm \om}{2}~,\nonumber
\end{eqnarray}
This is the extremal counterpart of the `symmetric D0-brane' in the
non-extremal NS5-brane background \eqn{symmetric D0}.

We can also consider the `half S-brane' counterpart by
taking the Hartle-Hawking contours depicted in the Figures
\ref{HH-future} and \ref{HH-past}. Namely, for the `absorbed brane',
we obtain
\begin{eqnarray}
 {}_{\msc{absorb}}\bra{B;\rho_0,t_0}
= \left\lb \int_0^{\infty}\frac{\dd p}{2\pi}
\int_0^{\infty}\frac{\dd \om}{2\pi} + \int_{-\infty}^0\frac{\dd
p}{2\pi} \int_{-\infty}^0\frac{\dd \om}{2\pi}\right\rb\,
\Psi(\rho_0,t_0;p,\om)\, \dbra{p,\om} ~, \label{falling D0 extremal}
\end{eqnarray}
and for the `emitted brane',
\begin{eqnarray}
 {}_{\msc{emitted}}\bra{B;\rho_0,t_0}
= \left\lb \int_0^{\infty}\frac{\dd p}{2\pi}
\int^0_{-\infty}\frac{\dd \om}{2\pi} + \int_{-\infty}^0 \frac{\dd
p}{2\pi} \int^{\infty}_0\frac{\dd \om}{2\pi}\right\rb\,
\Psi(\rho_0,t_0;p,\om)\, \dbra{p,\om} ~. \label{emitted D0 extremal}
\end{eqnarray}
They are regarded as the counterparts of \eqn{HH symm D0 1} and
\eqn{HH symm D0 2}.

The radiation rates were already evaluated in \cite{NST,Sahakyan}.\footnote{
In this paper, we scaled energy and momentum differently
from \cite{NST}. In light of normalization as in \eqn{on-shell
super}, $\om, p$ in this work should be read as $2 \sqrt{k}$ times
$\om, p$ in \cite{NST}.} Crucial differences from the non-extremal
case are the followings: We have the `forward radiations' ({\em
e.g.}, the incoming radiation for the absorbed D-brane \eqn{falling
D0 extremal}) only and no `backward radiations' ({\em e.g.}, the
outgoing radiation for the absorbed D-brane). This is because there
is no reflection relation in the extremal case. The forward
radiations behave in the completely same way as the non-extremal
case (that is, in a fermionic two-dimensional black hole with $k >1
$), giving rise to the Hagedorn-like ultraviolet divergence again.
At fixed but large $M$ before integrating over $p$, the partial radiation number distribution
takes again exactly the same asymptotic form as in \eqn{grey body}
except that now the coefficient $2 \pi \sqrt{2k}$ is {\sl not}
interpretable as the inverse Hawking temperature of the black hole.\footnote
{An obvious alternative interpretation could be that, even
for extremal background, the falling D0-brane excites the NS5-brane
above the extremality.} Again, this has to do with the peculiarity
that the Hawking temperature of the two-dimensional black hole is
set by the level $k$, not by the nonextremality $\mu$. On the other
hand, the absence of the backward radiation matches with the
extremality of the background; there is no Hawking radiation.


\section{More on Physical Interpretations :
Hartle-Hawking States}

We shall now revisit the boundary states we constructed in this work
and elaborate further on their physical interpretations with
particular emphasis on analogy with the rolling tachyon problem via
the radion-tachyon correspondence. We also elaborate on the fate of
R-R charge carried by the D0-brane. To be concrete, we shall focus
on the cases $k \geq 2$ admitting interpretation in terms of near
horizon geometry of black NS5 branes.

The boundary state \eqn{falling D0} describes the late-time rolling
($t \gg t_0$) of the D0-brane rolling into the black NS5 branes.
The relevant D0-brane has the initial condition $\rho=\rho_0$,
$\frac{d\rho}{dt}=0$ at $t=t_0$ and starts to roll down toward the
black hole. After sufficiently long coordinate time elapsed, the
D0-brane gets close to the future horizon (${\cal H}^+$). As
examined in section 4, almost all energy of the D0-brane is absorbed
by the black hole in the form of incoming radiation.
The incoming radiation is dominated by very massive, and hence
highly non-relativistic closed string excitations. Via the
radion-tachyon correspondence, these states are identifiable with
the `tachyon matter' in the rolling tachyon problem in flat
spacetime. On the other hand, we have seen that a small part of
energy escapes to the spatial infinity (${\cal I}^+$) as the
outgoing radiation. We have seen that the spectral distribution is
characterized by the Hawking temperature, and is necessarily
dominated by light modes. This interpretation is quite natural from
the viewpoint of the radion-tachyon correspondence for the extremal
NS5-brane background \cite{Kutasov}. Since we are now working with
the non-extremal NS5-brane background, our analysis may be
considered as an evidence that the correspondence is valid even at
finite temperature.


What about evolution in the far past $t < t_0$? Here, we face a
subtlety. Recall that the boundary condition defining \eqn{falling
D0} does not allow contributions from the past horizon (${\cal
H}^-$), namely, the basis of Ishibashi states $\dket{p,\om}^U$
does not reproduce the past half of the classical trajectory
\eqn{trajectory D0}. Rather, the NS-NS sector of the D0-brane
boundary wave function appears widely distributed in the space-time
in the far past. This may be interpreted as radiations imploding to
$\rho = \rho_0$ from spatial infinity, but then it is subtle to
trace the R-R charge carried by the D0-brane, created out of the
imploding radiation. Classically, the D0-brane charge density ought
to be localized along the classical trajectory \eqn{trajectory D0}
and hence emanates from the past horizon. Once stringy effects are
taken into account, the charge appears to originate from asymptotic
infinity along the light-cone coordinate. Complete understanding of
this curious feature is highly desirable but we shall relegate it to
future study. Here, instead, we present a simple prescription of
avoiding this subtlety: a version of `Hartle-Hawking' boundary
condition.

\begin{figure}[htbp]
    \begin{center}
    \includegraphics[width=0.5\linewidth,keepaspectratio,clip]
      {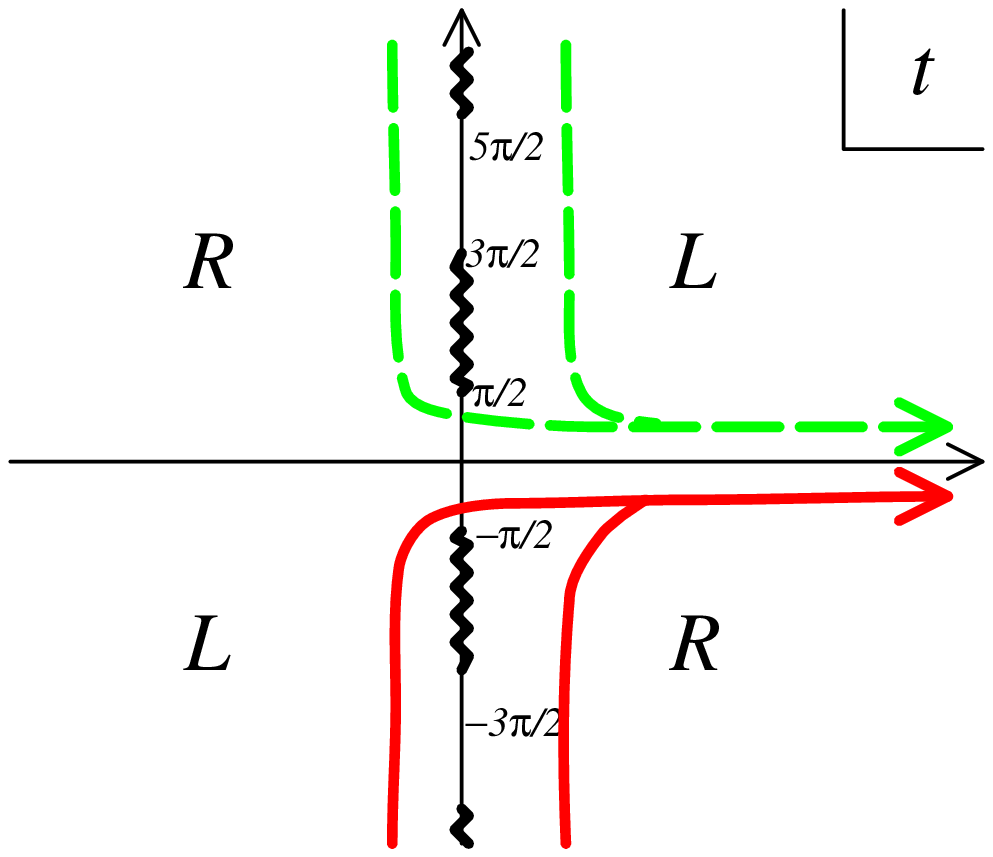}
    \end{center}
    \caption{`future Hartle-Hawking contour' : the red (green broken)
        line is the contour $\cC^+_{\msc{future}}$ for $\om > 0$
        ($\cC^-_{\msc{future}}$ for $\om <0$). The `$L$' (`$R$') contour
         should be used if calculating the overlap with
         $L^p_{\om}(\rho,t)$ ($R^p_{\om}(\rho,t)$) for
         the convergence of integral.}
    \label{HH-future}
\end{figure}
\begin{figure}[htbp]
    \begin{center}
    \includegraphics[width=0.5\linewidth,keepaspectratio,clip]
      {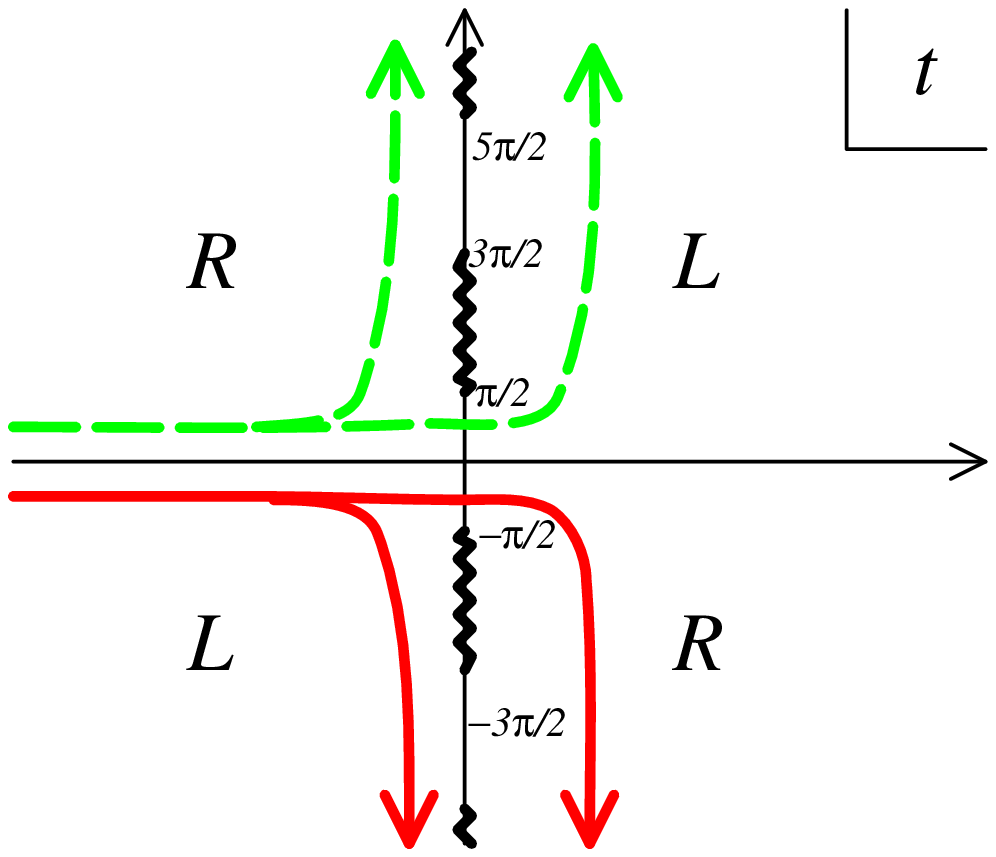}
    \end{center}
    \caption{`past Hartle-Hawking contour' : the red (green broken)
        line is the contour $\cC^+_{\msc{past}}$ for $\om > 0$
         ($\cC^-_{\msc{past}}$ for $\om <0$).}
    \label{HH-past}
\end{figure}

We shall first focus on the absorbed D0-brane boundary state
\eqn{falling D0}. Formally, by construction, we can regard
the boundary wave function specified by the time-integration over
the `real contour' $\cC= \br$ as in \eqn{Wick rotation 0}. Now, let
us discuss what happens if we choose the `Hartle-Hawking' type
contour instead of the real contour, which connect the Euclidean
time with the future or past half of real time axis at $t=t_0$:
\begin{eqnarray}
\cC_{\msc{future}}^{\pm}
= \left(t_0+i\br_{\mp}\right) \cup \left(t_0+\br_{+}\right)~,~~~
\cC_{\msc{past}}^{\pm} = \left(t_0+i\br_{\mp}\right) \cup
\left(t_0+\br_{-}\right)~.
\end{eqnarray}
More precisely, we should avoid suitably the branch cuts on
$t_0+i\br$ to render the integral convergent. See Figures
\ref{HH-future} and \ref{HH-past} for details. The superscript $+$
$(-)$ is associated with the positive (negative) energy sector. Note
that the phase-factor $e^{-i\om t}$ behaves well on the lower
(upper) half of complex $t$-plane if $\om $ is positive (negative).
Let us pick up $\cC_{\msc{future}}$. Following the traditional
interpretation of the Hartle-Hawking type wave function, we may
suppose that both the D0-brane and black NS5-brane are created from
`nothing' at $t=t_0$, and then the D0-brane starts to fall down
toward the future horizon along the classical trajectory
\eqn{trajectory D0}. In this prescription, the subtlety we mentioned
above is completely circumvented.

One may paraphrase the prescription as follows: choosing the
Hartle-Hawking contour $\cC_{\msc{future}}$, we explicitly obtain
\begin{eqnarray}
&& \hspace{-5mm} {}_{HH +,\,\msc{absorb}}\!\bra{B;\rho_0,t_0} =
\int_0^{\infty}\frac{\dd p}{2\pi}\, \left[
\int_{0}^{\infty}\frac{\dd \om}{2\pi}\,
\Psi_{\msc{symm}}(\rho_0,t_0;p,\om) + \int_{-\infty}^{0}\frac{\dd
\om}{2\pi}\, \cR(p,\om)\Psi^*_{\msc{symm}}(\rho_0,-t_0;p,\om)
\right] \, {}^{\widehat{U}}\!\dbra{p,\om}~, \nn && \label{HH falling
D0}
\end{eqnarray}
where $\Psi_{\msc{symm}}(\rho_0,t_0;p,\om)$ is defined in
\eqn{symmetric D0}. In fact, by taking $\cC_{\msc{future}}$, we are
only left with the $L^p_{\om}$ ($R^p_{\om}$)-part of the one-point
function for the $\om>0$ ($\om<0$) sector. See the figure
\ref{HH-future}. This boundary wave function is formally regarded as
the limit of \eqn{falling D0} under $t_0\,\rightarrow\, -\infty$,
$\rho_0\,\rightarrow\,+\infty$ while keeping $|\rho_0|/|t_0|$
finite. Note that the second (first) term $\propto e^{ip\rho_0-i\om
t_0}$ ($\propto e^{-ip\rho_0-i\om t_0}$) in \eqn{falling D0}
oscillates very rapidly in this limit for $\om >0$ ($\om <0$) and
hence drops off.\footnote
   {More precise argument would be as follows:
    The disk amplitude for a wave packet {\em e.g.}
    $\int \frac{\dd p}{2\pi} \int \frac{\dd \om}{2\pi}\, f(p,\om) \ket{L^p_{\om}}$ is evaluated
    as $
    \lim_{\rho_0\,\rightarrow\,+\infty , \,
    t_0\,\rightarrow\,-\infty}\,
    \int \frac{\dd p}{2\pi} \int \frac{\dd\om}{2\pi} f(p,\om) \Psi(\rho_0,t_0;p,\om)$.
    Then, the rapidly oscillating term in the boundary wave
     function $\Psi(\rho_0,t_0;p,\om)$ cannot contribute for any
     $L^2$-normalizable wave packet $f(p,\om)$.
} The limit just means that the D0-brane moving along the
trajectory \eqn{trajectory D0} is coming from the past infinity
$({\cal I}^-)$, and falling into the future horizon (${\cal H}^+$).
Everything is supposed to be localized over the classical trajectory
in this case.


Adopting the past Hartle-Hawking contour $\cC_{\msc{past}}$ for the
boundary state of emitted D0-brane \eqn{emitted D0} is completely
parallel. We take the time-reversal of the above:
\begin{eqnarray}
&& \hspace{-5mm} {}_{HH -,\,\msc{emit}}\!\bra{B;\rho_0,t_0} =
\int_0^{\infty}\frac{\dd p}{2\pi} \left[ \int_{0}^{\infty}\frac{\dd
\om}{2\pi}\, \Psi^*_{\msc{symm}}(\rho_0,-t_0;p,\om) +
\int_{-\infty}^{0}\frac{\dd \om}{2\pi}\,
\cR^*(p,\om)\Psi_{\msc{symm}}(\rho_0,t_0;p,\om) \right]
 {}^{\widehat{V}}\!\dbra{p,\om}~, \nn && \label{HH emitted D0}
\end{eqnarray}
which is regarded as the $t_0\,\rightarrow\,+\infty$,
$\rho_0\,\rightarrow\,+\infty$ limit of \eqn{emitted D0}. It
describes the trajectory of D0-brane emitted from the past horizon
${\cal H}^-$ and escaping to the future infinity ${\cal I}^+$.


Let us turn to the `symmetric' D0-brane \eqn{symmetric D0}. Naively,
it appears that the prescription is that
\begin{eqnarray}
&& \hspace{-5mm} {}_{HH +,\,\msc{symm}}\!\bra{B;\rho_0,t_0}' =
\int_0^{\infty}\frac{\dd p}{2\pi} \left[ \int_{0}^{\infty}\frac{\dd
\om}{2\pi}\, 2 \Psi_{\msc{symm}}(\rho_0,t_0;p,\om) \,
{}^L\!\dbra{p,\om} +\int_{-\infty}^{0}\frac{d\om}{2\pi}\, 2
\Psi^*_{\msc{symm}}(\rho_0,-t_0;p,\om) {}^R\!\dbra{p,\om} \right]
\nn && \label{HH symm D0 1}
\end{eqnarray}
for the future Hartle-Hawking contour $\cC_{\msc{future}}$, and
\begin{eqnarray}
&& \hspace{-5mm} {}_{HH -,\,\msc{symm}}\!\bra{B;\rho_0,t_0}' =
\int_0^{\infty}\frac{\dd p}{2\pi} \left[ \int_{0}^{\infty}\frac{\dd
\om}{2\pi}\, 2 \Psi^*_{\msc{symm}}(\rho_0,-t_0;p,\om)
{}^R\!\dbra{p,\om} +\int_{-\infty}^{0}\frac{d\om}{2\pi} 2
\Psi_{\msc{symm}}(\rho_0,t_0;p,\om) {}^L\!\dbra{p,\om} \right] \nn
&& \label{HH symm D0 2}
\end{eqnarray}
for the past Hartle-Hawking contour $\cC_{\msc{past}}$.
However, this cannot be the whole story. The existence of Euclidean
part of the Hartle-Hawking path-integral enforces the boundary
states to be expanded by the basis smoothly connected to the
Euclidean ones, while $\ket{L^p_{\om}}$, $\ket{R^p_{\om}}$ do not
possess such a property. Consequently, to achieve the correct
Hartle-Hawking states, we ought to make further the projection to
$\cH^U$, ($\widehat{\cH^U}$) for the contour $\cC_{\msc{future}}$,
and to $\cH^V$, ($\widehat{\cH^V}$) for $\cC_{\msc{past}}$. We thus
obtain as the correct Hartle-Hawking states:
\begin{eqnarray}
&& {}_{HH +,\,\msc{symm}}\!\bra{B;\rho_0,t_0} =  {}_{HH
+,\,\msc{symm}}\!\bra{B;\rho_0,t_0}' \widehat{P_U} \equiv {}_{HH
+,\,\msc{absorb}}\!\bra{B;\rho_0,t_0} ~, \nn && {}_{HH
-,\,\msc{symm}}\!\bra{B;\rho_0,t_0} =  {}_{HH
-,\,\msc{symm}}\!\bra{B;\rho_0,t_0}' \widehat{P_V} \equiv {}_{HH
-,\,\msc{emitted}}\!\bra{B;\rho_0,t_0} ~, \label{relation HH}
\end{eqnarray}
where the right-hand sides are already given in \eqn{HH falling D0},
\eqn{HH emitted D0}.


Remarkably, this feature resembles much that of the S-branes
discussed in \cite{LLM}. Namely, it was shown there that
\begin{equation}
\mbox{half S-brane} ~ \cong ~ \mbox{full S-brane with the
Hartle-Hawking contour} ~. \label{half full S}
\end{equation}
In our case, \eqn{symmetric D0} corresponds to the full S-brane,
while the Hartle-Hawking state \eqn{HH falling D0} (\eqn{HH emitted
D0}) is identifiable as the analogue of the half S-brane describing
unstable D-brane decay (creation) process. The equalities
\eqn{relation HH} suggest that we have roughly identical relation to
\eqn{half full S}.

Notice that the parameters $\rho_0$, $t_0$ appear just as phase
factors of boundary wave functions in \eqn{HH falling D0}, \eqn{HH
emitted D0} contrary to \eqn{falling D0}, \eqn{emitted D0}. Namely,
the choice of parameters $\rho_0$, $t_0$ does not cause any physical
difference for the Hartle-Hawking type states : They all can be
regarded as describing the D0-brane moving from ${\cal I}^-$ to
${\cal H}^+$ (from ${\cal H}^-$ to ${\cal I}^+$) for \eqn{HH falling
D0} (for \eqn{HH emitted D0}) irrespective of $\rho_0$, $t_0$. These
two parameters merely parameterize displacing the trajectory in
two-dimensional black hole background. Similar feature comes about
for the full S-brane with Hartle-Hawking contour as well: It is
equivalent to the half S-brane not depending on any shift of the
origin (the point connecting the real and imaginary times).

Finally, we remark a comment from the viewpoints of boundary
conformal field theory: in contrast to the original ones
\eqn{falling D0}, \eqn{emitted D0} and \eqn{symmetric D0}, the
Hartle-Hawking boundary states \eqn{relation HH} (or equivalently
\eqn{HH falling D0}, \eqn{HH emitted D0}) are not compatible with
the reflection relations. One may regard the boundary states
\eqn{falling D0} and \eqn{emitted D0} as the `completions' of the
Hartle-Hawking states \eqn{relation HH} so that they satisfy the
reflection relations.


~

\noindent
{\bf Note Added : } After this work was published, the work \cite{OR} was brought to our 
attention. However, the work \cite{OR} claimed results discrepant with ours 
concerning treatment of spectral amplitudes and the closed string
radiations. 
We made in another publication \cite{NRS2} further investigation from the
open string 
channel. The result of \cite{NRS2} completely supports the result given in 
this work, and clarifies where the errors originate in the work \cite{OR}. 


~


\acknowledgments{We thank Changrim Ahn, Dongsu Bak, Tohru Eguchi,
Yasuaki Hikida, Jaemo Park, Jongwon Park, Steve Shenker, Hiromitsu
Takayanagi, Tadashi Takayanagi and Jung-Tay Yee for useful
discussions. We are especially grateful to H.~Takayanagi for his
collaboration in the early stage of this work. SJR is supported in
part by the KRF BK-21 Physics Divison, KRF Leading Scientist Grant,
KOSEF SRC Program ``Center for Quantum Spacetime" (R11-2005-021), and
by F.W. Bessel Research Award from Alexander von Humboldt
Foundation. YN was supported in part by JSPS Research Fellowships
for Young Scientists. YS was supported by the Ministry of Education,
Culture, Sports, Science and Technology of Japan.}


\clearpage

\section*{Appendix A ~ Useful Formulae}
\setcounter{equation}{0}
\def\theequation{A.\arabic{equation}}

\begin{eqnarray}
&& \hspace{-5mm} \int_{-\frac{\pi}{2}}^{\frac{\pi}{2}} (2 \cos
\theta)^{a-1} e^{i b \theta} \dd \theta = \pi \frac{\Gamma(a)}
{\Gamma\left(\frac{1}{2}+\frac{a+b}{2}\right)
\Gamma\left(\frac{1}{2}+\frac{a-b}{2}\right)}~,~~~ (\Re\,
a>0~,~~\left|\Re\, b \right|< \Re\, a + 1)~,
\label{formula 1} \\
&& \hspace{-5mm} \int_{-\infty}^{\infty} (2 \cosh t)^{a-1} e^{i b t}
\dd t = \frac{1}{2} B\left(\frac{1}{2}-\frac{a+ib}{2},
\frac{1}{2}-\frac{a-ib}{2}\right) \equiv \frac{1}{2} \frac
{\Gamma\left(\frac{1}{2}-\frac{a+ib}{2}\right)
\Gamma\left(\frac{1}{2}-\frac{a-ib}{2}\right)} {\Gamma(1-a)}~,\nn &&
\hspace{10cm} (\Re\, a <1~,~~ \left|\Im \, b \right| < 1- \Re\, a)~.
\label{formula 2}
\end{eqnarray}
The integral \eqn{formula 2} follows from the more general formula:
\begin{eqnarray}
\int_0^{\infty} \frac{\cosh(2at)}{\cosh^{2\beta} (pt)} \dd t =
4^{\beta-1} p^{-1} B\left(\beta+\frac{a}{p},
\beta-\frac{a}{p}\right)~,~~~ (p>0,
~~\Re\,\left(\beta\pm\frac{a}{p}\right)>0)~,
\end{eqnarray}
given in \cite{transcendental}. It is also derivable from
\eqn{formula 1} by contour deformation, as was shown in \cite{NST}.


\section*{Appendix B ~ Proof of \eqn{evaluation overlap phi}\footnote{The results of this section is due to H. Takayanagi.}
}
\setcounter{equation}{0}
\def\theequation{B.\arabic{equation}}

Here we would like to evaluate explicitly the integral
\eqn{evaluation overlap phi} for any $\rho_0$ (strictly speaking, we
need to assume $\sinh \rho_0>1$). We begin with series expansion of
the hypergeometric function in $\phi^p_n(\rho,\theta)$:
\begin{eqnarray}
&&
F\left(\frac{1}{2}+\frac{ip+n}{2},\frac{1}{2}+\frac{ip-n}{2};
ip+1;
-\frac{\cos^2 \theta}{\sinh^2 \rho_0}\right)\nn
&&=\sum_{\ell=0}^\infty \frac{\Gamma(ip+1)}
{\Gamma(\frac{1}{2}+\frac{ip+n}{2})
\Gamma(\frac{1}{2}+\frac{ip-n}{2})}
\frac{\Gamma(\frac{1}{2}+\frac{ip+n}{2}+\ell)
\Gamma(\frac{1}{2}+\frac{ip-n}{2}+\ell)}
{\Gamma(ip+1+\ell)}\frac{(-1)^\ell}{\ell !}
\left(\frac{\cos \theta}{\sinh \rho_0}\right)^{2\ell}\ .
\label{expansion}
\end{eqnarray}
Using the formula (\ref{formula 1}), we can perform, in the
$\ell$-th sector, the integral (\ref{evaluation overlap phi}) as
\begin{eqnarray}
\Psi_\ell=g(\ell)\int_{-\frac{\pi}{2}}^{\frac{\pi}{2}} \dd\theta \,
e^{in\theta}(\cos \theta)^{ip-1+ 2\ell} =\frac{g(\ell)}{2^{ip-1
+2\ell}}\cdot \frac{\pi\Gamma(ip-1+2\ell+1)}
{\Gamma(\frac{1}{2}+\ell+\frac{ip+n}{2})
\Gamma(\frac{1}{2}+\ell+\frac{ip-n}{2})}\ , \nonumber
\end{eqnarray}
where $g(\ell)$ refers to
\begin{equation}
g(\ell)=\frac{(-1)^\ell}{\ell !}
(\sinh \rho_0)^{-ip -2\ell}\frac{\Gamma(ip+1)}
{\Gamma(\frac{1}{2}+\frac{ip+n}{2})
\Gamma(\frac{1}{2}+\frac{ip-n}{2})}
\frac{\Gamma(\frac{1}{2}+\frac{ip+n}{2}+\ell)
\Gamma(\frac{1}{2}+\frac{ip-n}{2}+\ell)}{\Gamma(ip+1+\ell)}\ .
\end{equation}
Then the total integral (\ref{evaluation overlap phi}) is
\begin{equation}
\sum_{\ell=0}^\infty \Psi_\ell
=\frac{\pi}{2^{ip-1}(\sinh \rho_0)^{ip}}\frac{\Gamma(ip+1)}
{\Gamma(\frac{1}{2}+\frac{ip+n}{2})
\Gamma(\frac{1}{2}+\frac{ip-n}{2})}\sum_{\ell=0}^\infty
\frac{(-1)^\ell}{\ell !}\frac{1}{2^{2\ell}(\sinh \rho_0)^{2\ell}}
\frac{\Gamma(ip+2\ell)}{\Gamma(ip+1+\ell)}\ .
\end{equation}
We can rewrite the summation into a hypergeometric function
by using
\begin{equation}
\Gamma(ip+2\ell)=\frac{2^{ip-1+2\ell}}{\sqrt{\pi}}
\Gamma\left(\frac{ip}{2}+\ell\right)
\Gamma\left(\frac{1}{2}+\frac{ip}{2}+\ell\right)~.
\end{equation}
and then obtain
\begin{equation}
\sum_{\ell=0}^\infty \Psi_\ell=\frac{\sqrt{\pi}}
{(\sinh \rho_0)^{ip}}\frac{\Gamma(\frac{ip}{2})
\Gamma(\frac{1}{2}+\frac{ip}{2})}
{\Gamma(\frac{1}{2}+\frac{ip+n}{2})
\Gamma(\frac{1}{2}+\frac{ip-n}{2})}
F\left(\f{ip}{2},\frac{1}{2}+\frac{ip}{2}; ip+1 ;
-\frac{1}{\sinh^2 \rho_0}\right)\ .
\label{calc1}
\end{equation}
Making use of the formula
\begin{eqnarray}
&&F\left(2\alpha,2\beta;\alpha+\beta+\frac{1}{2};z\right)=
F\left(\alpha,\beta;\alpha+\beta+\frac{1}{2};4z(1-z)\right)\ .
\label{calc2} \\
&& \hskip5cm |z|<\frac{1}{2},\quad |z(1-z)|<\frac{1}{4}~ \nonumber
\end{eqnarray}
we find that
\begin{equation}
\sum_{\ell=0}^\infty \Psi_\ell=\frac{\sqrt{\pi}}
{(\sinh \rho_0)^{ip}}\frac{\Gamma(\frac{ip}{2})
\Gamma(\frac{1}{2}+\frac{ip}{2})}
{\Gamma(\frac{1}{2}+\frac{ip+n}{2})
\Gamma(\frac{1}{2}+\frac{ip-n}{2})}
F\left(ip,ip+1;ip+1;\frac{1}{2}-
\frac{\cosh \rho_0}{2\sinh \rho_0}\right)\ .
\end{equation}
Note that the second and third arguments
of the hypergeometric function are the same.
The function is thus simplified as
\begin{equation}
F\left(ip,ip+1;ip+1;\frac{1}{2}-
\frac{\cosh \rho_0}{2\sinh \rho_0}\right)=
\left(\frac{\sinh \rho_0 +\cosh \rho_0}
{2\sinh \rho_0}\right)^{-ip}
=\left(2e^{-\rho_0}\sinh \rho_0\right)^{ip}~,
\end{equation}
because of the relation
\begin{equation}
(1-z)^{\alpha}=F(-\alpha,\beta;\beta;z)~.
\end{equation}
In this way, we finally obtain
\begin{eqnarray}
&& \sum_{\ell=0}^\infty \Psi_\ell=\sqrt{\pi}
e^{-ip \rho_0 }2^{ip}
\frac{\Gamma(\frac{ip}{2})\Gamma(\frac{1}{2}+\frac{ip}{2})}
{\Gamma(\frac{1}{2}+\frac{ip+n}{2})\Gamma(\frac{1}{2}
+\frac{ip-n}{2})}
= \frac{2\pi \,\Gamma(ip)}
{\Gamma(\frac{1}{2}+\frac{ip+n}{2})
\Gamma(\frac{1}{2}+\frac{ip-n}{2})}e^{-ip\rho_0}~,
\end{eqnarray}
and this is the desired formula.


\newpage



\begin{thebibliography}{99}
\small \baselineskip=16pt

\bibitem{Sen-RT}
A.~Sen,
JHEP {\bf 0204}, 048 (2002)
[arXiv:hep-th/0203211];
JHEP {\bf 0207}, 065 (2002)
[arXiv:hep-th/0203265];
Mod.\ Phys.\ Lett.\ A {\bf 17}, 1797 (2002)
[arXiv:hep-th/0204143].

\bibitem{2DBH}
E. Witten,
Phys. Rev. {\bf D44} (1991) 314;
G S. Elitzur, A. Forge and E. Rabinovici,
Nucl. Phys. {\bf B359} (1991) 581;
Mandal, A. Sengupta and S. Wadia,
Mod. Phys. Lett. {\bf A6} (1991) 1685;
I. Bars and D. Nemeschansky, 
Nucl. Phys. {\bf B348} (1991) 89.




\bibitem{DVV}
R.~Dijkgraaf, H.~Verlinde and E.~Verlinde,
Nucl.\ Phys.\ B {\bf 371}, 269 (1992).
\bibitem{Horowitz:1991cd}
  G.~T.~Horowitz and A.~Strominger,
  Nucl.\ Phys.\ B {\bf 360}, 197 (1991).



\bibitem{NS5brane}
S.~J.~Rey,
  Phys.\ Rev.\ D {\bf 43}, 526 (1991);
ibid. "{\sl Axionic String Instantons and Their Low-Energy
Implications}", UCSB-TH-89/49, in the proceedings of the "Workshop
on Superstrings and Particle Theory" pp. 291-300 (World Scientific
Pub., 1989).

\bibitem{CHS}
C.~G.~.~Callan, J.~A.~Harvey and A.~Strominger,
Nucl.\ Phys.\ B {\bf 359}, 611 (1991);
S.~J. Rey,  "{\sl On String Theory and Axionic Strings and
Instantons}", SLAC-PUB-5659, in the proceedings of "Particle and
Fields '91 Conference" pp. 876-881 (American Physical Society,
1991).

\bibitem{Kutasov}
D.~Kutasov,
arXiv:hep-th/0405058.

\bibitem{Kutasov2}
  D.~Kutasov,
  arXiv:hep-th/0408073.

\bibitem{Strominger}
A.~Strominger,
arXiv:hep-th/0209090.

\bibitem{LNT}
  F.~Larsen, A.~Naqvi and S.~Terashima,
  JHEP {\bf 0302}, 039 (2003)
  [arXiv:hep-th/0212248].


\bibitem{radion-DBI}
  H.~Yavartanoo,
  arXiv:hep-th/0407079;
  K.~L.~Panigrahi,
  Phys.\ Lett.\ B {\bf 601}, 64 (2004)
  [arXiv:hep-th/0407134];
  A.~Ghodsi and A.~E.~Mosaffa,
  Nucl.\ Phys.\ B {\bf 714}, 30 (2005)
  [arXiv:hep-th/0408015];
  J.~Kluson,
  JHEP {\bf 0411}, 013 (2004)
  [arXiv:hep-th/0409298];
  arXiv:hep-th/0501010;
  B.~Chen and B.~Sun,
  arXiv:hep-th/0501176;
  D.~Bak, S.~J.~Rey and H.~U.~Yee,
  JHEP {\bf 0412}, 008 (2004)
  [arXiv:hep-th/0411099];
  S.~Thomas and J.~Ward,
  JHEP {\bf 0502}, 015 (2005)
  [arXiv:hep-th/0411130];
  arXiv:hep-th/0501192,
  arXiv:hep-th/0502228;
  arXiv:hep-th/0504226;
  W.~H.~Huang,
  JHEP {\bf 0502}, 061 (2005)
  [arXiv:hep-th/0502023];
  J.~Kluson and K.~L.~Panigrahi,
  arXiv:hep-th/0506012.

\bibitem{Bars}
  I.~Bars and K.~Sfetsos,
  Phys.\ Rev.\ D {\bf 46}, 4495 (1992)
  [arXiv:hep-th/9205037];
  I.~Bars and J.~Schulze,
  Phys.\ Rev.\ D {\bf 51}, 1854 (1995)
  [arXiv:hep-th/9405156].




\bibitem{NST}
Y.~Nakayama, Y.~Sugawara and H.~Takayanagi,
JHEP {\bf 0407}, 020 (2004)
[arXiv:hep-th/0406173].
\bibitem{NPRT}
  Y.~Nakayama, K.~L.~Panigrahi, S.~J.~Rey and H.~Takayanagi,
  JHEP {\bf 0501}, 052 (2005)
  [arXiv:hep-th/0412038].
\bibitem{Sahakyan}
  D.~A.~Sahakyan,
  JHEP {\bf 0410}, 008 (2004)
  [arXiv:hep-th/0408070].


\bibitem{CLS}
  B.~Chen, M.~Li and B.~Sun,
  JHEP {\bf 0412}, 057 (2004)
  [arXiv:hep-th/0412022];

\bibitem{LapanLi}
  J.~M.~Lapan and W.~Li,
  arXiv:hep-th/0501054.


\bibitem{MS-blackNS5}
  J.~M.~Maldacena and A.~Strominger,
  JHEP {\bf 9712}, 008 (1997)
  [arXiv:hep-th/9710014];


\bibitem{RibS}
S.~Ribault and V.~Schomerus,
JHEP {\bf 0402}, 019 (2004)
[arXiv:hep-th/0310024].
\bibitem{KarMS}
J.~L.~Karczmarek, J.~Maldacena and A.~Strominger,
arXiv:hep-th/0411174.

\bibitem{GKRS}
  A.~Giveon, D.~Kutasov, E.~Rabinovici and A.~Sever,
  arXiv:hep-th/0503121.

\bibitem{maldacena}
J.~Maldacena,
  arXiv:hep-th/0503112.



\bibitem{Teschner-reflection}
J.~Teschner,
Nucl.\ Phys.\ B {\bf 546}, 390 (1999)
[arXiv:hep-th/9712256];
Nucl.\ Phys.\ B {\bf 571}, 555 (2000)
[arXiv:hep-th/9906215].


\bibitem{GK}
A.~Giveon and D.~Kutasov,
JHEP {\bf 9910}, 034 (1999)
[arXiv:hep-th/9909110];
A.~Giveon and D.~Kutasov,
JHEP {\bf 0001}, 023 (2000)
[arXiv:hep-th/9911039].


\bibitem{BP}
C.~Bachas and M.~Petropoulos,
JHEP {\bf 0102}, 025 (2001)
[arXiv:hep-th/0012234].

\bibitem{PST}
B.~Ponsot, V.~Schomerus and J.~Teschner,
JHEP {\bf 0202}, 016 (2002)
[arXiv:hep-th/0112198];
A.~Giveon, D.~Kutasov and A.~Schwimmer,
Nucl.\ Phys.\ B {\bf 615}, 133 (2001)
[arXiv:hep-th/0106005];
P.~Lee, H.~Ooguri and J.~W.~Park,
Nucl.\ Phys.\ B {\bf 632}, 283 (2002)
[arXiv:hep-th/0112188].

\bibitem{LVZ}
S.~L.~Lukyanov, E.~S.~Vitchev and A.~B.~Zamolodchikov,
Nucl.\ Phys.\ B {\bf 683}, 423 (2004)
[arXiv:hep-th/0312168].


\bibitem{ES-L}
T.~Eguchi and Y.~Sugawara,
JHEP {\bf 0401}, 025 (2004)
[arXiv:hep-th/0311141].

\bibitem{ASY}
C.~Ahn, M.~Stanishkov and M.~Yamamoto,
Nucl.\ Phys.\ B {\bf 683}, 177 (2004)
[arXiv:hep-th/0311169].

\bibitem{IPT2}
  D.~Israel, A.~Pakman and J.~Troost,
  Nucl.\ Phys.\ B {\bf 710}, 529 (2005)
  [arXiv:hep-th/0405259].


\bibitem{FNP}
  A.~Fotopoulos, V.~Niarchos and N.~Prezas,
  Nucl.\ Phys.\ B {\bf 710}, 309 (2005)
  [arXiv:hep-th/0406017].

\bibitem{Nakayama}
  Y.~Nakayama,
  Int.\ J.\ Mod.\ Phys.\ A {\bf 19}, 2771 (2004)
  [arXiv:hep-th/0402009].

\bibitem{Yogendran}
  K.~P.~Yogendran,
  JHEP {\bf 0501}, 036 (2005)
  [arXiv:hep-th/0408114].


\bibitem{GS1}
M.~Gutperle and A.~Strominger,
JHEP {\bf 0204}, 018 (2002)
[arXiv:hep-th/0202210].


\bibitem{excitedDdecay}
C.~G.~.~Callan and J.~M.~Maldacena,
  Nucl.\ Phys.\ B {\bf 472}, 591 (1996)
  [arXiv:hep-th/9602043];
S.~R.~Das and S.~D.~Mathur,
  Nucl.\ Phys.\ B {\bf 478}, 561 (1996)
  [arXiv:hep-th/9606185].






\bibitem{LLM}
  N.~Lambert, H.~Liu and J.~Maldacena,
  arXiv:hep-th/0303139.

\bibitem{KLMS}
J.~L.~Karczmarek, H.~Liu, J.~Maldacena and A.~Strominger,
JHEP {\bf 0311}, 042 (2003)
[arXiv:hep-th/0306132].

\bibitem{BH}
G.~'t Hooft,
Print-87-0775 (UTRECHT)
{\it Lectures given at Cargese Summer Inst. on Nonperturbative
Quantum Field Theory, Cargese, France, Jul 16-30, 1987};
C.~F.~E.~Holzhey and F.~Wilczek,
  Nucl.\ Phys.\ B {\bf 380}, 447 (1992)
  [arXiv:hep-th/9202014];
G.~T.~Horowitz and J.~Polchinski,
  Phys.\ Rev.\ D {\bf 55}, 6189 (1997)
  [arXiv:hep-th/9612146].

\bibitem{GKPS}
A.~Giveon, A.~Konechny, A.~Pakman and A.~Sever,
JHEP {\bf 0310}, 025 (2003)
[arXiv:hep-th/0309056].

\bibitem{KutS}
D.~Kutasov and N.~Seiberg,
Nucl.\ Phys.\ B {\bf 358}, 600 (1991).


\bibitem{thermal open}
  Y.~Sugawara,
  Nucl.\ Phys.\ B {\bf 650}, 75 (2003)
  [arXiv:hep-th/0209145],
  JHEP {\bf 0308}, 008 (2003)
  [arXiv:hep-th/0307034].


\bibitem{thermal closed}
J.~Polchinski,
Commun.\ Math.\ Phys.\  {\bf 104}, 37 (1986);
B.~Sathiapalan,
Phys.\ Rev.\ D {\bf 35}, 3277 (1987);
Y.~I.~Kogan,
JETP Lett.\  {\bf 45}, 709 (1987)
[Pisma Zh.\ Eksp.\ Teor.\ Fiz.\  {\bf 45}, 556 (1987)];
K.~H.~O'Brien and C.~I.~Tan,
Phys.\ Rev.\ D {\bf 36}, 1184 (1987);
J. J. Atick and E. Witten,
Nucl. Phys. B {\bf 310}, 291 (1988).



\bibitem{Kutasov:1990ua}
  D.~Kutasov and N.~Seiberg,
  Phys.\ Lett.\ B {\bf 251}, 67 (1990).


\bibitem{FZZ2}
V.~Fateev, A.~B.~Zamolodchikov and Al.~B.~Zamolodchikov,
unpublished;
  V.~Kazakov, I.~K.~Kostov and D.~Kutasov,
  Nucl.\ Phys.\ B {\bf 622}, 141 (2002)
  [arXiv:hep-th/0101011].

\bibitem{HK1}
  K.~Hori and A.~Kapustin,
  JHEP {\bf 0108}, 045 (2001)
  [arXiv:hep-th/0104202].


\bibitem{Friess:2004tq}
  J.~J.~Friess and H.~Verlinde,
  arXiv:hep-th/0411100.


\bibitem{KMS}
  I.~R.~Klebanov, J.~Maldacena and N.~Seiberg,
  JHEP {\bf 0307}, 045 (2003)
  [arXiv:hep-th/0305159].


\bibitem{GKP}
A.~Giveon, D.~Kutasov and O.~Pelc,
JHEP {\bf 9910}, 035 (1999)
[arXiv:hep-th/9907178].

\bibitem{AGK}
O.~Aharony, A.~Giveon and D.~Kutasov,
Nucl.\ Phys.\ B {\bf 691}, 3 (2004)
[arXiv:hep-th/0404016].

\bibitem{ES-BH}
T.~Eguchi and Y.~Sugawara,
JHEP {\bf 0405}, 014 (2004)
[arXiv:hep-th/0403193].

\bibitem{ES-con}
  T.~Eguchi and Y.~Sugawara,
  JHEP {\bf 0501}, 027 (2005)
  [arXiv:hep-th/0411041].



\bibitem{ABKS}
O.~Aharony, M.~Berkooz, D.~Kutasov and N.~Seiberg,
JHEP {\bf 9810}, 004 (1998)
[arXiv:hep-th/9808149].


\bibitem{HarO}
  T.~Harmark and N.~A.~Obers,
  Phys.\ Lett.\ B {\bf 485}, 285 (2000)
  [arXiv:hep-th/0005021].

\bibitem{BerRoz}
  M.~Berkooz and M.~Rozali,
  JHEP {\bf 0005}, 040 (2000)
  [arXiv:hep-th/0005047].

\bibitem{KutSah}
  D.~Kutasov and D.~A.~Sahakyan,
  JHEP {\bf 0102}, 021 (2001)
  [arXiv:hep-th/0012258].


\bibitem{OR}
  K.~Okuyama and M.~Rozali,
  arXiv:hep-th/0602060.
  JHEP {\bf 0603}, 071 (2006)
  [arXiv:hep-th/0602060].

\bibitem{NRS2}
  Y.~ Nakayama, S.~J.~Rey and Y. ~Sugawara,
  [arXiv:hep-th/0605013].



\bibitem{transcendental}
 H. Bateman,
{\em ``Higher Transcendental Functions,''} Vol. 1.
McGRAW-Hill BOOK COMPANY, INC. (1953)





\end{thebibliography}
\end{document}